%% file: krw_jhep_simple.tex
\documentclass[a4paper,11pt]{article}
\usepackage{jheppub}
\usepackage{amsthm,bm}
\usepackage{booktabs}
\usepackage{adjustbox}
\usepackage{tikz}
\usetikzlibrary{arrows.meta,positioning,calc,decorations.pathreplacing}
\usepackage{xcolor}
\newcommand{\wl}{\mathsf{w}}   
\allowdisplaybreaks

\title{A lattice construction of the Kormos--Runkel--Watts defect flows in minimal models}

\author[a]{Junchen Rong}
\affiliation[a]{CPHT, CNRS, \'Ecole Polytechnique, Institut Polytechnique de Paris, Palaiseau, France}
\emailAdd{junchen.rong@polytechnique.edu}

\abstract{Kormos, Runkel and Watts (KRW) showed that a topological defect of a diagonal minimal model $M(p,p+1)$, perturbed by a weakly relevant defect field, flows to a new conformal interface. We study the flow of the $(2,1)$ Verlinde defect $D_{(2,1)}$ driven by the spinful defect field $\phi_{(1,3)}\otimes\bar\phi_{(3,3)}$ on the lattice. 
The flow preserves the $(r,1)$ Verlinde lines. 
This preserved symmetry makes a dilute RSOS realisation possible. We study this flow numerically using both exact diagonalization and density matrix renormalization group (DMRG). 
At $p=4$, the flow ends on the invertible line $\eta=D_{(3,1)}$.
At $p=5$ it ends on a non-trivial conformal interface $C_5$ with $g=1.503(2)$ and transmission $\mathcal T=0.665(10)$, whose spectrum, leading irrelevant dimension and lattice $\beta$-function we report. Gauging $\eta$ maps $C_5$ to an interface of the three-state Potts model, which breaks $\mathbb{Z}_3$ but is a stable interface once the Fibonacci line $\tau_{\mathrm{Fib}}$ and charge conjugation are imposed. 
A loop-model formulation allows us to follow the family $C_p$ at continuous $4\le p\le5$: it moves smoothly from $C_5$ to $\eta$ as we lower $p$.}

\keywords{Conformal Field Theory, Conformal Interface, Renormalization Group}

\begin{document}
\maketitle
\section{Introduction}\label{sec:intro}

Topological defect lines are by now a standard tool in two-dimensional conformal field theory: they encode the symmetries of a theory, including the non-invertible ones, they act on bulk and boundary fields, and they constrain renormalisation-group flows \cite{ref2,ref3,ref4,ref5,ref13}. 
In a diagonal minimal model $M(p,p+1)$ the topological defects are the Verlinde lines $D_{(r,s)}$, labelled like the primary fields, and they form a fusion category. Kormos, Runkel and Watts (KRW) \cite{ref1} asked what happens when such a line is perturbed by one of its own relevant defect fields: the defect is no longer topological, and it flows, in the manner of a boundary RG flow, to a new conformal interface, which may be topological again or may be a genuinely transmitting, non-topological defect. 
Perturbed topological defects of minimal models have been studied from several angles---chiral perturbations, which preserve pure transmission while generally breaking conformal invariance away from fixed points \cite{ref8,ref9}; RG interfaces between the UV and IR theories of a bulk flow \cite{ref10,ref11}; the defect flows of the tricritical Ising model \cite{ref12}---but the non-topological endpoints of the KRW flows are known only through perturbation theory in $1/p$, truncated conformal space approach \cite{ref1,ref84,ref85} and through level-truncated string field theory \cite{ref86}. A first-principles, non-perturbative lattice determination of these interfaces has been missing.

In this paper, we study the flow of the $(2,1)$ defect $D_{(2,1)}$ driven by the non-chiral, spinful defect field $\phi_{(1,3)}\otimes\bar\phi_{(3,3)}$, at $p=4$ (tricritical Ising, $c=7/10$), at $p=5$ ($M(5,6)$, $c=4/5$), and at continuous $p$ in between. In the two-parameter space of KRW, $D_{(1,2)}$ perturbed by $\lambda_l\,\phi_{(1,3)}\otimes\mathbf{1}+\lambda_r\,\mathbf{1}\otimes\bar\phi_{(1,3)}$, the defect $D_{(2,1)}$ is the endpoint of the purely chiral flow and $\phi_{(1,3)}\otimes\bar\phi_{(3,3)}$ is the first-order image of the anti-chiral field along it; perturbing $D_{(2,1)}$ by it with one sign leads to the new conformal defect $C$ (the endpoint of the diagonal $\lambda_l=\lambda_r>0$ flow from $D_{(1,2)}$), with the other sign to the identity defect. Our main results are the following. At $p=5$ the flow ends on the genuinely conformal interface $C_5$, whose $g$-function, transmission, operator content and leading irrelevant coupling we determine; this is the first non-perturbative lattice determination of a KRW defect. 
At $p=4$, where KRW's truncated-conformal-space analysis could not locate the fixed point, we find that the flow ends on the invertible topological line $\eta=D_{(3,1)}$, an endpoint that had not been proposed before, so that the tricritical Ising model has no non-topological KRW defect at all. 
Following the family $C_p$ at continuous $p$ then shows how the two cases are connected: $C_p$ moves smoothly from $C_5$ to $\eta$ as $p\to4$. Notice the $\eta=D_{(1,4)}$ at $p=4$. The $D_{(1,4)}$ interface exists at any p. When $p\rightarrow 4$, the $C_p$ interface merges with $D_{(1,4)}$ because the relevant operator driving the RG flow between them becomes a null-state.

Two facts make a lattice construction possible, and they are the two themes of the paper. The first is a property of the flow itself: it preserves the $(r,1)$ Verlinde lines of $M(p,p+1)$, which form a fusion category $\mathcal C_p$ with the fusion rules of $SU(2)_{p-2}$ \cite{ref5,ref13}\footnote{The fusion rules are those of $SU(2)_{p-2}$; the $F$-symbols, braiding and twists of the $(r,1)$ lines differ from those of the $SU(2)_{p-2}$ Wess--Zumino--Witten category by a $\mathbb Z_2$ twist (at $p=4$ they are those of the Ising fusion category, not of $SU(2)_2$). The categorical conventions and the braid and twist data used here are described in \S\ref{sec:2.2}.} ($\{\mathbf{1},\sigma'=D_{(2,1)},\eta=D_{(3,1)}\}$, the Ising category, at $p=4$; $\mathrm{Fib}\boxtimes\mathbb{Z}_2$ with $D_{(2,1)}=\wl=\tau_{\mathrm{Fib}}\otimes\eta$ of quantum dimension $\varphi=\tfrac{1+\sqrt5}{2}$ at $p=5$). We make this statement precise in Section~\ref{sec:cft}: a line $x$ wrapping the defect has to be attached to it by a junction, and since $x\otimes D_{(2,1)}$ has two fusion channels there are two inequivalent ways of doing so, the over- and under-crossing ``lassos''. The deformation selects one of them---the lasso that reads the holomorphic label of the defect fields, under which $\phi_{(1,3)}\otimes\bar\phi_{(3,3)}$ is neutral---and we explain, in the three-dimensional Reshetikhin--Turaev description of the theory on $\Sigma\times[-1,1]$, why an over-crossing reads the holomorphic label and an under-crossing the antiholomorphic one. This is a point that does not arise in KRW's original setting, where the flow starts from $D_{(1,2)}$ and the junction space is one-dimensional. The second fact is a property of the lattice: the dilute $A_{p-1}$ RSOS chain \cite{ref14,ref15} is precisely an \emph{anyon chain} \cite{ref16,ref17} of $\mathcal C_p$, for every integer $p$ considered here. Its height variables are $SU(2)_{p-2}$ labels, its domain walls are $\wl$ anyons\footnote{Throughout we write $\wl$ for the $(2,1)$ anyon, the generating object of $\mathcal C_p$, labelled by spin $\tfrac12$ in the $SU(2)_{p-2}$ fusion rules; the symbol $\tau$ is reserved for the Fibonacci anyon $\tau_{\mathrm{Fib}}$, which at $p=5$ is a different object ($D_{(2,1)}=\tau_{\mathrm{Fib}}\otimes\eta$).}, and its Hamiltonian is built from the fusion data. A $(r,1)$ defect line is therefore just one more anyon of the same category threaded through the chain, and it can be made \emph{exactly} topological on the lattice by coupling it to the chain with the same $F$- and $R$-moves that define the category, in the spirit of the lattice topological defects of \cite{ref20,ref21} and of the Temperley--Lieb constructions of \cite{ref22,ref23,ref24}. (For the special case $p=4$, where $\mathcal C_4$ is the Ising fusion category, a closely related one-dimensional model built directly on the fusion spaces of the Ising category, with an explicit vacancy label and the categorical symmetry realised exactly, has recently been studied by Roychowdhury and Wang \cite{ref18} within the fusion-category lattice models of \cite{ref19}.) Because the symmetry preserved along the flow is the one the lattice realises exactly, the whole flow can be followed with the symmetry imposed, sector by sector; and the perturbing field has a transparent meaning in this language, as a quantum-dimension-weighted combination of the projectors onto the two fusion channels of the defect anyon and its neighbouring domain wall.

The paper is organised as follows. Section~\ref{sec:cft} reviews the KRW flows and establishes, in the continuum, which symmetry the flow $D_{(2,1)}\to C$ preserves and why. Section~\ref{sec:lattice} builds the lattice model: the dilute height chain, the topological seam, the lattice symmetry operators, the clean deformation operator, the symmetry-resolved numerics and the observables used to characterise an interface. Sections~\ref{sec:4} and \ref{sec:5} report the results at $p=4$ and $p=5$. Section~\ref{sec:loop} reformulates the construction as a loop model valid at continuous $p$, and Section~\ref{sec:nonint} follows the family $C_p$ for $4\le p\le5$ and discusses its endpoint at $p=4$. Section~\ref{sec:concl} summarises and compares with conformal perturbation theory, integrable defect flows and string field theory.

\section{The KRW flow and the symmetry it preserves}\label{sec:cft}

\subsection{Perturbed topological defects: the work of Kormos, Runkel and Watts}\label{sec:krw}

A topological defect $D$ of a conformal field theory is a line that commutes with both copies of the Virasoro algebra, $[L_n,D]=[\bar L_n,D]=0$; it can be deformed freely as long as it does not cross field insertions, and its action on the bulk fields is encoded in a defect operator on the Hilbert space of the theory. In a diagonal minimal model the elementary topological defects are the Verlinde lines $D_{(r,s)}$ \cite{ref2,ref3}, one for each Kac label, which act on the primary field $\phi_{(r',s')}$ by the Verlinde ratio $S_{(r,s),(r',s')}/S_{(1,1),(r',s')}$ and fuse according to the minimal-model fusion rules. Their quantum dimension $d_{(r,s)}=S_{(r,s),(1,1)}/S_{(1,1),(1,1)}$ is the $g$-function of the line \cite{ref42}, and only the $(1,1)$ line and the invertible line $\eta=D_{(p-1,1)}\cong D_{(1,p)}$ have $g=1$. A defect $D$ also carries its own fields, the \emph{defect fields}, which live on the line; their space is the twisted sector of $D\otimes\bar D$ (Section~\ref{sec:defops}), and it contains in general chiral fields of weight $(h,0)$ and $(0,h)$ as well as non-chiral ones.

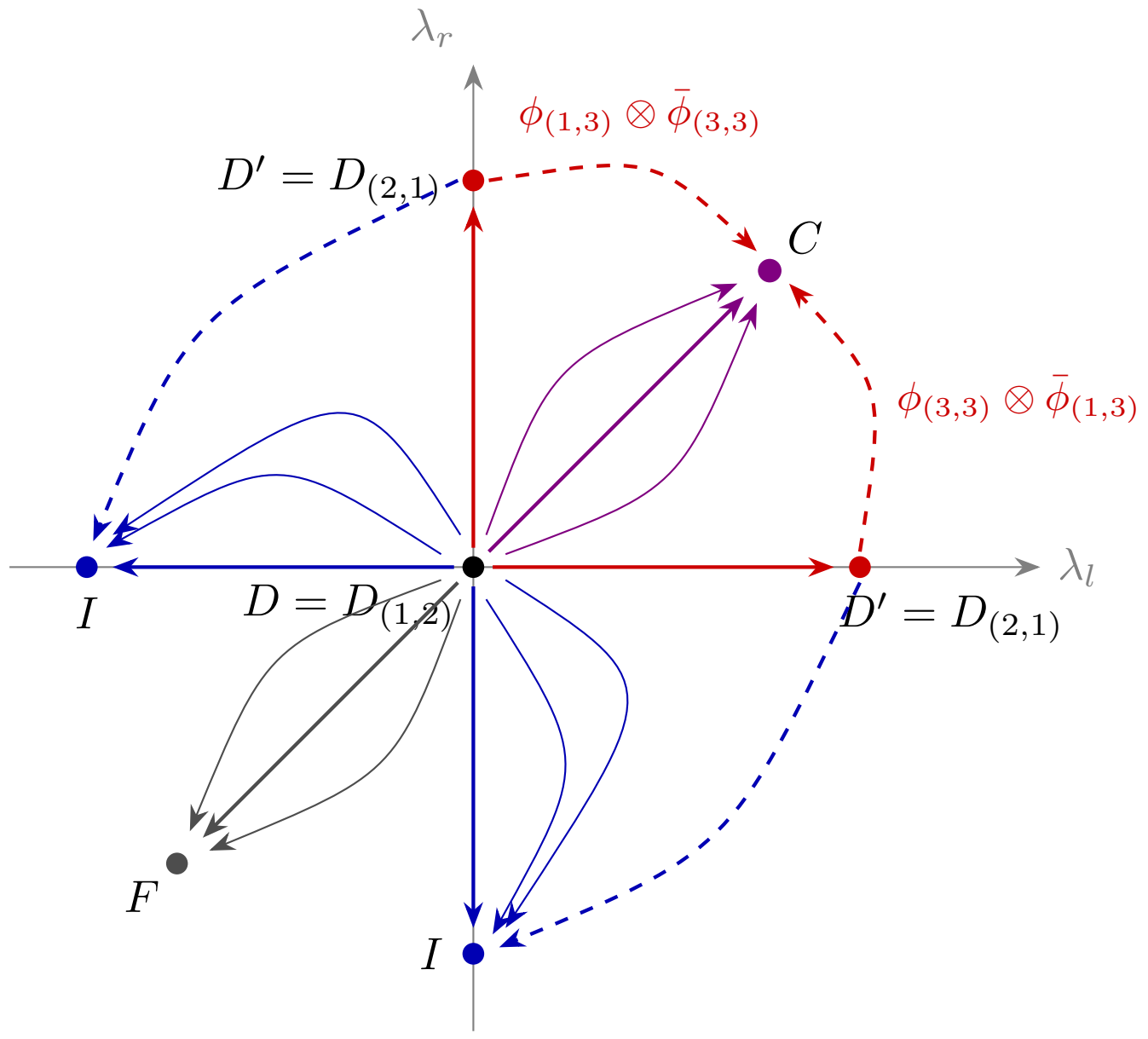
\begin{figure}[ht]
\centering
\begin{tikzpicture}[x=1cm,y=1cm, every node/.style={font=\small}, >={Stealth[length=2mm]}]
  \draw[->, gray] (-3.6,0) -- (4.4,0) node[right] {$\lambda_l$};
  \draw[->, gray] (0,-3.6) -- (0,3.9) node[above left] {$\lambda_r$};
  \fill[black] (0,0) circle (2.4pt); \node[below left] at (0,0) {$D=D_{(1,2)}$};
  \fill[red!80!black] (3.0,0) circle (2.4pt); \node[anchor=north west, inner sep=2pt] at (2.75,-0.12) {$D'=D_{(2,1)}$};
  \fill[red!80!black] (0,3.0) circle (2.4pt); \node[left] at (-0.1,3.0) {$D'=D_{(2,1)}$};
  \fill[blue!70!black] (-3.0,0) circle (2.4pt); \node[below] at (-3.0,-0.1) {$I$};
  \fill[blue!70!black] (0,-3.0) circle (2.4pt); \node[left] at (-0.1,-3.0) {$I$};
  \fill[violet] (2.3,2.3) circle (2.6pt); \node[above right] at (2.3,2.3) {$C$};
  \fill[gray!60!black] (-2.3,-2.3) circle (2.4pt); \node[below left] at (-2.3,-2.3) {$F$};
  \draw[->, red!80!black, thick] (0.15,0) -- (2.8,0);
  \draw[->, red!80!black, thick] (0,0.15) -- (0,2.8);
  \draw[->, blue!70!black, thick] (-0.15,0) -- (-2.8,0);
  \draw[->, blue!70!black, thick] (0,-0.15) -- (0,-2.8);
  \draw[->, violet, thick] (0.12,0.12) -- (2.1,2.1);
  \draw[->, violet] (0.1,0.25) .. controls (0.6,1.6) .. (2.05,2.2);
  \draw[->, violet] (0.25,0.1) .. controls (1.6,0.6) .. (2.2,2.05);
  \draw[->, red!80!black, thick, dashed] (3.0,0.12) .. controls (3.2,1.4) .. (2.45,2.2);
  \draw[->, red!80!black, thick, dashed] (0.12,3.0) .. controls (1.4,3.2) .. (2.2,2.45);
  \node[red!80!black, anchor=west, font=\footnotesize] at (3.15,1.3) {$\phi_{(3,3)}\otimes\bar\phi_{(1,3)}$};
  \node[red!80!black, anchor=south, font=\footnotesize] at (1.3,3.2) {$\phi_{(1,3)}\otimes\bar\phi_{(3,3)}$};
  \draw[->, blue!70!black] (-0.25,0.1) .. controls (-1.5,0.9) .. (-2.85,0.15);
  \draw[->, blue!70!black] (-0.1,0.25) .. controls (-0.9,1.5) .. (-2.8,0.25);
  \draw[->, blue!70!black] (0.25,-0.1) .. controls (1.5,-0.9) .. (0.25,-2.8);
  \draw[->, blue!70!black] (0.1,-0.25) .. controls (0.9,-1.5) .. (0.15,-2.85);
  \draw[->, blue!70!black, thick, dashed] (3.0,-0.12) .. controls (2.0,-2.2) .. (0.2,-2.95);
  \draw[->, blue!70!black, thick, dashed] (-0.12,3.0) .. controls (-2.2,2.0) .. (-2.95,0.2);
  \draw[->, gray!60!black, thick] (-0.12,-0.12) -- (-2.1,-2.1);
  \draw[->, gray!60!black] (-0.1,-0.25) .. controls (-0.6,-1.6) .. (-2.05,-2.2);
  \draw[->, gray!60!black] (-0.25,-0.1) .. controls (-1.6,-0.6) .. (-2.2,-2.05);
\end{tikzpicture}
\caption{The space of flows of the $(1,2)$ defect perturbed by $\lambda_l\phi+\lambda_r\bar\phi$, after figure~1 of Kormos, Runkel and Watts \cite{ref1} (schematic; $p>3$). The chiral flows along the axes end on the topological defects $D'=D_{(2,1)}$ (positive couplings) and $I$ (negative couplings). All flows in the first quadrant end on the new conformal defect $C$, those in the second and fourth quadrants on $I$, and the negative diagonal on the factorising defect $F$. The dashed lines are the boundaries of the basins: flows from $D_{(2,1)}$ to $C$, driven by the spinful field $\phi_{(1,3)}\otimes\bar\phi_{(3,3)}$ or its mirror image, and flows from $D_{(2,1)}$ to $I$. The flow $D_{(2,1)}\to C$ along the dashed boundary of the first quadrant is the one studied in this paper.}
\label{fig:krwplane}
\end{figure}

Kormos, Runkel and Watts \cite{ref1} considered the perturbation of a topological defect by its relevant defect fields and asked where the resulting flow ends. Their starting point is the defect $D_{(1,2)}$ of $M(p,p+1)$, whose relevant defect fields are the two chiral fields $\phi=\phi_{(1,3)}\otimes\mathbf 1$ and $\bar\phi=\mathbf 1\otimes\bar\phi_{(1,3)}$ of weights $(h_{1,3},0)$ and $(0,h_{1,3})$, $h_{1,3}=(p-1)/(p+1)$, so that the perturbed defect is
\begin{equation}
D_{(1,2)}(\lambda_l,\lambda_r)=D_{(1,2)}\,\exp\Big(\lambda_l\!\int\!\phi+\lambda_r\!\int\!\bar\phi\Big),
\label{eq:krwdef}
\end{equation}
with $y=1-h_{1,3}=2/(p+1)$ the RG eigenvalue of both couplings. Their analysis of this two-parameter space of flows combines three tools: exact arguments for the chiral flows, conformal perturbation theory in $y$, and the truncated conformal space approach for the non-chiral directions. The resulting flow diagram, their figure~1, is reproduced schematically in Fig.~\ref{fig:krwplane}, and three of its features organise everything that follows.

First, a purely chiral perturbation ($\lambda_r=0$ or $\lambda_l=0$) preserves pure transmission and the chiral algebra on the unperturbed side \cite{ref8,ref9}, but need not preserve conformal invariance along the flow. A conformal endpoint of such a flow is topological; matching the action on the boundary state $\|1,1\rangle\!\rangle$ fixes the endpoints (\cite{ref1}, \S3.1): the positive chiral flows of $D_{(1,2)}$ end on $D'=D_{(2,1)}$ and the negative ones on the identity defect $I$. The chiral flows have recently been put on a systematic footing in \cite{refAKRW, refAKRW2}. 
In the present paper the perturbation is non-chiral and the endpoint is a genuinely conformal defect (or, at $p=4$, a topological line reached by a non-chiral flow), a situation outside the scope of that analysis. Second, a non-chiral perturbation, $\lambda_l\lambda_r\ne0$, destroys the commutation with both Virasoro algebras and produces a genuinely conformal, non-topological interface. KRW find that every flow in the second and fourth quadrants ends on $I$, that the negative diagonal $\lambda_l=\lambda_r<0$ ends on a factorising defect $F$, the superposition of the $(r,1)$ boundary conditions, and that the whole first quadrant flows to a new conformal defect $C$, the subject of this paper; near $c=1$ they thus find six fixed points, five of which are known defects and one, $C$, a new one (\cite{ref1}, abstract and \S6). This picture is established for large $p$ and confirmed by TCSA down to $M(5,6)$; for $M(4,5)$ they could not locate the non-chiral fixed point and left its nature open (\cite{ref1}, \S6.2.4). For $C$ they computed the $g$-function and the reflection coefficient in conformal perturbation theory in powers of $y$, i.e.\ at large $p$: $\ln g_C-\ln g_{D_{(2,1)}}=-\tfrac{\pi^2}{4}y^3+O(y^4)$ \cite{ref1,ref86} and $R_C=1-\mathcal T=\tfrac98\pi^2y^4+O(y^5)$ \cite{ref84}. Third, and most relevant here, the boundaries of these basins of attraction are themselves flows out of $D_{(2,1)}$: the boundary of the first quadrant consists of flows from $D'$ to $C$, generated by the spinful defect field $\phi_{(1,3)}\otimes\bar\phi_{(3,3)}$ on the side $\lambda_r>\lambda_l$ and by its mirror image $\phi_{(3,3)}\otimes\bar\phi_{(1,3)}$ on the side $\lambda_l>\lambda_r$, while the boundary of the second and fourth quadrants is a flow from $D_{(2,1)}$ to $I$ (\cite{ref1}, \S6.2.3). In other words, once $D_{(1,2)}$ has reached $D_{(2,1)}$ along a chiral axis, the remaining perturbation is the first-order image of the other chiral field, the spinful field $\phi_{(1,3)}\otimes\bar\phi_{(3,3)}$ of dimension $\Delta=h_{1,3}+h_{3,3}=1-2(p-1)/(p(p+1))$, and $D_{(2,1)}$ perturbed by it flows to $C$ for one sign of the coupling and to $I$ for the other. It is this flow, $D_{(2,1)}\to C$, that we realise on the lattice.

KRW also observed that the perturbed defects \eqref{eq:krwdef} commute with the topological lines $D_{(r,1)}$ for all values of the couplings (their \S2.4.2), and used this commutation to constrain the possible infrared fixed points (their \S3.2). For the flow out of $D_{(2,1)}$ the corresponding statement is both true and subtle, and since it is what allows the lattice construction we devote the rest of this section to it: the preserved symmetry is not ``the line $D_{(r,1)}$'' but a specific way of attaching that line to the defect, which the deformation selects.

\subsection{Defect operators of $D_{(2,1)}$}\label{sec:defops}

In radial quantisation about a point of the ring, the states of a ring crossed by a line $D$ at two antipodal points are the operators sitting on a straight $D$ line through the origin, i.e. the \emph{defect operators} of $D$. For a topological $D$ this space is \cite{ref3,ref5}
\begin{equation}
\mathcal{H}_{D \times D} = \bigoplus_k N_{DD}^{k}\, \mathcal{H}_k, \qquad
\mathcal{H}_k = \bigoplus_{i,j} N_{ki}^{j}\, V_i \otimes \bar V_j,
\label{eq:23}
\end{equation}
where $\mathcal{H}_k$ is the $k$-twisted sector of the bulk theory ($\mathcal{H}_{\mathbf{1}}$ is the diagonal spectrum), $i = (r',s)$, $j = (r'',s)$ label Virasoro representations and $L$ are the fusion multiplicities. For $D = D_{(2,1)}$, $D \times D = D_{(1,1)} + D_{(3,1)}$, so the ring carries $\mathcal{H}_{\mathbf{1}} \oplus \mathcal{H}_{(3,1)}$: the diagonal fields $V_i \otimes \bar V_i$ and the fields $V_{(r',s)} \otimes \bar V_{(r'',s)}$ with $r'' \in r' \otimes 3$. A ring crossed only once by $D$ carries instead the twisted sector $\mathcal{H}_D = \bigoplus_{i,j} N_{Di}^{j} V_i \otimes \bar V_j$, the space of operators at which the line \emph{ends}; this is what the single-strand computation of Sec.~\ref{sec:4} uses.

\subsection{Lassos: a crossing is a choice}\label{sec:lassos}

A wrapped $(r,1)$ line $x$ on the ring must cross the defect line at every strand, and a crossing is not canonical. It is a junction $u \in \mathrm{Hom}(x \otimes D,\, D \otimes x)$, a space whose dimension is the number of fusion channels of $x\otimes D$: it is spanned by the fuse-and-split junctions $u_{D'}$, in which $x$ and $D$ fuse into an intermediate line $D' \in x \otimes D$ and split again. For the KRW starting point $D_{(1,2)}$ and $x=(r,1)$ the fusion $(r,1)\otimes(1,2)=(r,2)$ has a single channel, so the junction is unique up to normalisation. For our starting point $D_{(2,1)}=\wl$ and $x=\wl$ one has $\wl\otimes\wl=\mathbf 1\oplus\mathrm{adj}$, so the junction space is two-dimensional (for $x=\eta$ it is again one-dimensional, so the crossing is unique up to a scalar; the over- and under-crossings can still differ by a monodromy phase). A loop of $x$ with a chosen junction at each crossing is a \emph{lasso}, an element of the tube algebra of $D$ \cite{ref13,ref39}. As long as $D$ is topological every lasso commutes with the Hamiltonian, but different lassos are different operators. Two of them are distinguished: the over- and under-crossings, in which $x$ passes in front of or behind $D$ at the crossing. They are the two half-braidings of $x$ with $D$---the junctions compatible with moving $x$ past the fusion vertices of the defect---and this is what makes them special among all junctions: a lasso built from either one acts diagonally on the defect fields, with eigenspaces the objects $(i,\bar j)$ of the Drinfeld centre $\mathcal{C} \boxtimes \bar{\mathcal{C}}$, the irreducible representations of the tube algebra \cite{ref35,ref39}, whereas a generic junction need not act diagonally within an irreducible tube-algebra representation or on the corresponding defect multiplicity spaces. It does not mix distinct irreducible representation blocks. We denote the two half-braidings by $B$ and $B^{-1}$; on the lattice they are the braid matrix $B=FRF^{-1}$ of Section~\ref{sec:lattice} and its inverse, and their explicit expansion in the channel basis is given in Section~\ref{sec:latlasso}, eq.~\eqref{eq:Bexp}. On a defect field $V_i \otimes \bar V_j$ the over-crossing lasso acts by the Verlinde eigenvalue of the \emph{left} label, $S_{xi}/S_{0i}$, and the under-crossing lasso by that of the right label, $S_{xj}/S_{0j}$.

The reason a crossing reads one chiral label is most transparent in the three-dimensional description of the diagonal theory \cite{ref4,ref5}, Fig.~\ref{fig:lasso3d}. There the CFT on a surface $\Sigma$ is the Reshetikhin--Turaev topological field theory on the slab $\Sigma\times[-1,1]$: the holomorphic degrees of freedom live on the top sheet $\Sigma\times\{+1\}$, the antiholomorphic ones on the bottom sheet $\Sigma\times\{-1\}$, and all defect lines, the symmetry lines $x$ as well as the defect $D$, are Wilson lines in the middle layer $\Sigma\times\{0\}$. A field with labels $(i,\bar j)$ is the meeting point of a Wilson line $i$ coming down from the top sheet and a Wilson line $j$ coming up from the bottom sheet; for a bulk field of the diagonal theory $i=j$, for a defect field of $D$ the two labels are independent and the lines meet on $D$. A loop of $x$ in the middle layer around the insertion point crosses $D$ twice, and ``over'' and ``under'' at a crossing are literally the two sides of the middle layer in the third direction. If $x$ passes over $D$ at both crossings the loop can be lifted off the middle layer into the holomorphic half of the slab, where it links only the $i$-line and contracts to $S_{xi}/S_{0i}$; if it passes under at both crossings it can be pushed into the antiholomorphic half, where it links only the $j$-line and gives $S_{xj}/S_{0j}$. A loop with one over- and one under-crossing can be moved to neither side, links both lines at once, and need not act as a scalar within a tube-algebra representation; this does not imply mixing between distinct irreducible representation blocks. For a bulk field $i=j$ and the two lassos agree, which is why the action of a Verlinde line on bulk fields never requires a junction to be specified; for a defect field with $i\ne j$ they differ. On the lattice, moving the wrapped line through the strand vertex with the bare $F$-move keeps it in front of the strand in the fusion tree, which is the over-crossing, while passing behind the strand requires the ribbon twists $\theta_a$ of the labels (Section~\ref{sec:latlasso}); which of the two reads the holomorphic label is a matter of orientation conventions, fixed here empirically by the undeformed-ring check of Table~\ref{tab:3}.

\begin{figure}[ht]
\centering
\resizebox{\textwidth}{!}{%
\begin{tikzpicture}[x=0.95cm,y=0.95cm, every node/.style={font=\small}]
\def\W{5.2}\def\sk{1.6}\def\hh{1.1}
\newcommand{\sheet}[4]{
  \fill[#3, opacity=0.55] (#1,#2) -- (#1+\W,#2) -- (#1+\W+\sk,#2+\hh) -- (#1+\sk,#2+\hh) -- cycle;
  \draw[gray!70] (#1,#2) -- (#1+\W,#2) -- (#1+\W+\sk,#2+\hh) -- (#1+\sk,#2+\hh) -- cycle;
  \node[anchor=west, gray!80!black] at (#1+\W+\sk+0.1,#2+0.5*\hh) {#4};}
\newcommand{\lassopanel}[2]{
\begin{scope}[xshift=#1]
  \sheet{0}{0}{blue!10}{$\Sigma\times\{-1\}$ (antiholomorphic)}
  \sheet{0}{2.2}{gray!10}{$\Sigma\times\{0\}$ (defect lines)}
  \sheet{0}{4.4}{red!10}{$\Sigma\times\{+1\}$ (holomorphic)}
  \coordinate (P) at (3.0,2.81);
  \node[red, anchor=east] at (0.35,2.5) {$D$};
  \ifnum#2=0
     \draw[red, very thick] (0.4,2.55) -- ($(P)+(-1.37,-0.135)$);
     \draw[red, very thick] ($(P)+(-1.06,-0.104)$) -- ($(P)+(1.06,0.104)$);
     \draw[red, very thick] ($(P)+(1.37,0.135)$) -- (6.5,3.15);
  \else
     \draw[red, very thick] (0.4,2.55) -- (6.5,3.15);
  \fi
  \draw[red!80!black, thick] (P) -- ++(0,2.2) node[pos=0.75, anchor=west] {$i$};
  \draw[blue!80!black, thick] (P) -- ++(0,-2.2) node[pos=0.75, anchor=west] {$\bar j$};
  \fill[red!80!black] ($(P)+(0,2.2)$) circle (1.6pt); \fill[blue!80!black] ($(P)+(0,-2.2)$) circle (1.6pt);
  \ifnum#2=0
     \draw[blue, thick] (P) ellipse (1.25 and 0.5);
  \else
     \draw[blue, thick] ($(P)+({1.25*cos(20.8)},{0.5*sin(20.8)})$) arc[start angle=20.8, end angle=186.8, x radius=1.25, y radius=0.5];
     \draw[blue, thick] ($(P)+({1.25*cos(200.8)},{0.5*sin(200.8)})$) arc[start angle=200.8, end angle=366.8, x radius=1.25, y radius=0.5];
  \fi
  \node[blue, anchor=east] at ($(P)+(-1.3,0.12)$) {$x$};
  \fill[black] (P) circle (2.2pt); \node[anchor=north west, inner sep=1pt] at (P) {$P$};
  \ifnum#2=0
     \draw[blue, thick, dashed, -{Stealth[length=2mm]}] ($(P)+(1.25,0.1)$) .. controls ($(P)+(1.7,0.8)$) .. ($(P)+(1.3,1.45)$);
     \draw[blue, thick, dashed] ($(P)+(0,1.5)$) ellipse (1.25 and 0.5);
     \node[blue, anchor=west, font=\footnotesize] at ($(P)+(1.4,1.75)$) {lifted up: bonds $i$ only};
     \node[blue, anchor=west, font=\footnotesize] at ($(P)+(1.4,1.35)$) {$\Rightarrow\ S_{xi}/S_{0i}$};
     \node[anchor=north, font=\footnotesize] at (3.6,-0.3) {(a) over-crossing lasso $B$: reads the holomorphic label $i$};
  \else
     \draw[blue, thick, dashed, -{Stealth[length=2mm]}] ($(P)+(1.25,-0.1)$) .. controls ($(P)+(1.7,-0.8)$) .. ($(P)+(1.3,-1.45)$);
     \draw[blue, thick, dashed] ($(P)+(0,-1.5)$) ellipse (1.25 and 0.5);
     \node[blue, anchor=west, font=\footnotesize] at ($(P)+(1.4,-1.3)$) {pushed down: bonds $\bar j$ only};
     \node[blue, anchor=west, font=\footnotesize] at ($(P)+(1.4,-1.7)$) {$\Rightarrow\ S_{xj}/S_{0j}$};
     \node[anchor=north, font=\footnotesize] at (3.6,-0.3) {(b) under-crossing lasso $B^{-1}$: reads the antiholomorphic label $\bar j$};
  \fi
\end{scope}}
\lassopanel{0cm}{0}
\begin{scope}[yshift=-7.6cm]\lassopanel{0cm}{1}\end{scope}
\end{tikzpicture}}
\caption{Why a lasso reads one chiral label. In the Reshetikhin--Turaev description on $\Sigma\times[-1,1]$ the holomorphic data live on the top sheet, the antiholomorphic data on the bottom sheet, and all defect lines in the middle layer; a defect field $(i,\bar j)$ at $P$ is the meeting point of a Wilson line $i$ from above and $\bar j$ from below. (a) A loop of $x$ passing above $D$ at both crossings can be lifted into the holomorphic half, where it links only $i$ and gives $S_{xi}/S_{0i}$. (b) Passing below at both crossings it can be pushed into the antiholomorphic half, where it links only $\bar j$ and gives $S_{xj}/S_{0j}$.}
\label{fig:lasso3d}
\end{figure}

\subsection{The deformation fixes the lasso}\label{sec:fixlasso}

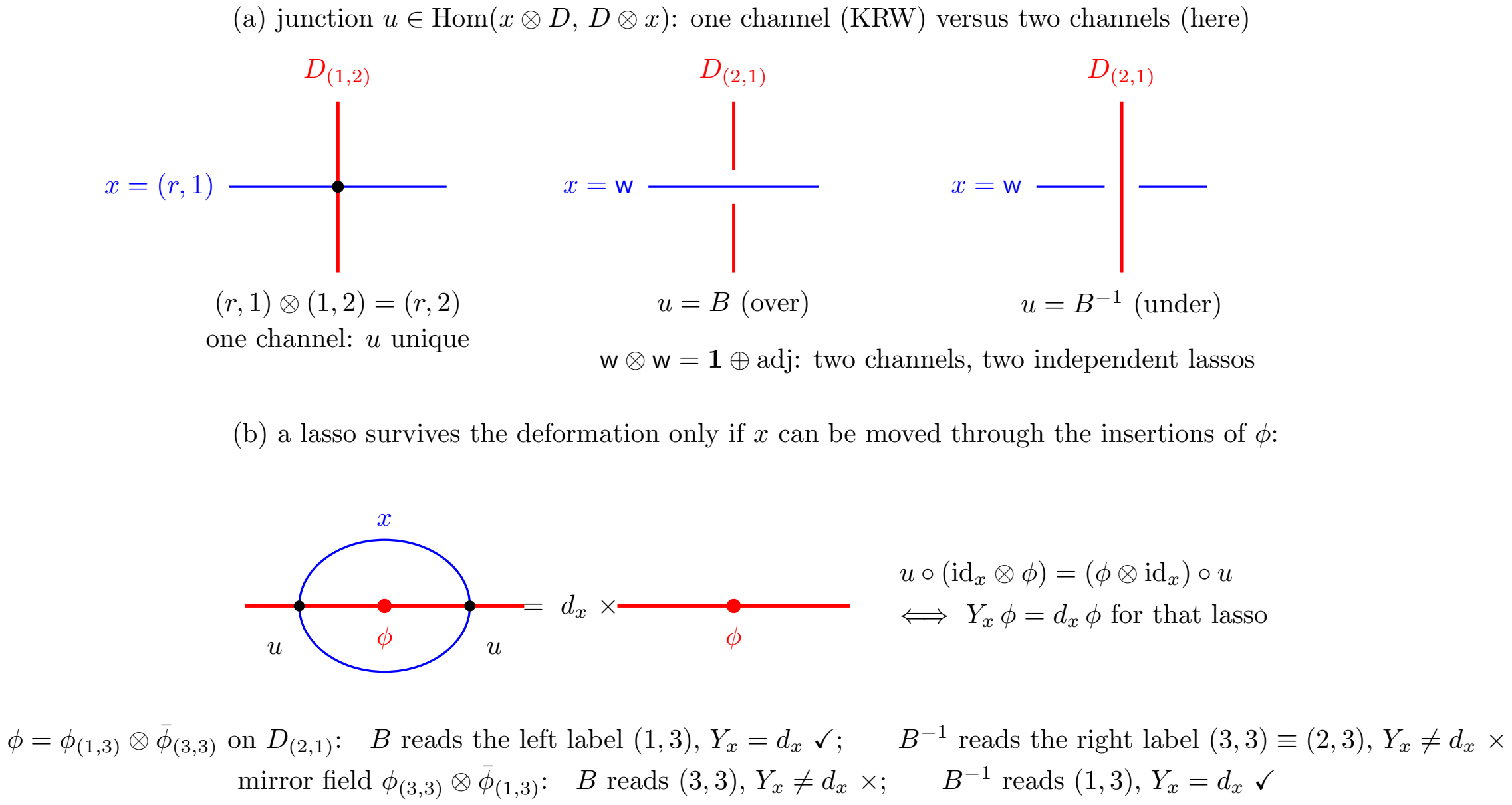
\begin{figure}[ht]
\centering
\resizebox{\textwidth}{!}{%
\begin{tikzpicture}[every node/.style={font=\small}]
\node[anchor=west] at (-0.3,3.35) {(a)\ junction $u\in\mathrm{Hom}(x\otimes D,\,D\otimes x)$: one channel (KRW) versus two channels (here)};
\begin{scope}[xshift=1.2cm, yshift=1.2cm]
  \draw[red, very thick] (0,-1.1) -- (0,1.1);
  \draw[blue, thick] (-1.4,0) -- (1.4,0);
  \fill[black] (0,0) circle (2.2pt);
  \node[red, anchor=south] at (0,1.15) {$D_{(1,2)}$};
  \node[blue, anchor=east] at (-1.45,0) {$x=(r,1)$};
  \node[anchor=north, align=center] at (0,-1.2) {$(r,1)\otimes(1,2)=(r,2)$\\ one channel: $u$ unique};
\end{scope}
\begin{scope}[xshift=6.3cm, yshift=1.2cm]
  \draw[red, very thick] (0,-1.1) -- (0,-0.22);
  \draw[red, very thick] (0,0.22) -- (0,1.1);
  \draw[blue, thick] (-1.1,0) -- (1.1,0);
  \node[red, anchor=south] at (0,1.15) {$D_{(2,1)}$};
  \node[blue, anchor=east] at (-1.15,0) {$x=\wl$};
  \node[anchor=north] at (0,-1.2) {$u=B$ (over)};
\end{scope}
\begin{scope}[xshift=11.3cm, yshift=1.2cm]
  \draw[red, very thick] (0,-1.1) -- (0,1.1);
  \draw[blue, thick] (-1.1,0) -- (-0.22,0);
  \draw[blue, thick] (0.22,0) -- (1.1,0);
  \node[red, anchor=south] at (0,1.15) {$D_{(2,1)}$};
  \node[blue, anchor=east] at (-1.15,0) {$x=\wl$};
  \node[anchor=north] at (0,-1.2) {$u=B^{-1}$ (under)};
\end{scope}
\node[anchor=north, align=center] at (8.8,-0.75) {$\wl\otimes\wl=\mathbf 1\oplus\mathrm{adj}$: two channels, two independent lassos};
\node[anchor=west] at (-0.3,-2.0) {(b)\ a lasso survives the deformation only if $x$ can be moved through the insertions of $\phi$:};
\begin{scope}[yshift=-4.2cm]
  \begin{scope}[xshift=1.8cm]
    \draw[red, very thick] (-1.8,0) -- (1.8,0);
    \draw[blue, thick] (0,0) ellipse (1.1 and 0.85);
    \fill[black] (-1.1,0) circle (2pt); \fill[black] (1.1,0) circle (2pt);
    \fill[red] (0,0) circle (2.6pt); \node[red, anchor=north] at (0,-0.12) {$\phi$};
    \node[blue, anchor=south] at (0,0.9) {$x$};
    \node[anchor=west] at (-1.65,-0.55) {$u$}; \node[anchor=east] at (1.65,-0.55) {$u$};
  \end{scope}
  \node at (4.2,0) {$=\ d_x\ \times$};
  \begin{scope}[xshift=6.3cm]
    \draw[red, very thick] (-1.5,0) -- (1.5,0);
    \fill[red] (0,0) circle (2.6pt); \node[red, anchor=north] at (0,-0.12) {$\phi$};
  \end{scope}
  \node[anchor=west, align=left] at (8.3,0.15) {$u\circ(\mathrm{id}_x\otimes\phi)=(\phi\otimes\mathrm{id}_x)\circ u$\\[2pt]
     $\Longleftrightarrow\ Y_x\,\phi=d_x\,\phi$ for that lasso};
\end{scope}
\node[anchor=north, align=center] at (6.6,-5.6) {%
  $\phi=\phi_{(1,3)}\otimes\bar\phi_{(3,3)}$ on $D_{(2,1)}$:\quad
  $B$ reads the left label $(1,3)$, $Y_x=d_x$ \checkmark;\qquad
  $B^{-1}$ reads the right label $(3,3)\equiv(2,3)$, $Y_x\neq d_x$ $\times$\\[2pt]
  mirror field $\phi_{(3,3)}\otimes\bar\phi_{(1,3)}$:\quad $B$ reads $(3,3)$, $Y_x\neq d_x$ $\times$;\qquad $B^{-1}$ reads $(1,3)$, $Y_x=d_x$ \checkmark\\[2pt]};
\end{tikzpicture}}
\caption{Why the deformation selects a lasso. (a) The junction $u\in\mathrm{Hom}(x\otimes D, D\otimes x)$: unique for the KRW starting point $D_{(1,2)}$, where $(r,1)\otimes(1,2)=(r,2)$ has one channel, and two-dimensional for $D_{(2,1)}=\wl$, where $\wl\otimes\wl=\mathbf 1\oplus\mathrm{adj}$, with the over- and under-crossings $B$, $B^{-1}$ as the distinguished choices. (b) A lasso survives the deformation only if $x$ can be moved through every insertion of $\phi$ with its junction, eq.~\eqref{eq:lassocond}, i.e.\ only if $\phi$ is neutral under it. For $\phi_{(1,3)}\otimes\bar\phi_{(3,3)}$ only the over-crossing lasso, which reads the neutral left label, survives; for the mirror field only the under-crossing one.}
\label{fig:lasso}
\end{figure}

Once $D$ is deformed by a defect field $\phi$, a lasso stays a symmetry only if the line $x$ can be moved through the insertions of $\phi$ along $D$ with the chosen junction,
\begin{equation}
u \circ (\mathrm{id}_x \otimes \phi) = (\phi \otimes \mathrm{id}_x) \circ u ,
\label{eq:lassocond}
\end{equation}
read with time running along $D$: the insertion of $\phi$ on $D$ is a point operator $\phi:D\to D$, and $\mathrm{id}_x\otimes\phi$ (resp.\ $\phi\otimes\mathrm{id}_x$) denotes this insertion while $x$ runs to the left (resp.\ right) of $D$, i.e.\ before (resp.\ after) the crossing $u:x\otimes D\to D\otimes x$. Both sides are maps $x\otimes D\to D\otimes x$, and the equation states that the crossing can be slid along $D$ past the insertion point of $\phi$. Closing $x$ into a small loop around the insertion shows that this holds exactly when $\phi$ is neutral under that lasso, $Y_x\phi=d_x\phi$: the two crossings of the loop can then be brought together and cancel, leaving a free loop worth $d_x$ (Fig.~\ref{fig:lasso}). Since the over-crossing reads the left label of $\phi$ and the under-crossing its right label, the rule is simply that a deformation selects the lasso that reads its neutral label. For the driving field $\phi = \phi_{(1,3)} \otimes \bar\phi_{(3,3)}$ on $D_{(2,1)}$ (holomorphic label first) the criterion is decisive: the left label $(1,3)$ has $r'=1$ and hence the same eigenvalue $d_x$ as the identity under every $(r,1)$ line, so $\phi$ is neutral under the over-crossing lasso; the right label $(3,3)$ has $Y_{(2,1)} = -\varphi^{-1} \neq \varphi$ at $p=5$, where $(3,3)\equiv(2,3)$, and $Y_{(2,1)}=-\sqrt2\neq\sqrt2$ at $p=4$, where $(3,3)\equiv(1,2)$, so the under-crossing lasso is broken. The symmetry preserved along the flow $D_{(2,1)}\to C$ is therefore not ``the line $x$'', which can be attached to the defect in two ways, but the specific lasso that crosses \emph{over} the defect at every intersection. The mirror field $\phi_{(3,3)} \otimes \bar\phi_{(1,3)}$, charged on the left and neutral on the right, fixes the under-crossing lasso and breaks the over-crossing one; the two deformations are exchanged by the swap of the two sides of the defect, which is why the flow $D_{(2,1)} \to C$ is not swap-symmetric while both its endpoints are. The two fields are the two members of the driving doublet at $\Delta=11/15$ in Table~\ref{tab:3}, which the over-crossing lasso places in different sectors ($a+$ and $c+$): the lasso preserved by the deformation is the one that distinguishes the deformation from its mirror image, and a deformation by the sum of the two fields would preserve neither lasso.

With two strands deformed by the same field, the preserved operator is the lasso with the over-crossing at both strands, and with this choice \eqref{eq:23} is diagonalised by the left labels:
\begin{equation}
Y_{(r,1)}\big|_{V_{(r',s)} \otimes \bar V_{(r'',s)}} \;=\; \frac{S_{(r,1),(r',s)}}{S_{(1,1),(r',s)}} \;=\; (-1)^{(r-1)s}\, \frac{\sin\!\big(\pi r r' \tfrac{p+1}{p}\big)}{\sin\!\big(\pi r' \tfrac{p+1}{p}\big)},
\label{eq:24}
\end{equation}
independently of $r''$ and of the sector $\mathcal{H}_k$ the field comes from. The eigenvalues are the quantum dimensions of $SU(2)_{p-2}$ up to signs: at $p=4$, $Y_{(2,1)} \in \{+\sqrt2, 0, -\sqrt2\}$ and $\eta = Y_{(3,1)} = \pm1$; at $p=5$, $Y_{(2,1)} = \pm\varphi, \pm\varphi^{-1}$ and $\eta = Y_{(4,1)} = \pm1$, with $Y_{(2,1)} = Y_{\tau_{\mathrm{Fib}}}\, \eta$. Since the preserved lasso commutes with the deformed Hamiltonian, these eigenvalues are conserved along the whole flow: every operator of the infrared interface inherits a $(Y_{(2,1)}, \eta)$ label from the ultraviolet defect operator it descends from, so that the spectrum of $C$ can be organised, and its finite-size extrapolation performed, sector by sector (Sections~\ref{sec:4}, \ref{sec:5}); at the endpoint $\eta$, $x\otimes\eta$ is simple, so the crossing junction is unique up to normalization. Any scalar difference between the over- and under-crossings must still be tracked; in the two-crossing defect-operator geometry the corresponding monodromy factors cancel.

Kormos, Runkel and Watts prove, by a matrix-element argument (\cite{ref1}, \S2.4.2), that the perturbed defect \eqref{eq:krwdef} commutes with all the lines $D_{(r,1)}$ for every value of the couplings, and use this to constrain the infrared fixed points (\cite{ref1}, \S3.2). In their setting the statement is unambiguous for two reasons that the preceding analysis makes explicit: the junction space of $x=(r,1)$ with $D_{(1,2)}$ is one-dimensional, so there is only one lasso to speak of, and the chiral perturbing fields carry the labels $(1,3)\otimes(1,1)$ and $(1,1)\otimes(1,3)$, both of $(1,s)$ type and hence neutral under $D_{(r,1)}$ whichever side of the defect the line reads, so that \eqref{eq:lassocond} holds automatically. The junction never has to be discussed. Our flow starts from $D_{(2,1)}$, a member of the same family as the symmetry lines, so that the junction space is two-dimensional, and it is driven by a spinful field with one charged label, so that the two lassos are inequivalent on the deformed defect. The requirement that the deformation commute with the symmetry therefore does more than it does in KRW: it selects one element of the junction space, and this is the precise sense in which ``the $(r,1)$ symmetry is preserved along the flow $D_{(2,1)}\to C$''.

This is also what makes a lattice realisation possible. The preserved symmetry is a line of the fusion category $\mathcal C_p$ attached to the defect by a definite junction, so it can be realised exactly on a lattice whose degrees of freedom are the fusion spaces of that category, with the junction transcribed into the $F$-moves that carry the wrapped line across the defect. The dilute $A_{p-1}$ RSOS chain is such a lattice, and the next section constructs on it the defect, the wrapped lines with the over-crossing junction, and the deformation; the conserved charges \eqref{eq:24} are then imposed exactly, sector by sector, along the whole flow.

\section{The lattice model}\label{sec:lattice}

\subsection{The dilute RSOS chain}\label{sec:2.1}

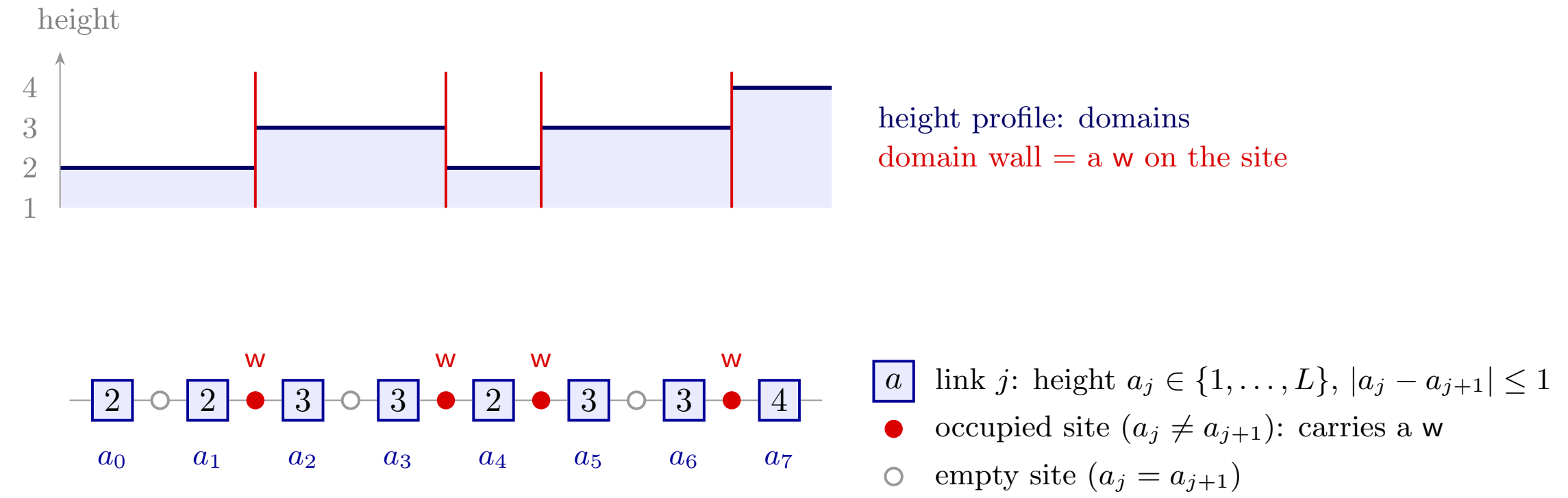
\begin{figure}[ht]\centering
\resizebox{\textwidth}{!}{\input{figs/fig1_tikz.tex}}
\caption{The dilute $A_{p-1}$ RSOS chain (here $p=5$, four heights). Bottom: each bond $j$ carries a height $a_j\in\{1,\dots,p-1\}$ (boxes), neighbours differing by at most one; the site between two bonds is occupied by a $\wl$ (filled red dot) when the heights differ and empty (open circle) when they are equal. Top: the same configuration drawn as a height profile---the heights are domains and every domain wall carries a $\wl$. The Hamiltonian \eqref{eq:7} hops these walls and creates/annihilates them in pairs.}\label{fig:1}
\end{figure}

\paragraph{Degrees of freedom.} We use the following convention throughout the paper: the \emph{height} variables live on the \emph{bonds} of the chain, and the \emph{anyons} (the domain walls, or their absence) live on the \emph{sites} between consecutive bonds. Consider a ring of $L$ bonds. Each bond $j$ carries an integer height
\begin{equation}
a_j\in\{1,\dots,p-1\}\qquad(\text{three values for tricritical Ising},\ \text{four for }M(5,6)),
\label{eq:1}
\end{equation}
subject to the \emph{admissibility} rule that neighbouring heights differ by at most one,
\begin{equation}
|a_j-a_{j+1}|\le 1\qquad\text{at every site.}
\label{eq:2}
\end{equation}
Between two consecutive bonds there is a \emph{site}. A site carries no variable of its own; we call it \emph{empty} if the two heights around it are equal and \emph{occupied} if they differ,
\begin{equation}
\text{site between bonds } a,b:\qquad a=b\ \Rightarrow\ \text{empty},\qquad a\neq b\ \Rightarrow\ \text{occupied}.
\label{eq:3}
\end{equation}
An occupied site is a \emph{domain wall} between two height domains. The word ``dilute'' refers to the fact that empty sites are allowed (in the ordinary, dense $A_{p-1}$ RSOS model every site would be a wall)  see Fig.~\ref{fig:1}. The dense $A_{p-1}$ models are those of Andrews, Baxter and Forrester \cite{ref25}, whose critical points realise the unitary minimal models \cite{ref26,ref27}; the dilute $A_{p-1}$ models were constructed by Warnaar, Nienhuis and Seaton \cite{ref14} from the $O(n)$ loop model of \cite{ref28,ref29}.

\paragraph{Local Hilbert space.} The heights are the only degrees of freedom, so the local Hilbert space of a bond is
\begin{equation}
\mathcal{H}_j=\mathrm{span}\{\,|a_j\rangle:\ a_j=1,\dots,p-1\,\},\qquad \dim\mathcal{H}_j=p-1,
\label{eq:4}
\end{equation}
a three-state space at $p=4$ and a four-state space at $p=5$. The many-body space is not the full tensor product $\bigotimes_j\mathcal{H}_j$ but its \emph{admissible} subspace, spanned by the height configurations obeying \eqref{eq:2} at every site. Every operator introduced below (the Hamiltonian, the seam, the wrapped lines, the deformation) preserves this subspace. In exact diagonalisation we enumerate the admissible configurations and work in that basis; in DMRG we keep the full $(p-1)$-dimensional space on each bond and let the Hamiltonian enforce the constraint (forbidden height jumps have no matrix elements, so an admissible state never leaves the admissible subspace).

\paragraph{Anyon-chain reading.} The same data have a second, equivalent description that will be essential for the defect. Identify a height $a$ with an object of $\mathcal C_p$, using the $SU(2)_{p-2}$ spin label $j=\frac{a-1}{2}$, and let $\wl$ denote the spin-$\frac12$ object (category label 1), of quantum dimension
\begin{equation}
d_\wl=2\cos\frac{\pi}{p}=\frac{S_2}{S_1},\qquad S_a\equiv\sin\frac{\pi a}{p}
\label{eq:5}
\end{equation}
($d_\wl=\sqrt2$ at $p=4$, $d_\wl=\varphi=1.618$ at $p=5$; $S_a$ is the vector of quantum dimensions, which will appear in all the weights below). Then the admissibility rule \eqref{eq:2} says exactly that $a_{j+1}\in a_j\otimes(\mathbf{1}\oplus\wl)$: an empty site fuses the trivial object into the chain, an occupied site fuses a $\wl$. In other words, the height configuration $\{a_j\}$ is the \emph{fusion-tree basis} of a chain of $\wl$ anyons sitting on the occupied sites, with the heights recording the intermediate total charge as one moves along the ring. ``A $\wl$ lives on every domain wall'' is the precise statement, not just a picture. The two kinds of labels play different categorical roles, and it is worth stating them precisely, since this is the structure the defect construction will exploit. The site labels $\mathbf 1$ and $\wl$ (and later the defect labels $x$) are objects of the fusion category $\mathcal C=\mathcal C_p$: they are the things that fuse, braid and carry quantum dimensions. The heights, on the other hand, are objects of a \emph{module category} $\mathcal M$ over $\mathcal C$: they are acted upon by the objects of $\mathcal C$ ($a\mapsto a\otimes\wl$), and the admissibility graph of the height model is the fusion graph of this action. For the $A_{p-1}$ models $\mathcal M$ is $\mathcal C$ itself, acting on itself by fusion (the regular module category), which is why the heights can be identified with $SU(2)_{p-2}$ labels at all; in the $D$- and $E$-type height models of Pasquier \cite{ref27}, the heights instead label vertices of other ADE graphs, illustrating why the action on heights must be specified separately from the fusion rules, and the module-category structure is the general statement that an anyon chain is defined by a pair $(\mathcal C,\mathcal M)$ \cite{ref16,ref17}. This is the lattice counterpart of the correspondence between module categories and conformal boundary conditions, or equivalently between module categories and the ways a chiral algebra can be completed to a full CFT \cite{refOstrik,refFRS1}; on the lattice it is the organising principle of the fusion-category lattice models of \cite{ref19,ref21} and of the general framework of dualities and non-invertible symmetries in one-dimensional quantum lattice models developed by Lootens, Delcamp, Ortiz and Verstraete \cite{refLDOV,refLDV}, which is reviewed in Delcamp's IHES lectures \cite{refDelcampIHES}. 

\begin{figure}[ht]\centering
\resizebox{\textwidth}{!}{\input{figs/fig2_tikz.tex}}
\caption{The six face-weight moves of \eqref{eq:9}, each drawn as a ``before $\to$ after'' configuration of the three bonds $(l,a,r)\to(l,a',r)$. Boxes are heights, the red box is the middle height after the move, and the dots between boxes show whether the site is a domain wall ($\wl$, filled) or empty (open). From left to right and top to bottom: no walls (empty); one wall (it stays); two walls (they stay); a wall hops across the middle; a wall pair is created/annihilated ($l=r$); and the Temperley--Lieb generator $e$ ($l=r$, both sites occupied), which is also the operator used to build the deformation of Sec.~\ref{sec:2.3}.}\label{fig:2}
\end{figure}
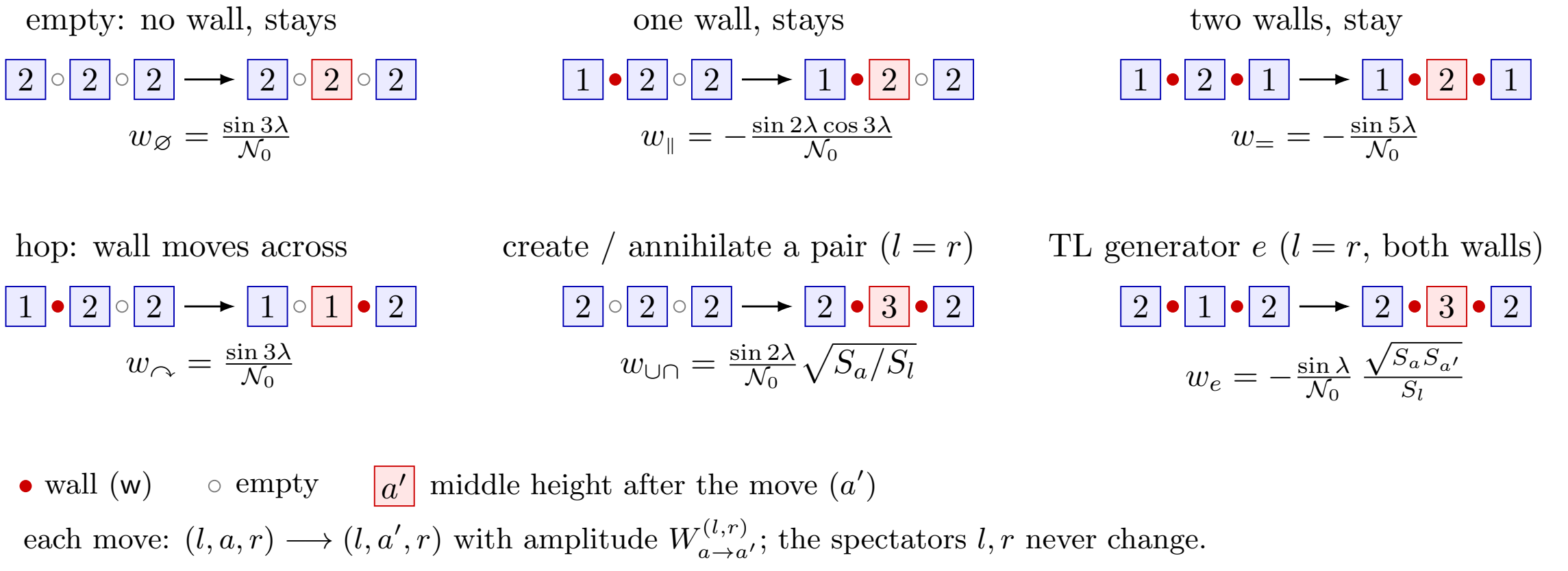

\paragraph{The Hamiltonian.} The critical dynamics is a nearest-neighbour Hamiltonian
\begin{equation}
H_0=-\sum_{j=1}^{L}h_j,
\label{eq:6}
\end{equation}
where $h_j$ acts on three consecutive bonds---a left spectator $l=a_{j-1}$, the middle bond $a=a_j$ and a right spectator $r=a_{j+1}$---and only ever changes the middle height, $a\to a'$, keeping the result admissible. With $P_l=|l\rangle\langle l|$ the projector on a spectator height,
\begin{equation}
h_j\;=\;\sum_{l,r}\sum_{a,a'}W^{(l,r)}_{a\to a'}\,P_l\left(|a'\rangle\langle a|\right)P_r,
\label{eq:7}
\end{equation}
so $h_j$ is a $(p-1)^3\times(p-1)^3$ matrix specified by the numbers $W^{(l,r)}_{a\to a'}$. These are the Hamiltonian limit of the dilute $A_{p-1}$ face weights of Nienhuis \cite{ref28} and of Warnaar, Nienhuis and Seaton \cite{ref14,ref15}, taken at the critical branch with crossing parameter
\begin{equation}
\lambda=\frac{\pi(p-1)}{4p}\qquad(\lambda=\tfrac{3\pi}{16}\ \text{at }p=4,\quad \lambda=\tfrac{\pi}{5}\ \text{at }p=5).
\label{eq:8}
\end{equation}
They depend only on the \emph{shape} of the move, i.e.\ on which of the two internal sites are domain walls before ($o_{\mathrm{L}}=[l\neq a]$, $o_{\mathrm{R}}=[a\neq r]$) and after ($o'_{\mathrm{L}}=[l\neq a']$, $o'_{\mathrm{R}}=[a'\neq r]$) the move, and they take one of six values:
\begin{equation}
\begin{aligned}
&\text{empty }(a{=}a',\ o_{\mathrm{L}}{=}o_{\mathrm{R}}{=}0):&& w_\varnothing=\frac{\sin 3\lambda}{\mathcal N_0},\\
&\text{one wall }(a{=}a',\ o_{\mathrm{L}}{\neq}o_{\mathrm{R}}):&& w_{\parallel}=-\frac{\sin 2\lambda\cos 3\lambda}{\mathcal N_0},\\
&\text{two walls }(a{=}a',\ o_{\mathrm{L}}{=}o_{\mathrm{R}}{=}1):&& w_{=}=-\frac{\sin 5\lambda}{\mathcal N_0},\\
&\text{hop }(o_{\mathrm{L}}o_{\mathrm{R}}=0,\ \text{wall moves side}):&& w_{\curvearrowright}=\frac{\sin 3\lambda}{\mathcal N_0},\\
&\text{create/annihilate }(l{=}r,\ \text{one bump}):&& w_{\cup\cap}=\frac{\sin 2\lambda}{\mathcal N_0}\sqrt{\frac{S_m}{S_l}},\\
&\text{TL }(l{=}r,\ \text{both walls, any }a'):&& w_e=-\frac{\sin\lambda}{\mathcal N_0}\frac{\sqrt{S_aS_{a'}}}{S_l},
\end{aligned}
\label{eq:9}
\end{equation}
with $\mathcal N_0=\sin 2\lambda\,\sin 3\lambda$ a convenient overall normalisation. The six moves are drawn in Fig.~\ref{fig:2}. Two points of reading are worth making explicit, since they decide the matrix. In the creation/annihilation weight $m$ is the \emph{bump} height, the middle height of the configuration in which both sites are walls, whichever side of the move it is on: $W^{(l,l)}_{l\to m}=W^{(l,l)}_{m\to l}=\frac{\sin2\lambda}{\mathcal N_0}\sqrt{S_m/S_l}$ for $|m-l|=1$, which makes the Hermiticity manifest. The Temperley--Lieb weight applies to every pair of configurations with $l=r$ and both sites occupied, \emph{including} $a'=a$, so that on such a configuration it adds to the diagonal: $W^{(l,l)}_{a\to a}=w_{=}-\frac{\sin\lambda}{\mathcal N_0}\frac{S_a}{S_l}$. (These are the Hamiltonian limits of the dilute face weights, and the resulting matrix is the one used in all our numerics.) In the anyon-chain reading $h_j$ is simply a sum of projectors on the fusion channels of the two site anyons (the dilute generalisation of the golden-chain Hamiltonian $H=-\sum_jP^{\mathbf 1}_{j,j+1}$ of \cite{ref16}, in which every site carries a $\wl$); the channel projectors are introduced in \S\ref{sec:2.3}, where they are needed to build the deformation. The first three are diagonal (the middle height does not change): they are the energies of an empty site, of an isolated wall, and of two adjacent walls. The last three are off-diagonal: a wall hops across the middle bond; a pair of walls is created from, or annihilated into, an empty stretch (this needs $l=r$, since the two new walls must carry opposite height steps); and the \emph{Temperley--Lieb move} (with the diagonal part just mentioned), in which two adjacent walls with $l=r$ are replaced by another pair with the same end heights. The square-root factors in the last two lines are what makes $h_j$ Hermitian in the height basis and are the only place where the quantum dimensions $S_a$ enter; they are inherited from the loop-model origin of the weights, where a closed domain wall (a loop) carries a fugacity $d_\wl$. With periodic boundary conditions the low-energy spectrum of $H_0$ is that of $M(p,p+1)$, with $c=7/10$, $v=1.778$ at $p=4$ and $c=4/5$, $v=5/3$ at $p=5$ \cite{ref15,ref30}; we use these sound velocities to convert all gaps into scaling dimensions, $\Delta=(E-E_0)\,L/(2\pi v)$ \cite{ref31,ref32,ref33}.

\subsection{The topological $(r,1)$ interfaces as braided seams}\label{sec:2.2}

\paragraph{What we want.} A defect line is \emph{topological} if it can be deformed freely without changing any correlation function \cite{ref3,ref4,ref5}. In radial quantisation on the ring, the defect is a static worldline that pierces the spatial circle at one point, between two sites; on the lattice ``topological'' means that this insertion commutes with the Hamiltonian in the sense that the strand can be moved from one site to the next without changing the spectrum, and more strongly that the \emph{wrapped} version of the same line (the operator that transports it once around the ring) commutes with $H$ exactly. We now build the $(r,1)$ lines with this property.

\paragraph{The line as an anyon.} In the anyon-chain reading of \S\ref{sec:2.1}, the $(r,1)$ Verlinde line is the object of $\mathcal C_p$ with label $x=r-1$ (so $x=1$ is $\wl=D_{(2,1)}$ and $x=p-2$ is the invertible line $\eta=D_{(p-1,1)}$). Inserting the defect at a point of the ring means threading a static anyon $x$ through the chain there. In the fusion-tree basis this is done by adding \emph{one extra bond} $c_j$ between $a_j$ and $a_{j+1}$, whose height records the total charge after fusing $x$ into the chain,
\begin{equation}
c_j\in a_j\otimes x\qquad(\text{for }x=\wl:\ |c_j-a_j|=1).
\label{eq:11}
\end{equation}
So the strand of anyon $x$ is one more height variable, subject to a stricter adjacency rule than an ordinary bond (for $x=\wl$ the site between $a_j$ and $c_j$ is always a domain wall), and the next ordinary site, between $c_j$ and $a_{j+1}$, obeys the usual dilute rule \eqref{eq:2}. Away from the strand nothing changes.

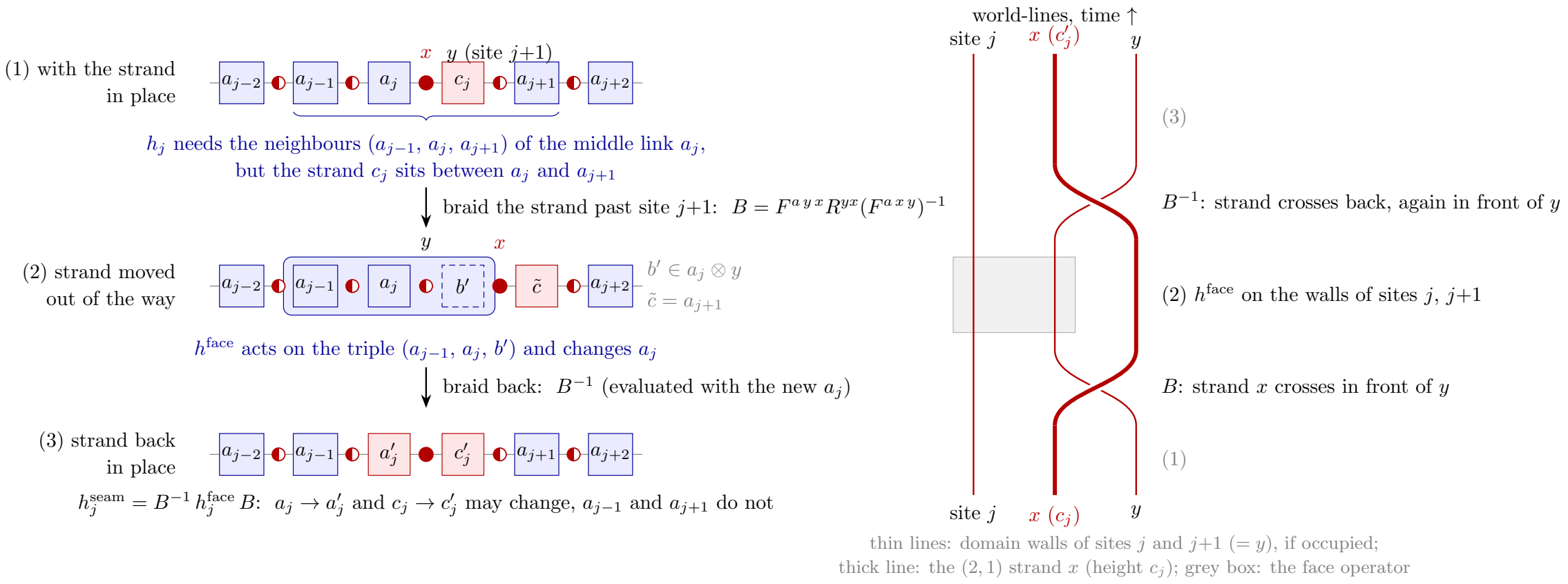
\begin{figure}[ht]\centering
\resizebox{\textwidth}{!}{\input{figs/fig3_tikz.tex}}
\caption{How the Hamiltonian acts across the strand. (1) The face operator $h_j$ needs the physical neighbours $(a_{j-1},a_j,a_{j+1})$ of the middle bond $a_j$, but the strand bond $c_j$ sits between $a_j$ and $a_{j+1}$. (2) The strand $x$ is braided past the anyon $y$ on site $j{+}1$ with the unitary $B$ of \eqref{eq:12}; the label between $a_j$ and the strand is now $b'\in a_j\otimes y$, an ordinary-looking bond, and the ordinary face operator acts on the triple $(a_{j-1},a_j,b')$. (3) The strand is braided back with $B^{-1}$. The net effect, $h_j^{\mathrm{seam}}=B^{-1}h_j^{\mathrm{face}}B$, changes $a_j$ and the strand height $c_j$ but leaves $a_{j\pm1}$ untouched, exactly as an ordinary $h_j$ would leave its spectators. Right: the same three steps drawn as world-lines in the skein picture of Fig.~\ref{fig:15}(b), time upwards: the thick line is the $(2,1)$ strand, the thin lines are the domain walls of sites $j$ and $j{+}1$ (present when the site is occupied); the strand $x$ crosses in front of the $y$ line ($B$), the face operator acts on the now adjacent walls of sites $j$ and $j{+}1$ (grey box), and the strand crosses back, again in front of $y$ ($B^{-1}$, the inverse braid), so that the two crossings cancel and the world-lines are not linked. The over/under choice is the orientation of the braid discussed in \S\ref{sec:lassos}.}\label{fig:3}
\end{figure}

\paragraph{Face operators across the strand: braiding.} The only question is how the Hamiltonian acts near the strand (Fig.~\ref{fig:3}). The face operator $h_j$ of \eqref{eq:7} needs the two \emph{physical} neighbours of the middle bond $a_j$, namely $a_{j-1}$ and $a_{j+1}$; but the bond immediately to the right of $a_j$ is now the strand $c_j$, which is not a physical bond, and $h_j$ knows nothing about anyon $x$. The face operator must therefore skip $c_j$ and act on the next physical bond $a_{j+1}$, and it must do so in a way that keeps the strand consistent with the fusion rules. The resolution is to move the strand out of the way \emph{before} acting and to put it back afterwards.

Moving an anyon $x$ past the anyon $y$ on the neighbouring site (here $y=\mathbf{1}$ if site $j{+}1$ is empty and $y=\wl$ if it is occupied) is a braiding, which in the fusion-tree basis is the unitary
\begin{equation}
B\;=\;F^{a\,y\,x}\,R^{yx}\left(F^{a\,x\,y}\right)^{-1},
\label{eq:12}
\end{equation}
where $F$ is the associator ($6j$-symbol) of the category, $R^{yx}$ the diagonal braiding phase \cite{ref34,ref35}, and the matrices act on the intermediate label---i.e.\ on the strand height $c_j$, which after the braid becomes an ordinary-looking label $b'\in a_j\otimes y$. The seam face operator is then
\begin{equation}
h_j^{\mathrm{seam}}\;=\;B^{-1}\left(h_j^{\mathrm{face}}\right)B,
\label{eq:13}
\end{equation}
where $h_j^{\mathrm{face}}$ is the ordinary operator \eqref{eq:7} acting on the triple $(a_{j-1},a_j,b')$. Because $B$ is unitary, $h_j^{\mathrm{seam}}$ is Hermitian and has the same spectrum as $h_j^{\mathrm{face}}$; the strand bond is never the middle bond of any face operator, so the Hamiltonian still has exactly $L$ terms, one of which is braided. For $p=4$ the ingredients are completely explicit: with $x=x^\prime=\wl=\sigma^\prime$ and $\eta$ the invertible line,
\[
F^{\wl\wl\wl}_{\wl}=\frac{1}{\sqrt2}\begin{pmatrix}1&1\\1&-1\end{pmatrix},\qquad R^{\wl\wl}_{\mathbf{1}}=e^{-7i\pi/8},\quad R^{\wl\wl}_{\eta}=e^{+5i\pi/8},
\]
so \eqref{eq:12} is a $2\times2$ unitary on the two fusion channels $\wl\times\wl=\mathbf{1}\oplus\eta$. For $p=5$ the same formulas use $d_\wl=\varphi$ and $R$-phases that are tenth roots of unity, $R^{\wl\wl}_{\mathbf{1}}=e^{-i4\pi/5}$, $R^{\wl\wl}_{\mathrm{adj}}=e^{+3i\pi/5}$. Nothing else is needed: no extra couplings, no fine tuning.

\paragraph{Which category.}
Fusion rules do not determine a category, and it is worth stating precisely which data the construction uses, since the seam operator is sensitive to them. The $(r,1)$ lines have the fusion rules of $SU(2)_{p-2}$, label $a=r-1=2j$, but they are the Wilson lines of the chiral modular category of $M(p,p+1)$ restricted to its $(r,1)$ sector, whose twists are $\theta_{(r,1)}=e^{2\pi i h_{(r,1)}}$ rather than $e^{2\pi i\,j(j+1)/p}$: the two differ by the factor $i^{[a]}$, $[a]=a \bmod 2$, since $h_{(2,1)}=\frac{3}{4p}+\frac14$. Accordingly the $F$-symbols are the quantum $6j$-symbols of $SU(2)_{p-2}$ multiplied by the $\mathbb Z_2$ $3$-cocycle $(-1)^{[a][b][c][d]}$ (which flips the Frobenius--Schur indicator of the half-integer-spin objects: at $p=4$ this is the Ising fusion category, with $F^{\wl\wl\wl}_{\wl}=\frac{1}{\sqrt2}\left(\begin{smallmatrix}1&1\\1&-1\end{smallmatrix}\right)$ up to vertex gauge, and not $SU(2)_2$), expressed in a real orthogonal vertex gauge compatible with the positive height-basis projectors of \eqref{eq:17}; commutation of the wrapped lines \eqref{eq:25} with $H_0$ is a consistency check on this common choice, rather than a definition of the gauge; and the $R$-symbols are
\begin{equation}
R^{ab}_c=i^{[a][b]}\,(-1)^{(a+b-c)/2}\,q^{\,j_c(j_c+1)-j_a(j_a+1)-j_b(j_b+1)},\qquad q=e^{i\pi/p},
\label{eq:Rsym}
\end{equation}
which solve the hexagon equations with these $F$-symbols and reproduce the minimal-model twists ($\theta_\wl=\tfrac1{d_\wl}\sum_c d_c R^{\wl\wl}_c=e^{2\pi i h_{(2,1)}}$). The prefactor $i^{[a][b]}$ is not a convention that can be dropped. The braid \eqref{eq:12} is $1$ when one of the two slots is empty and $R^{\wl\wl}_c$ on the channels when both are occupied, and since the dilute Hamiltonian creates and annihilates pairs of walls the relative phase between these two blocks enters $h^{\mathrm{seam}}=B^{-1}hB$: with the $SU(2)_{p-2}$ phases $-q^{-3/2}$, $q^{1/2}$ in place of \eqref{eq:Rsym} the one-strand ring at $p=4$, $L=6$, has lowest scaled levels $0.298$, $0.354$, $0.757$ instead of the $D_{(2,1)}$ values $0.1375$, $0.4375$, $0.6375$, which \eqref{eq:Rsym} reproduces to $10^{-3}$ at this size. In the loop-model language of Section~\ref{sec:loop} the same data appear as the Kauffman braid $A\mathbf 1+A^{-1}e$ with $A=iq^{1/2}$, eq.~\eqref{eq:45}, which agrees with \eqref{eq:Rsym} channel by channel ($A=iq^{1/2}=R^{\wl\wl}_{\mathrm{adj}}$, $A+A^{-1}d_\wl=-iq^{-3/2}=R^{\wl\wl}_{\mathbf 1}$). All numerical results of the paper, in both formulations, use this one set of data; explicit evaluations of the associators, braid operators and channel projectors must use the same vertex gauge. The braid phases and twists are given above, while \eqref{eq:17} fixes the height-basis normalization of the identity-channel projector.

\begin{figure}[ht]\centering
\includegraphics[width=0.9\textwidth]{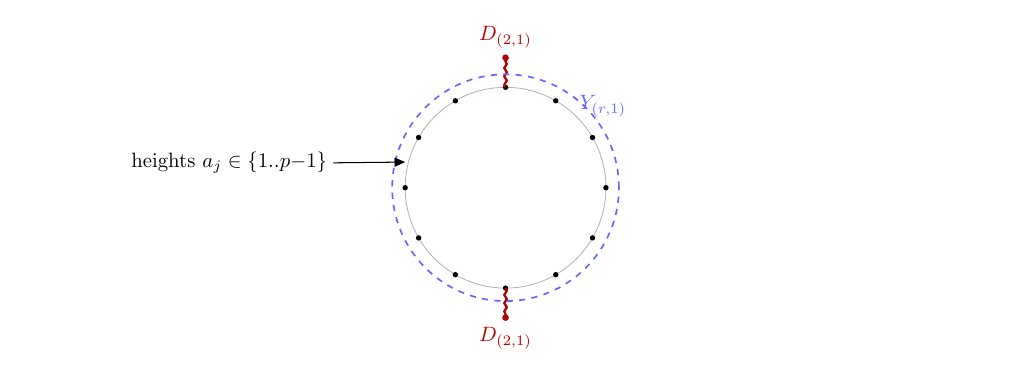}
\caption{Two-interface ring. Heights live on the bonds (black dots mark the sites between them); two $(2,1)$ strands (red), built with the same braiding, are inserted at antipodal points, and the dashed line is a wrapped Verlinde line $Y_{(r,1)}$ used to resolve the fusion and $\mathbb{Z}_2$ charges. Radial quantisation about a strand turns the ring eigenstates into the defect operators of the interface.}\label{fig:4}
\end{figure}

\paragraph{Why this is topological: the wrapped lines.} The construction inherits the consistency of the category. The operator $Y_x$ that transports anyon $x$ once around the ring is assembled from the same $F$-symbols, one per slot,
\begin{equation}
\left(Y_x\right)_{\{b\}\{a\}}=\prod_k F^{\,x\,a_{k-1}\,\ell_k}_{\,b_k\,b_{k-1}\,a_k},\qquad \ell_k=\text{anyon on slot }k\ (\mathbf{1},\ \wl,\ \text{or }x\text{ at the strand}),
\label{eq:14}
\end{equation}
and it commutes with $H_0$ by the pentagon identity and with every braided term \eqref{eq:13} by the hexagon identity \cite{ref34,ref35}; this is the same mechanism that makes the wrapped lines of \cite{ref20,ref21,ref36} exact lattice symmetries. There is a duality here with the dense RSOS models: there it is the $(1,s)$ lines that can be made exactly topological on the lattice, as centralisers of the affine Temperley--Lieb algebra \cite{ref22,ref24}, while the $(r,1)$ lines are only topological in the scaling limit; in the dilute algebra the roles are exchanged, as noted in \cite{ref24}, and the $(r,1)$ lines are the exact ones. So the wrapped $(r,1)$ lines are exact symmetries of the lattice Hamiltonian with a strand, which is the lattice statement that the strand is a topological defect of the same category. We verify this numerically on the ring used throughout (two $(2,1)$ strands, Fig.~\ref{fig:4}):
\[
\bigl\|[\,Y_{(2,1)},H\,]\bigr\|_\infty,\ \ \bigl\|[\,Y_{(p-1,1)},H\,]\bigr\|_\infty\ <\ 10^{-14}.
\]
The commuting lines are also conserved \emph{charges}: the simultaneous eigenvalues of $Y_{(2,1)}$ and $Y_{(p-1,1)}=\eta$ label every eigenstate by its fusion sector and its $\mathbb{Z}_2$ parity, and this labelling survives the deformation of \S\ref{sec:2.3}, which commutes with the same lines.

\paragraph{Check against the defect spectrum.} At zero deformation the ring with two strands is the lattice version of $M(p,p+1)$ on a cylinder crossed twice by $D_{(2,1)}$, so its low spectrum must be the space of operators living on the line \cite{ref3}, $\mathcal{H}_{D\times D}=\mathcal{H}_{\mathbf{1}}\oplus\mathcal{H}_{(3,1)}$ (the untwisted spectrum plus the $(3,1)$-twisted sector, since $D_{(2,1)}\times D_{(2,1)}=D_{(1,1)}+D_{(3,1)}$). This is what we find; the comparison, level by level and charge by charge, is made in \S\ref{sec:lassos} (Table~\ref{tab:3}). At $p=4$ the $g$-function of the undeformed strand also reproduces $g_{D_{(2,1)}}=\sqrt2$ (Sec.~\ref{sec:4}).

\paragraph{The two-strand ring.} All computations use a ring of $L$ physical sites with two $D_{(2,1)}$ strands at antipodal points (Fig.~\ref{fig:4}), both built with the same braiding $B$. Two strands are needed for the total fusion charge of the ring to contain the identity, so that the vacuum sector is present ($D_{(2,1)}$ is self-conjugate, and $D_{(2,1)}\times D_{(2,1)}\ni\mathbf{1}$). Radial quantisation about one strand then turns the ring eigenstates into the defect operators of the interface, and the two-strand geometry is also what allows the transmission coefficient to be measured from a current--current correlator across a single strand (\S\ref{sec:5.2}).

\paragraph{Antiunitary reflection symmetry.}
The reflection $P$ of the ring about the midpoint between the two strands exchanges the two sides of each strand and maps the braiding $B$ to its complex conjugate, so $PHP^{-1}=H^*$. Writing complex conjugation in the height basis as $\mathsf K$, the combination $P\mathsf K$ is therefore an antiunitary symmetry of $H$. This symmetry constrains the eigenstates but does not provide an additional invariant $\pm$ quantum number: rephasing a state changes its apparent antiunitary eigenphase. The sector labels used below are the eigenvalues of the unitary or Hermitian symmetry operators, in particular $(Y_{(2,1)},\eta)$. In the real $F$-gauge, $Y_x$ is real and $[P\mathsf K,Y_x]=[P,Y_x]\mathsf K$; consequently complex conjugation alone cannot turn a failure of $P$ to commute with $Y_x$ into commutation. The geometric exchange of holomorphic and antiholomorphic labels must be accompanied by the corresponding change of lasso junction. (A variant of the ring with $B^*$ at the second seam makes $P$ itself a unitary symmetry, with the corresponding change in the wrapped-line junctions; we used it only as a check.)

\subsection{The wrapped lines on the lattice}\label{sec:latlasso}

\paragraph{The junction compatible with the seam.}
Section~\ref{sec:lassos} showed that a line $x$ crossing the strand needs a junction $u\in\mathrm{Hom}(x\otimes x',x'\otimes x)$, with $x'$ the strand label, and that for $x=x'=\wl$ there are two independent ones, the over- and under-crossings $B$ and $B^{-1}$, of which the deformation preserves the over-crossing (\S\ref{sec:fixlasso}, Fig.~\ref{fig:lasso}). We now explain which junction the lattice realises and why it is the right one. The wrapped line $Y_x$ is built as follows (Fig.~\ref{fig:latlasso}). The ring is a fusion tree: a chain of trivalent vertices, one per slot, at which the site anyon $\ell_k$ ($\mathbf 1$ or $\wl$, and $x'$ at the strand slot) is fused into the running label, the heights $a_k$ being the labels of the segments between the vertices and the chain closing on itself. The line $x$ is a closed loop running parallel to the ring; it is fused into the tree at one point of one segment, which splits that segment into a part still carrying the old label and a part carrying the new label $b\in a\otimes x$ (Fig.~\ref{fig:latlasso}(a), top: $x$ attached to segment $5$, with $a_5$ on one side of the attachment point and $b_5$ on the other). The attachment point is then moved along the tree one vertex at a time. Moving it across the vertex of slot $k$ is the associator of the category, Fig.~\ref{fig:latlasso}(b): the fusion tree in which $x$ is fused into the label $a_k$ after the vertex is re-expanded in the basis in which $x$ is fused into the label $a_{k-1}$ before it, with amplitude $F^{\,x\,a_{k-1}\,\ell_k}_{\,b_k\,b_{k-1}\,a_k}$, so that the segment $k$ now carries $b_k$ throughout and the segment $k-1$ is split into $a_{k-1}$ and the new label $b_{k-1}\in a_{k-1}\otimes x$ (Fig.~\ref{fig:latlasso}(a), bottom). In the real gauge used here the $F$-symbols are orthogonal matrices, so the same symbol is the amplitude for the move in either direction. After the attachment point has gone once around the ring every segment carries a new label $b_k$, the two ends of the $x$ line meet again on the segment they started from, and the loop closes; the matrix element of $Y_x$ between the configurations $\{a\}$ and $\{b\}$ is the product of the $F$-symbols of all the moves, with the indices read cyclically, and the closure of the loop is automatic because the sum over the new labels in \eqref{eq:25} below includes the label of the starting segment. The strand is simply one more slot of the tree, with site anyon $\ell_k=x'$ and outgoing label $c_j$, so the $x$ line is moved across it by the \emph{same} $F$-move as across an ordinary site, the strand height changing as $c_j\to c_j'\in c_j\otimes x$; no $R$-matrix appears anywhere in $Y_x$. This pure-$F$ crossing is nevertheless a definite junction and not a third option. In the channel basis of $x\otimes x'$ (the channel projectors $P_{D'}$ of \S\ref{sec:2.3} are the junctions $u_{D'}$ of \S\ref{sec:lassos}) the two half-braidings expand with the $R$-matrix eigenvalues as coefficients, $B=\sum_{D'}R^{x x'}_{D'}\,u_{D'}$ and $B^{-1}=\sum_{D'}(R^{x x'}_{D'})^{-1}u_{D'}$; for $x=x'=\wl$, with the $R$-symbols \eqref{eq:Rsym} and $q=e^{i\pi/p}$,
\begin{equation}
B=-i\,q^{-3/2}\,u_{\mathbf 1}+i\,q^{1/2}\,u_{\mathrm{adj}},\qquad B^{-1}=i\,q^{3/2}\,u_{\mathbf 1}-i\,q^{-1/2}\,u_{\mathrm{adj}},
\label{eq:Bexp}
\end{equation}
so the two junctions differ by the relative phase $q^{\pm2}=e^{\pm2\pi i/p}$ between the two channels, Fig.~\ref{fig:latlasso}(c), which is exactly the phase that distinguishes passing in front of the strand from passing behind it. Why does a line built from $F$-moves alone commute with the braided seam term \eqref{eq:13}, and which of the two junctions is it? Both questions are answered by noticing what the planar $F$-move does and does not record. Place the trunk of the fusion tree and the line $x$ in the plane of Fig.~\ref{fig:latlasso}; the strand leaf $x'$ joins the trunk at its vertex $V$, and in three dimensions it may come down onto the plane from above or from below. The $F$-move that slides the attachment point of $x$ past $V$ is a planar operation on the trunk and on $x$, and it is the same matrix in both cases. The two cases differ in the side of $x'$ on which $x$ passes---under $x'$ if the leaf comes from above, over $x'$ if it comes from below---and they are related by one full passage of $x'$ through the plane around $x$. Figure~\ref{fig:twist} shows the two situations, before and after the move, flattened into the plane: to compare the move with a junction in $\mathrm{Hom}(x\otimes x',x'\otimes x)$ the leaf is brought down to the side of $x$, and it then crosses the trunk segment $a_{k-1}$ either over it (the leaf comes from above the plane) or under it (from below). In the fusion-tree basis the passage from one picture to the other is a number: the two differ by the double braiding of $x'$ with the trunk label it is fused to, which is the same as one full turn of the leaf $x'$ about its vertex $V$. On the lattice this full turn is the diagonal operator $U$ with eigenvalue $\theta_{c}/(\theta_{a}\theta_{x'})$ on a configuration whose strand vertex carries $(a,x'\to c)$, with $\theta_i=e^{2\pi i h_i}$ the twist of the $(r,1)$ label $i$. Since $Y_x$ is an operator from the configuration before the move to the one after it, passing to the other reading means applying the full turn on both sides, $Y'_x=U\,Y_x\,U^{-1}$. For the move of Fig.~\ref{fig:latlasso}(b), in which $x$ passes $V$ from right to left, the configuration before the move has $x$ to the right of $V$ and the trunk at $V$ carries $(a_{k-1},x'\to a_k)$, so $U^{-1}$ contributes the inverse of $\theta_{a_k}/(\theta_{a_{k-1}}\theta_{x'})$; the configuration after it has $x$ to the left of $V$, the trunk at $V$ carries $(b_{k-1},x'\to b_k)$ with $b_{k-1}\in a_{k-1}\otimes x$, and $U$ contributes $\theta_{b_k}/(\theta_{b_{k-1}}\theta_{x'})$. The $\theta_{x'}$ cancels and what remains is the ratio $\theta_{a_{k-1}}\theta_{b_k}/(\theta_{a_k}\theta_{b_{k-1}})$, the full turn of $x'$ around $a_{k-1}\otimes x$ divided by the full turn around $a_{k-1}$ alone, i.e.\ the monodromy of $x'$ with $x$; it is precisely what distinguishes the two junctions $B$ and $B^{-1}=B^{-2}B$ of \eqref{eq:Bexp}. ($U$ itself does not commute with $H$, so the fact that $UY_xU^{-1}$ is again a symmetry is not automatic; it is the lattice statement that both lassos are symmetries of a topological defect.) So the one planar operator corresponds to two three-dimensional situations, and the two lassos of \S\ref{sec:lassos} are realised on the lattice by the pure-$F$ line $Y_x$ of \eqref{eq:25} and by the same line with the $F$-symbol at the strand vertex multiplied by that ratio,
\begin{equation}
Y'_x=UY_xU^{-1}:\qquad F^{\,x\,a_{k-1}\,x'}_{\,b_k\,b_{k-1}\,a_k}\ \longrightarrow\ F^{\,x\,a_{k-1}\,x'}_{\,b_k\,b_{k-1}\,a_k}\ \frac{\theta_{a_{k-1}}\theta_{b_k}}{\theta_{a_k}\theta_{b_{k-1}}}\quad\text{at the strand slot}
\label{eq:Ydressed}
\end{equation}
(with the twists $\theta$ of the category whose $R$-matrix builds the seam \eqref{eq:12}; for a seam built from $B^*$ the conjugate twists are used). The planar line commutes with the seam term for either orientation of the braid, since a line that passes on one fixed side of two leaves is indifferent to how those two leaves pass each other (the naturality of the braiding, which in $F$- and $R$-symbols is the hexagon identity); the dressed line commutes with the seam whose orientation matches its dressing. We checked all of this on the undeformed single-strand rings at $p=4$ and $p=5$ ($L=6$), to machine precision: $Y_x$ commutes with the seam built from $B$, with the one built from $B^*$ and with a ring carrying one of each; $Y'_x$ of \eqref{eq:Ydressed} commutes with $H_0$ and with $Y_x$, is normal with exactly the Verlinde spectrum ($\{\pm\varphi,\pm\varphi^{-1}\}$ at $p=5$), and reads the \emph{right} label where $Y_x$ reads the left one: on the mirror pair $(1,2)\otimes\overline{(2,2)}$, $(2,2)\otimes\overline{(1,2)}$ at $\Delta=\tfrac{3}{20}$, $Y_{(2,1)}$ takes the values $(-\varphi,+\varphi^{-1})$ and $Y'_{(2,1)}$ the values $(+\varphi^{-1},-\varphi)$ (for the invertible line $\eta$ the two coincide at $p=5$; at $p=4$, its monodromy with a single $\sigma^\prime$ strand is $\theta_{\eta\otimes\sigma^\prime}/(\theta_\eta\theta_{\sigma^\prime})=-1$, giving $Y^\prime_\eta=-Y_\eta$ on the single-strand ring and $Y^\prime_\eta=Y_\eta$ on the two-strand ring). The lattice thus reproduces \S\ref{sec:fixlasso} literally: on the undeformed ring both lassos are exact symmetries, because the strand is topological, and the deformation decides between them.

\paragraph{Which lasso the deformation preserves, and why.}
The clean field $\Phi_{\mathrm{cl}}$ of \S\ref{sec:2.3} is built from the channel projectors $P^{\mathbf 1}$, $P^{\mathrm{adj}}$ of the pair formed by the strand and its neighbouring domain wall, eq.~\eqref{eq:17}. These projectors are planar objects of the fusion tree: they are written in the same basis as $Y_x$, with the strand leaf treated exactly like a site leaf, i.e.\ attached to the trunk from the same side of the plane. The closed wrapped categorical line can be moved past these compatible local fusion morphisms by naturality and the pentagon identity, so $[Y_x,\Phi_{\mathrm{cl}}]=0$ holds exactly, in the same way as $[Y_x,H_0]=0$. The dressed line $Y'_x=UY_xU^{-1}$ passes the strand on the other side of the plane, and $\Phi_{\mathrm{cl}}$ ties the strand to its neighbour: $[Y'_x,\Phi_{\mathrm{cl}}]=0$ would require $[Y_x,U^{-1}\Phi_{\mathrm{cl}}U]=0$, but the full turn $U$ of the strand leaf does not commute with a projector that fuses the strand with another leaf ($U^{-1}\Phi_{\mathrm{cl}}U$ is not planar), and indeed $\|[Y'_x,\Phi_{\mathrm{cl}}]\|=O(1)$ while $\|[Y_x,\Phi_{\mathrm{cl}}]\|<10^{-14}$. In the language of \S\ref{sec:fixlasso} this is the statement that the field created by $\Phi_{\mathrm{cl}}$ is neutral under one lasso and charged under the other: on the two-strand ring at $p=5$ ($L=6$ and $8$) the state $\Phi_{\mathrm{cl}}|\mathrm{vac}\rangle$ (vacuum component removed) is an eigenvector of both lassos to numerical precision at the tested sizes, $\|(Y_{(2,1)}-\varphi)\Phi_{\mathrm{cl}}|\mathrm{vac}\rangle\|<10^{-12}$ and $\|(Y'_{(2,1)}+\varphi^{-1})\Phi_{\mathrm{cl}}|\mathrm{vac}\rangle\|<10^{-12}$, with eigenvalue $+\varphi$ under $Y_{(2,1)}$, the neutral value, and $-\varphi^{-1}$ under $Y'_{(2,1)}$, the value of the right label $(3,3)\equiv(2,3)$, i.e.\ it is the driving doublet member $(1,3)\otimes\overline{(3,3)}$ of Table~\ref{tab:3} read by its left label under $Y_x$ and by its right label under $Y'_x$. The deformation therefore preserves $Y_x$ and breaks $Y'_x$, exactly as eq.~\eqref{eq:lassocond} requires of the junction that reads the deformation-neutral label. Which of the two sides of the plane is called the holomorphic one is a convention, the same convention that names the lattice deformation $\phi_{(1,3)}\otimes\bar\phi_{(3,3)}$ rather than its mirror image; with it, the pure-$F$ line is the over-crossing lasso of \S\ref{sec:fixlasso}, the one that survives the deformation, which is why $[Y_x,H]=0$ holds exactly (numerically to $10^{-14}$) at every coupling, for $\Phi_{\mathrm{cl}}$ as well as for $H_0$, and why the charges it assigns are those of the left label (Table~\ref{tab:3} below).

\begin{figure}[htp]\centering
\resizebox{0.8\textwidth}{!}{\input{figs/fig_latlasso_tikz.tex}}
\caption{The wrapped line on the lattice. (a) The ring as a fusion tree (eight slots, the strand at slot $3$ with its anyon $x'$, red, and strand height $c$; the chain closes on itself, slot $7$ being followed by slot $0$), with the symmetry line $x$ (blue) fused into one segment: top, $x$ attached to segment $5$, which carries $a_5$ on one side of the attachment point and $b_5\in a_5\otimes x$ on the other, the segments already passed ($6$, $7$) carrying their new labels; bottom, the attachment point moved one vertex to the left, across the vertex of slot $5$. (b) The move itself, the associator of the category: the tree with $x$ fused into the label after the vertex equals the sum over $b_{k-1}$ of the trees with $x$ fused into the label before it, with amplitude $F^{\,x\,a_{k-1}\,\ell_k}_{\,b_k\,b_{k-1}\,a_k}$; at the strand slot $\ell_k=x'$ and the label that changes is the strand height, and the move is the same. Repeating the move around the whole ring changes every segment label, $a_k\to b_k$ ($c\to c'$ at the strand), closes the loop, and gives the cyclic product of $F$-symbols of eq.~\eqref{eq:25}. (c) The crossing of $x$ with the strand decomposed in the fusion channels of $x\otimes x'$, eq.~\eqref{eq:Bexp}: the pure-$F$ move of (b) is one of the two junctions, $B$ in the convention of the text, and the other, $B^{-1}$, has the conjugate coefficients and is realised by the twist-dressed line \eqref{eq:Ydressed}.}\label{fig:latlasso}
\end{figure}
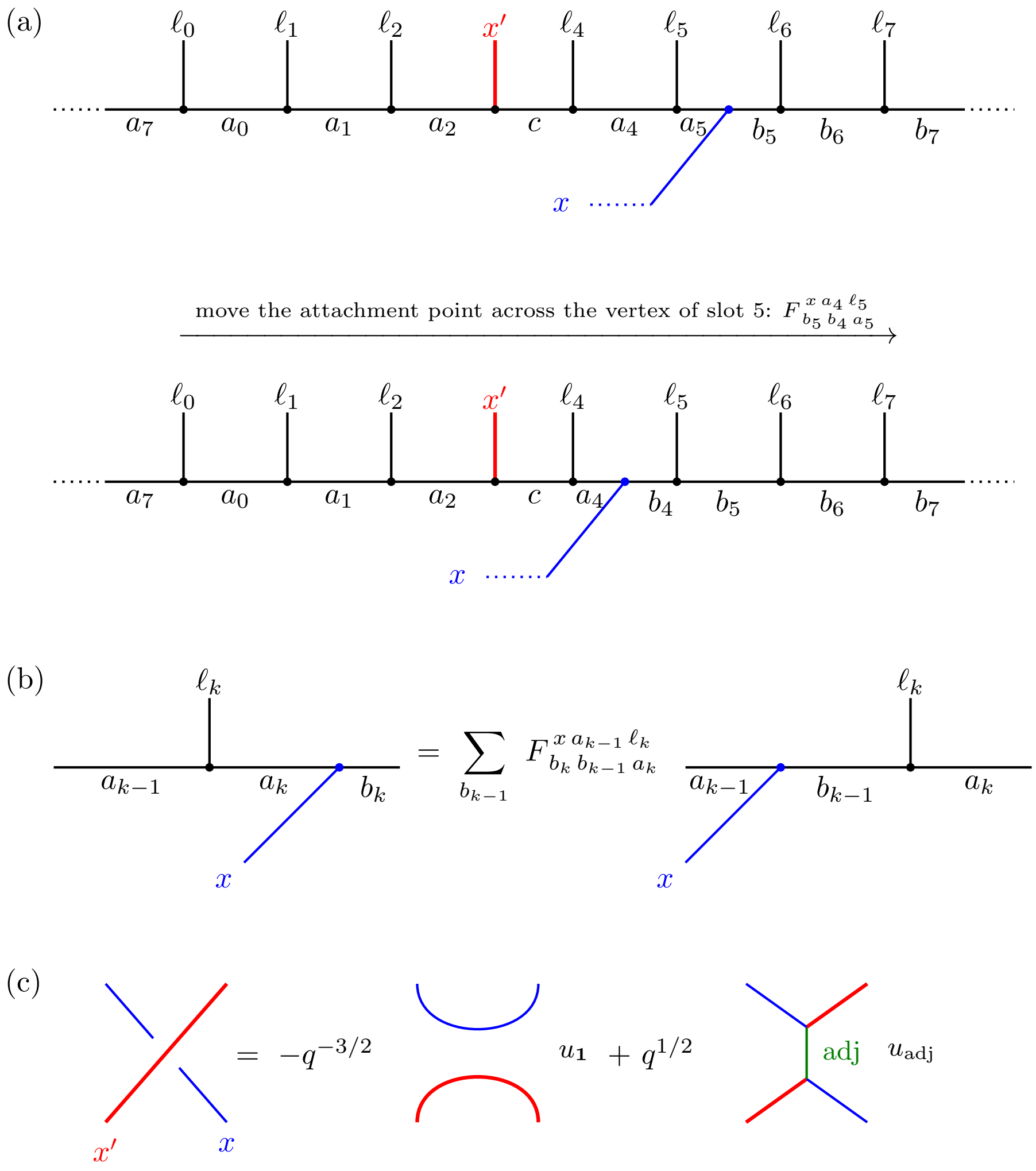

\begin{figure}[ht]\centering
\resizebox{0.9\textwidth}{!}{\input{figs/fig_twist_tikz.tex}}
\caption{The two junctions as one planar move. The trunk and the line $x$ lie in the plane; the strand leaf $x'$, attached at the vertex $V$, comes down onto the plane from above (left column) or from below (right column). Top row: the configuration before the move of Fig.~\ref{fig:latlasso}(b), with $x$ attached to the right of $V$ and the trunk at $V$ carrying $a_{k-1}$ and $a_k$; bottom row: after the move, with $x$ attached to the left of $V$ and the trunk at $V$ carrying $b_{k-1}\in a_{k-1}\otimes x$ and $b_k$. The planar $F$-move connecting the two rows is the same in both columns; the columns differ in the side of $x'$ on which $x$ passes. Flattened into the plane (the leaf brought down to the side of $x$ so that the move can be compared with a junction in $\mathrm{Hom}(x\otimes x',x'\otimes x)$), the leaf crosses the trunk over it (left) or under it (right), and in each row the two columns differ by one full turn of $x'$ about $V$, the double braiding of $x'$ with the trunk label it is fused to: $\theta_{a_k}/(\theta_{a_{k-1}}\theta_{x'})$ before the move and $\theta_{b_k}/(\theta_{b_{k-1}}\theta_{x'})$ after it. The left column is the pure-$F$ line $Y_x$, the right column the dressed line $Y'_x=UY_xU^{-1}$ of \eqref{eq:Ydressed}, whose matrix element carries the ratio of the two factors.}\label{fig:twist}
\end{figure}
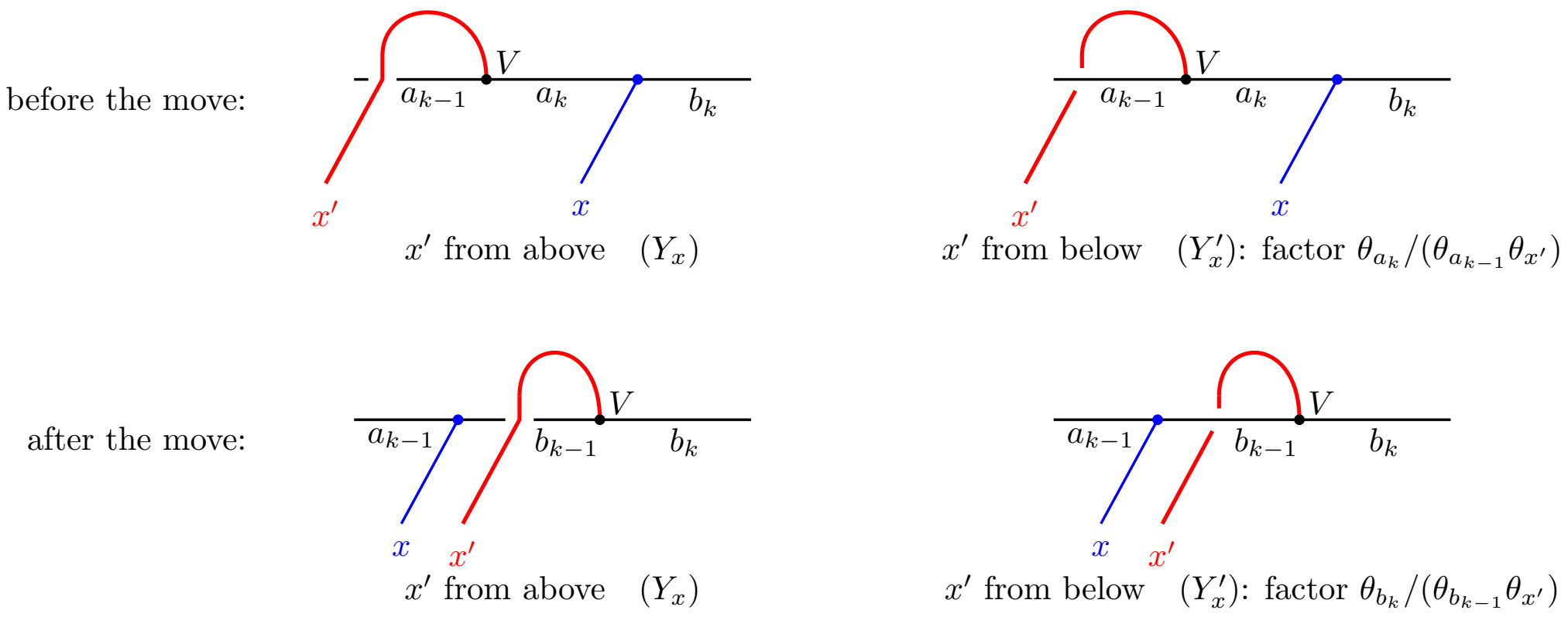

\paragraph{The lattice symmetry operators.}
Explicitly, the wrapped line $x$ is the operator on height configurations
\begin{equation}
Y_x\, |a_0, a_1, \dots\rangle = \sum_{\{b\}} \prod_k F^{\,x\ a_{k-1}\ \ell_k}_{\,b_k\ b_{k-1}\ a_k}\ |b_0, b_1, \dots\rangle, \quad
b_k \in a_k \otimes x, \quad
\ell_k = \begin{cases} \mathbf{1} & \text{site } k \text{ empty} \\ \wl & \text{site } k \text{ occupied} \\ x' & \text{strand slot,} \end{cases}
\label{eq:25}
\end{equation}
the product running over all slots of the ring, strand slots included; it is obtained by placing the $x$ line parallel to the ring and moving it across the tree one vertex at a time, each vertex contributing the associator $F$. For the invertible line $\eta = Y_{(p-1,1)}$ the fusion $a \otimes \eta$ has a single channel, $b = p{-}a$, and $Y_\eta$ is exactly the height reflection $a_k \mapsto p{-}a_k$ on every bond (strand bond included): all the $F$-symbols in \eqref{eq:25} multiply to $+1$ on every admissible configuration, so $Y_\eta$ is a pure permutation matrix. These are the operators used throughout the numerics: the ED charges and the $\eta$-block reduction of \S\ref{sec:2.5} use \eqref{eq:25} directly, and the DMRG uses its matrix-product form. For the exact diagonalisation these are sparse matrices in the admissible basis; the charges of the eigenstates are read off from $\langle\psi|Y_{(2,1)}|\psi\rangle$ and $\langle\psi|\eta|\psi\rangle$ after diagonalising $Y$ inside each degenerate multiplet of $H$ (the lines commute with $H$ and with each other, so this yields a simultaneous eigenbasis), and $\eta$, being a permutation of the basis, is also used to block-diagonalise $H$ (\S\ref{sec:2.5}). For the DMRG the same operators are written as matrix-product operators: $\eta$ has bond dimension 1 (a product of local reflections), and $Y_{(2,1)}$ has bond dimension $(p-1)^2$, the bond carrying the pair $(b_{k-1}, a_{k-1})$ needed by the next $F$-symbol, with the ring closure handled by summing over the boundary value of the bond index; they are used both to project on a charge sector (by adding a multiple of them to $H$) and to verify the charges of the converged states.

\paragraph{Check on the undeformed ring.}
Table~\ref{tab:3} tests \eqref{eq:24} on the undeformed two-strand ring at $p=5$: each of the ten lowest levels of $\mathcal{H}_{\mathbf{1}} \oplus \mathcal{H}_{(3,1)}$ lands in the sector predicted from its left label---including the two members of the driving doublet, $(1,3) \otimes \overline{(3,3)}$ in $a+$ and $(3,3) \otimes \overline{(1,3)}$ in $c+$, which have different left labels and are therefore distinguished by the preserved lasso, as they must be for the deformation to couple to only one of them.

\begin{table}[ht]
\caption{The undeformed two-strand ring at $p=5$ ($K = \lambda = 0$, $L=10$): lowest levels, their identification in $\mathcal{H}_{\mathbf{1}} \oplus \mathcal{H}_{(3,1)}$, the charges predicted by \eqref{eq:24} from the left label, and the sector in which the level is found numerically ($a+ = (\varphi,+)$, $b- = (\varphi^{-1},-)$, $c+ = (-\varphi^{-1},+)$, $d- = (-\varphi,-)$ in the notation of \S\ref{sec:5.3}). Weights of $M(5,6)$: $h_{1,2} = \tfrac18$, $h_{1,3} = \tfrac23$, $h_{2,1} = \tfrac25$, $h_{2,2} = \tfrac{1}{40}$, $h_{2,3} = \tfrac{1}{15}$, $h_{3,2} = \tfrac{21}{40}$, $h_{3,3} = \tfrac{1}{15}$, $h_{4,2} = \tfrac{13}{8}$.}\label{tab:3}
\centering
\adjustbox{max width=\textwidth}{\begin{tabular}{clccc}
\toprule
$\Delta$ (exact) & field $V_i \otimes \bar V_j$ & from & $(Y_{(2,1)}, \eta)$ predicted & sector, $\Delta$ at $L{=}10$ \\
\midrule
$0$ & $(1,1) \otimes \overline{(1,1)}$ & $\mathcal{H}_{\mathbf{1}}$ & $(+\varphi, +)$ & $a+$, 0.000 \\
$\tfrac{1}{20}$ & $(2,2) \otimes \overline{(2,2)}$ & $\mathcal{H}_{\mathbf{1}}$ and $\mathcal{H}_{(3,1)}$ & $(+\varphi^{-1}, -)$ & $b-$, 0.050 ($\times 2$) \\
$\tfrac{2}{15}$ & $(2,3) \otimes \overline{(2,3)}$ & $\mathcal{H}_{\mathbf{1}}$ and $\mathcal{H}_{(3,1)}$ & $(-\varphi^{-1}, +)$ & $c+$, 0.133 ($\times 2$) \\
$\tfrac{1}{4}$ & $(1,2) \otimes \overline{(1,2)}$ & $\mathcal{H}_{\mathbf{1}}$ & $(-\varphi, -)$ & $d-$, 0.250 \\
$\tfrac{13}{20}$ & $(1,2) \otimes \overline{(3,2)}$ & $\mathcal{H}_{(3,1)}$ & $(-\varphi, -)$ & $d-$, 0.651 \\
$\tfrac{13}{20}$ & $(3,2) \otimes \overline{(1,2)}$ & $\mathcal{H}_{(3,1)}$ & $(+\varphi^{-1}, -)$ & $b-$, 0.651 \\
$\tfrac{11}{15}$ & $(1,3) \otimes \overline{(3,3)}$ (driving field) & $\mathcal{H}_{(3,1)}$ & $(+\varphi, +)$ & $a+$, 0.734 \\
$\tfrac{11}{15}$ & $(3,3) \otimes \overline{(1,3)}$ (its mirror) & $\mathcal{H}_{(3,1)}$ & $(-\varphi^{-1}, +)$ & $c+$, 0.734 \\
$\tfrac{4}{5}$ & $(2,1) \otimes \overline{(2,1)}$ & $\mathcal{H}_{\mathbf{1}}$ and $\mathcal{H}_{(3,1)}$ & $(-\varphi^{-1}, +)$ & $c+$, 0.802 ($\times 2$) \\
$\tfrac{4}{3}$ & $(1,3) \otimes \overline{(1,3)}$ & $\mathcal{H}_{\mathbf{1}}$ & $(+\varphi, +)$ & $a+$, 1.340 \\
\bottomrule
\end{tabular}}
\end{table}

\paragraph{A single crossing.}
On a ring crossed once by $D$ the lattice $Y_x$ is a single lasso on the twisted sector $\mathcal{H}_D$, and the same over-crossing rule applies. We checked this on the undeformed single-strand rings: at $p=5$ all twelve lowest levels of $\mathcal{H}_{D_{(2,1)}}$ ($\Delta = \tfrac{2}{15}, \tfrac{3}{20}, \tfrac25, \tfrac{11}{20}, \tfrac{11}{15}, \dots$, e.g. $(1,2) \otimes \overline{(2,2)}$ and $(2,2) \otimes \overline{(1,2)}$ both at $\tfrac{3}{20}$) carry exactly the charges \eqref{eq:24} of their left label; at $p=4$ the same holds for both lines with the $\mathcal C_4$ $F$-symbols and the junction normalization used here. A change of sign in the normalization of the $(2,1)$ crossing junction flips $Y_{(2,1)}$ on a ring with a single crossing and cancels on a ring with an even number of crossings; this junction normalization is distinct from the categorical distinction between $\mathcal C_4$ and $SU(2)_2$. The single-strand spectra of Sec.~\ref{sec:4} are thus labelled by the Verlinde values of the left label: the $Y_{(2,1)} = -\sqrt2$ sector holds the fields with left label $(1,2)$, the $+\sqrt2$ sector those with left label $(1,1)$ and $(1,3)$, and $Y_{(2,1)}=0$ those with left label $(2,s)$.

\subsection{The clean deformation operator}\label{sec:2.3}

\paragraph{What the continuum asks for.}
The KRW flow \cite{ref1} is triggered by a local perturbation on the defect. The field that drives it is the non-chiral defect field $\phi_{(1,3)} \otimes \bar\phi_{(3,3)}$ of $D_{(2,1)}$, with $(h,\bar h) = (h_{1,3}, h_{3,3})$ and scaling dimension
\begin{equation}
\Delta_\phi = 1 - \varepsilon, \qquad \varepsilon = \frac{2(p-1)}{p(p+1)} \quad (\Delta_\phi = 0.7 \text{ at } p=4,\ \ \tfrac{11}{15} \text{ at } p=5),
\label{eq:15}
\end{equation}
so it is relevant; it carries spin $h_{1,3} - h_{3,3}$ ($\tfrac12$ at $p=4$, $\tfrac35$ at $p=5$), which is allowed for a field living on a line. Its mirror image $\phi_{(3,3)} \otimes \bar\phi_{(1,3)}$, of the same dimension and opposite spin, is also a defect field of $D_{(2,1)}$, but it sits in a different charge sector of the wrapped $(2,1)$ line (\S\ref{sec:2.2}): on the undeformed ring the two appear as one level at $\Delta = \tfrac{11}{15}$ in the sector $a+$ and one in $c+$ (in the notation of \S\ref{sec:5.3}). Among the nonidentity defect primaries in the symmetry sector of the identity, $\phi_{(1,3)} \otimes \bar\phi_{(3,3)}$ is the \emph{only} relevant one. A lattice operator that commutes with the wrapped lines, as ours does, connects the vacuum only to that sector, so the lattice flow is KRW's single-field perturbation of $D_{(2,1)}$; it does not preserve the exchange of the two sides of the defect, in agreement with the fact that the KRW trajectory from $D_{(2,1)}$ to $C$ lies off the symmetric diagonal $\lambda_l = \lambda_r$ of the $D_{(1,2)}$ flows (only the endpoints, $D_{(2,1)}$ and $C$, are symmetric). On the lattice we therefore seek a local operator at the strand with a nonzero overlap with this primary. Its remaining nonidentity relevant content is excluded by the preserved symmetry; identity and irrelevant contributions are allowed and do not obstruct the intended scaling limit. The identity shifts the defect energy, while irrelevant fields produce corrections to scaling, the leading symmetry-allowed primary being $\phi_{(1,3)} \otimes \bar\phi_{(1,3)}$ of dimension $2h_{1,3}$. Symmetry determines which fields may occur, but does not guarantee a nonzero overlap with the driving field. The overlaps below establish that overlap for our chosen operator and motivate a combination that suppresses the displayed admixtures.

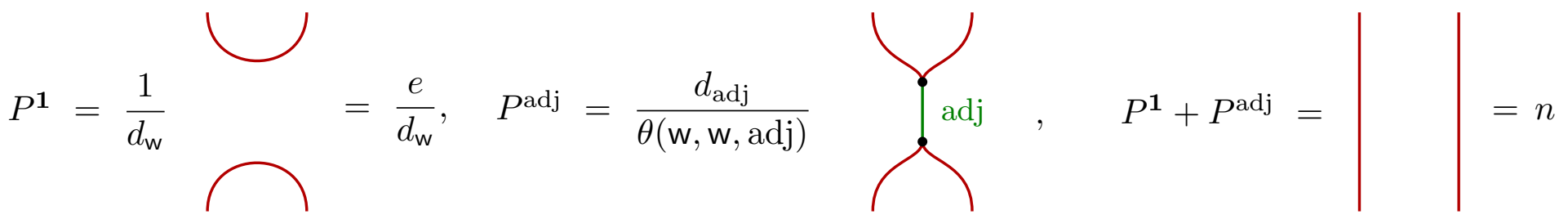
\begin{figure}[ht]\centering
\resizebox{0.85\textwidth}{!}{\input{figs/fig_proj_tikz.tex}}
\caption{The fusion-channel projectors of \eqref{eq:17} as fusion diagrams ($\wl$ lines in red). The projector on the channel $c$ fuses the two $\wl$'s (the domain wall on the site and the strand) into a $c$ line and splits them again, with the normalisation $d_c/\theta(\wl,\wl,c)$ that makes the diagram idempotent. For $c=\mathbf 1$ the channel line is trivial and $P^{\mathbf 1}$ is the cap--cup diagram divided by the loop value $d_\wl$, i.e.\ the Temperley--Lieb generator $e$ up to normalisation; in the height basis the cap and the cup each contribute one $F$-symbol, $F^{\,l\,\wl\,\wl}_{\,l;\,m\,\mathbf 1}=\sqrt{S_m/(d_\wl S_l)}$, which gives the weights $\sqrt{S_mS_{m'}}/(d_\wl S_l)$ of \eqref{eq:17}. For $c=\mathrm{adj}$ the cap and the cup are joined by an adjoint line (green). Their sum is the identity on two parallel strands, the occupation $n$ of \eqref{eq:18}.}\label{fig:proj}
\end{figure}

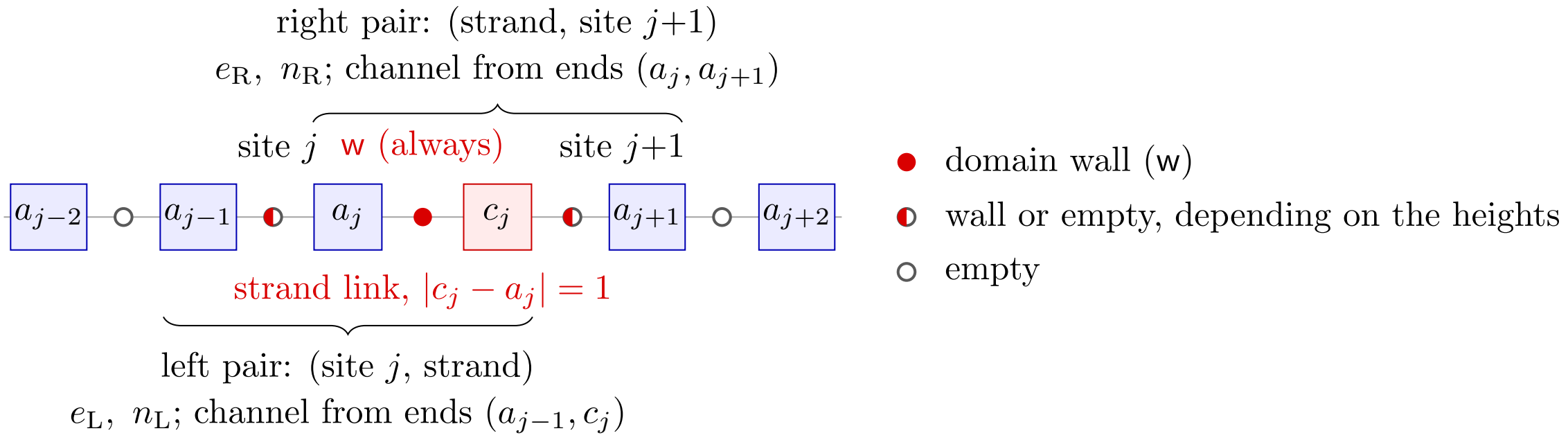
\begin{figure}[ht]\centering
\resizebox{\textwidth}{!}{\input{figs/fig5_tikz.tex}}
\caption{The heights around a strand. The strand is an extra bond $c_j$ (red box) inserted between $a_j$ and $a_{j+1}$; it is always a domain wall ($|c_j - a_j| = 1$), so the site between $a_j$ and $c_j$ always carries a $\wl$. The two neighbouring sites may or may not be occupied, depending on the heights. $e_{\mathrm{L}}, n_{\mathrm{L}}$ act on the left pair (site $j$, strand), whose fusion channel is read from the end heights $(a_{j-1}, c_j)$; $e_{\mathrm{R}}, n_{\mathrm{R}}$ act on the right pair (strand, site $j+1$) with end heights $(a_j, a_{j+1})$.}\label{fig:5}
\end{figure}

\paragraph{Two $\wl$'s next to each other.}
Fix a strand $x = \wl$ inserted after site $j$. Reading from left to right the relevant heights are (Fig.~\ref{fig:5})
\[
a_{j-1} \xrightarrow{\ \text{site } j\ } a_j \xrightarrow{\ \text{strand}\ } c_j \xrightarrow{\ \text{site } j+1\ } a_{j+1}.
\]
The strand bond $a_j$--$c_j$ is always a domain wall, i.e. always carries a $\wl$ (the defect anyon itself). Site $j$ carries a $\wl$ iff $a_{j-1} \neq a_j$, site $j+1$ iff $c_j \neq a_{j+1}$. Whenever a neighbouring site is occupied there are \emph{two} $\wl$'s side by side---the site's domain wall and the strand---and two $\wl$'s fuse in two ways,
\begin{equation}
\wl \otimes \wl = \mathbf{1} \oplus \mathrm{adj},
\label{eq:16}
\end{equation}
where ``adj'' is the object of $\mathcal C_p$ with the adjoint (spin-1) fusion label, the $\varepsilon$-like label ($\eta=(3,1)$ at $p=4$; $\tau_{\mathrm{Fib}}=(3,1)$ at $p=5$). In the fusion-tree basis the channel is read off from the heights at the two ends of the pair: for the left pair (site $j$, strand) the ends are $(a_{j-1}, c_j)$, for the right pair (strand, site $j+1$) they are $(a_j, a_{j+1})$. A path $l \to m \to r$ with $|m-l| = |r-m| = 1$ has $r \in l \otimes \wl \otimes \wl$, and
\[
r = l \pm 2 \ \Rightarrow\ \text{channel adj only}, \qquad r = l \ \Rightarrow\ \text{channels } \mathbf{1} \text{ and adj both possible}.
\]
On the $r = l$ block the identity channel is a single state, the quantum-dimension-weighted superposition of the middle height. This defines the two \emph{channel projectors} $P^{\mathbf{1}}$ and $P^{\mathrm{adj}}$ on a triple $(l,m,r)$:
\begin{equation}
P^{\mathbf{1}} \,|l,m,r\rangle = \delta_{l,r} \sum_{m'} \frac{\sqrt{S_m S_{m'}}}{d_\wl\, S_l}\, |l,m',r\rangle, \qquad
P^{\mathrm{adj}} = \big[\, l \neq m \text{ and } m \neq r \,\big] - P^{\mathbf{1}},
\label{eq:17}
\end{equation}
where the sum runs over heights $m'$ adjacent to both $l$ and $r$, $S_a$ and $d_\wl$ are as in \eqref{eq:5}, and $[\cdots]$ is the indicator that both sites of the triple are domain walls (i.e. that there really are two $\wl$'s to fuse). The identity $\sum_{m\,\mathrm{adj}\,l} S_m = d_\wl S_l$ gives $(P^{\mathbf{1}})^2 = P^{\mathbf{1}}$; $P^{\mathbf{1}} P^{\mathrm{adj}} = 0$; and $P^{\mathbf{1}} + P^{\mathrm{adj}}$ is the projector on the whole two-$\wl$ subspace. They are the height-basis images of the fusion-tree projectors $|\wl\wl;\mathbf{1}\rangle\langle\wl\wl;\mathbf{1}|$ and $|\wl\wl;\mathrm{adj}\rangle\langle\wl\wl;\mathrm{adj}|$, drawn in Fig.~\ref{fig:proj}. In the diagrammatic calculus the projector on the channel $c$ is the diagram in which the two $\wl$'s are fused into a $c$ line and split again, normalised by $d_c/\theta(\wl,\wl,c)$ with $\theta$ the theta-net of the vertex. For $c=\mathbf 1$ the connecting line is invisible and the diagram is the cap--cup divided by the loop value, $P^{\mathbf 1}=e/d_\wl$, whose matrix elements are the products of the two $F$-symbols $F^{\,l\,\wl\,\wl}_{\,l;\,m\,\mathbf 1}=\sqrt{S_m/(d_\wl S_l)}$ of \eqref{eq:17}; for $c=\mathrm{adj}$ the cap and the cup are joined by an adjoint line. The sum of the two projectors is the identity on two strands, i.e.\ the occupation $n$ of \eqref{eq:18} below, which is how $P^{\mathrm{adj}}$ is computed in practice. They commute with the wrapped lines $Y_{(r,1)}$.

\paragraph{The two natural building blocks.}
Two familiar local operators are linear combinations of these projectors: the Temperley--Lieb generator $e$ \cite{ref37}, which is $d_\wl$ times the identity-channel projector, and the occupation $n$, which is 1 exactly when the site next to the strand is occupied,
\begin{equation}
e \;=\; d_\wl\, P^{\mathbf{1}}, \qquad\qquad n \;=\; P^{\mathbf{1}} + P^{\mathrm{adj}}.
\label{eq:18}
\end{equation}
Each can be placed on either side of the strand. Written out on the four heights $(a_{j-1},$ $a_j,$ $c_j,$ $a_{j+1})$,
\begin{equation}
\begin{aligned}
n_{\mathrm{L}} &= \big[\, a_{j-1} \neq a_j \,\big], \qquad\qquad n_{\mathrm{R}} = \big[\, c_j \neq a_{j+1} \,\big], \\[4pt]
e_{\mathrm{L}}\, |a_{j-1}, a_j, c_j, a_{j+1}\rangle &= \delta_{a_{j-1},\,c_j}\ \big[\, a_{j-1} \neq a_j \,\big] \sum_{m' \neq a_{j-1}} \frac{\sqrt{S_{a_j} S_{m'}}}{S_{a_{j-1}}}\ |a_{j-1}, m', c_j, a_{j+1}\rangle, \\[4pt]
e_{\mathrm{R}}\, |a_{j-1}, a_j, c_j, a_{j+1}\rangle &= \delta_{a_j,\,a_{j+1}}\ \big[\, c_j \neq a_{j+1} \,\big] \sum_{m' \neq a_j} \frac{\sqrt{S_{c_j} S_{m'}}}{S_{a_j}}\ |a_{j-1}, a_j, m', a_{j+1}\rangle,
\end{aligned}
\label{eq:19}
\end{equation}
with $m'$ restricted to heights adjacent to the two fixed ends. Thus $e_{\mathrm{L}}$ acts on the triple (bond, site-$j$ wall, strand) and replaces the ordinary height $a_j$; $e_{\mathrm{R}}$ acts on the triple (strand, site-$(j{+}1)$ wall, bond) and replaces the \emph{strand} height $c_j$; $n_{\mathrm{L}}, n_{\mathrm{R}}$ are diagonal. The matrix of $e_{\mathrm{L}}, e_{\mathrm{R}}$ is the TL move $w_e$ of \eqref{eq:9} and Fig.~\ref{fig:2}, normalised so that $e^2 = d_\wl e$: it is the same Temperley--Lieb generator that builds the bulk energy operator, now straddling the strand. None of these four operators braids anything---they act on the strand bond $c_j$ exactly as on an ordinary bond---which is why they are local and Hermitian without any $B$-conjugation, and why they commute with the wrapped lines.

\paragraph{Measuring the field content.}
Which continuum fields does a given lattice operator $\mathcal{O}$ at the strand contain? Radial quantisation answers this directly: the low-lying eigenstates $|\Psi\rangle$ of the undeformed two-strand ring are the defect operators of $D_{(2,1)}$, so the matrix elements $\langle\Psi|\mathcal{O}|\mathrm{vac}\rangle$ are, up to normalisation, the coefficients of the corresponding fields in the scaling limit of $\mathcal{O}$. Table~\ref{tab:1} gives the absolute values of these overlaps for the natural candidates at $p=4$, where the defect spectrum is completely known, and Table~\ref{tab:2} repeats the exercise at $p=5$. The tables display magnitudes; linear combinations and cancellation tests use the signed or complex matrix elements before taking absolute values. The result is clear-cut: $n_{\mathrm{L}} + n_{\mathrm{R}}$ contains \emph{no} driving field at all---only the identity (a constant defect tension) and the irrelevant $\phi_{(1,3)} \otimes \bar\phi_{(1,3)}$; while $e_{\mathrm{L}} + e_{\mathrm{R}}$ does contain the driving field but drags along the same two admixtures.

\paragraph{The clean combination.}
Since both contaminations sit in the same two operators, and in the same ratio (Tables~\ref{tab:1} and \ref{tab:2}), a single combination of the two channel projectors removes both. It is the \emph{traceless} combination, the projectors weighted by their inverse quantum dimensions,
\begin{equation}
\frac{P^{\mathbf{1}}}{d_{\mathbf{1}}} - \frac{P^{\mathrm{adj}}}{d_{\mathrm{adj}}} \;\propto\; d_{\mathrm{adj}}\, P^{\mathbf{1}} - P^{\mathrm{adj}} \;=\; d_\wl\, e - n,
\label{eq:20}
\end{equation}
where we used \eqref{eq:18} and $d_\wl^2 = 1 + d_{\mathrm{adj}}$ (from $\wl \otimes \wl = \mathbf{1} \oplus \mathrm{adj}$). Summing the two sides of the strand,
\begin{equation}
\boxed{\;
\Phi_{\mathrm{cl}} \;=\; \Big( d_{\mathrm{adj}} P^{\mathbf{1}} - P^{\mathrm{adj}} \Big)_{\mathrm{L}} + \Big( d_{\mathrm{adj}} P^{\mathbf{1}} - P^{\mathrm{adj}} \Big)_{\mathrm{R}} \;=\; d_\wl \big( e_{\mathrm{L}} + e_{\mathrm{R}} \big) - \big( n_{\mathrm{L}} + n_{\mathrm{R}} \big)
\;}
\label{eq:21}
\end{equation}
At $p=4$ the adjoint is the invertible $\eta = (3,1)$, $d_{\mathrm{adj}} = 1$ and $d_\wl = \sqrt2 = 2/d_\wl$, so \eqref{eq:21} is simply the projector difference $P^{\mathbf{1}} - P^{\mathrm{adj}} = \tfrac{2}{d_\wl}(e_{\mathrm{L}} + e_{\mathrm{R}}) - (n_{\mathrm{L}} + n_{\mathrm{R}})$; at $p=5$ the adjoint is the Fibonacci object $\tau_{\mathrm{Fib}}=(3,1)$ with $d_{\mathrm{adj}} = \varphi$ and the clean field is $\varphi P^{\mathbf{1}} - P^{\mathrm{adj}} = \varphi(e_{\mathrm{L}} + e_{\mathrm{R}}) - (n_{\mathrm{L}} + n_{\mathrm{R}})$, which is \emph{not} the projector difference. The bottom rows of Tables~\ref{tab:1} and \ref{tab:2} show what this buys on the undeformed $D_{(2,1)}$ ring, at both values of $p$: among the three lowest primaries of the neutral sector, $\Phi_{\mathrm{cl}}$ connects the vacuum to the driving field only, its identity and $\phi_{(1,3)} \otimes \bar\phi_{(1,3)}$ components vanish, whereas at $p=5$ the naive difference $P^{\mathbf{1}} - P^{\mathrm{adj}}$ retains both (Table~\ref{tab:2}). The vanishing of the identity component is not an accident of the table: in the topological ground state the two channels of the strand and its neighbouring wall are populated in the ratio of their quantum dimensions, $\langle n\rangle=d_\wl\langle e\rangle$, so that $\langle\mathrm{vac}|\Phi_{\mathrm{cl}}|\mathrm{vac}\rangle=0$ holds to machine precision ($10^{-15}$) at every size we have checked, $L\le14$, at both $p$. This numerical observation is consistent with the interpretation of the identity component as a quantum trace, which vanishes for the channel asymmetry $d_{\mathrm{adj}}P^{\mathbf 1}-P^{\mathrm{adj}}$ since $d_{\mathrm{adj}}\,d_{\mathbf 1}-d_{\mathrm{adj}}=0$. We have not attempted a general proof relating this categorical trace to the interacting lattice ground-state expectation value at arbitrary size. Two qualifications are in order. First, like any lattice operator, $\Phi_{\mathrm{cl}}$ also has components on the descendants of the driving field and on higher primaries of the same lasso sector (the state $\Phi_{\mathrm{cl}}|\mathrm{vac}\rangle$ is an eigenvector of both lassos to numerical precision at the tested sizes, \S\ref{sec:latlasso}, which fixes the sector but not a single scaling field); these are irrelevant corrections, and what the construction needs is only that the relevant content of the perturbation be the driving field alone. Second, the symmetry argument of \S\ref{sec:fixlasso} says which fields a symmetric lattice operator \emph{may} contain, not that every symmetric operator contains the driving field; that $\Phi_{\mathrm{cl}}$ does, with the amplitudes of the tables, is a numerical fact. The meaning of \eqref{eq:21} is transparent. On each side of the strand it asks whether the defect anyon and its neighbouring domain wall fuse to the identity or to the adjoint, weighting the answer so that the trivial (channel-blind) part cancels: it is a Kondo-type coupling \cite{ref38,ref8} that distinguishes the two screening channels of the impurity anyon and realises the driving perturbation up to irrelevant corrections. 
The main structural result is that this channel asymmetry realises the KRW perturbation in the scaling limit, with the unwanted identity and leading irrelevant-primary overlaps vanishing to numerical precision at the tested sizes.

\begin{table}[ht]
\caption{Defect-field content of the candidate lattice operators on the $p=4$ $D_{(2,1)}$ ring ($|\langle\Psi|\mathcal{O}|\mathrm{vac}\rangle|$, $L=10$; absolute overlap magnitudes). Only $\Phi_{\mathrm{cl}}$ has vanishing identity and $\phi_{(1,3)}\otimes\bar\phi_{(1,3)}$ components; these overlaps vanish to numerical precision at the tested sizes (see text).}\label{tab:1}
\centering
\adjustbox{max width=\textwidth}{\begin{tabular}{lccc}
\toprule
$\mathcal{O}$ & $\mathbf{1}$ ($\Delta{=}0$) & $\phi_{(1,3)}{\otimes}\bar\phi_{(3,3)}$ $(0.7)$ & $\phi_{(1,3)}{\otimes}\bar\phi_{(1,3)}$ $(1.2)$ \\
\midrule
$e_{\mathrm{L}} + e_{\mathrm{R}}$ & 0.64 & 0.42 & 0.24 \\
$n_{\mathrm{L}} + n_{\mathrm{R}}$ & 0.91 & 0 & 0.34 \\
$\Phi_{\mathrm{cl}} = \tfrac{2}{d_\wl}(e_{\mathrm{L}}{+}e_{\mathrm{R}}) - (n_{\mathrm{L}}{+}n_{\mathrm{R}}) = \sqrt2(e_{\mathrm{L}}{+}e_{\mathrm{R}}) - (n_{\mathrm{L}}{+}n_{\mathrm{R}})$ & 0 & 0.59 & 0 \\
\bottomrule
\end{tabular}}
\end{table}

\begin{table}[ht]
\caption{Absolute overlap magnitudes, as in Table~\ref{tab:1}, at $p=5$ ($D_{(2,1)}$ two-strand ring, $L=10$, neutral sector). The identity and the irrelevant field enter $e_{\mathrm{L}}{+}e_{\mathrm{R}}$ and $n_{\mathrm{L}}{+}n_{\mathrm{R}}$ in the same ratio $1.62=\varphi=d_\wl$ (at $p=4$, Table~\ref{tab:1}, the ratio is $1.41=\sqrt2=d_\wl$), so the combination $d_\wl(e_{\mathrm{L}}{+}e_{\mathrm{R}}) - (n_{\mathrm{L}}{+}n_{\mathrm{R}})$ with $d_\wl = \varphi$ removes both; the projector difference $\tfrac{2}{d_\wl}(e_{\mathrm{L}}{+}e_{\mathrm{R}}) - (n_{\mathrm{L}}{+}n_{\mathrm{R}})$ does not.}\label{tab:2}
\centering
\adjustbox{max width=\textwidth}{\begin{tabular}{lccc}
\toprule
$\mathcal{O}$ & $\mathbf{1}$ ($\Delta{=}0$) & $\phi_{(1,3)}{\otimes}\bar\phi_{(3,3)}$ $(\tfrac{11}{15})$ & $\phi_{(1,3)}{\otimes}\bar\phi_{(1,3)}$ $(\tfrac{4}{3})$ \\
\midrule
$e_{\mathrm{L}} + e_{\mathrm{R}}$ & 0.64 & 0.54 & 0.19 \\
$n_{\mathrm{L}} + n_{\mathrm{R}}$ & 1.04 & 0 & 0.31 \\
$\tfrac{2}{d_\wl}(e_{\mathrm{L}}{+}e_{\mathrm{R}}) - (n_{\mathrm{L}}{+}n_{\mathrm{R}})$ & 0.25 & 0.66 & 0.07 \\
$\Phi_{\mathrm{cl}} = d_\wl(e_{\mathrm{L}}{+}e_{\mathrm{R}}) - (n_{\mathrm{L}}{+}n_{\mathrm{R}}) = \varphi(e_{\mathrm{L}}{+}e_{\mathrm{R}}) - (n_{\mathrm{L}}{+}n_{\mathrm{R}})$ & 0 & 0.87 & 0 \\
\bottomrule
\end{tabular}}
\end{table}

\paragraph{The deformed Hamiltonian.}
The model studied in the rest of the paper is
\begin{equation}
H \;=\; H_0 \;+\; \sum_{\text{strands}} \Big[\, K\, \Phi_{\mathrm{cl}} + \lambda\, \big( n_{\mathrm{L}} + n_{\mathrm{R}} \big) \Big],
\label{eq:22}
\end{equation}
with the two seams built as in \S\ref{sec:2.2}. The coupling $K$ drives the KRW flow; the second coupling $\lambda$ multiplies the channel-blind occupation, which contains only the identity and irrelevant fields and is used as a counterterm to tune the finite-size spectrum onto the fixed point (the ``improved point'' of Secs.~\ref{sec:4}, \ref{sec:5}). Both terms commute with the wrapped lines, so every eigenstate of \eqref{eq:22} keeps its $(Y_{(2,1)}, \eta)$ labels along the whole flow.

\subsection{Symmetry-resolved numerics}\label{sec:2.5}

The admissible Hilbert space grows quickly with $L$ (about $2.7 \times 10^6$ states at $p=4$, $L=16$, and $1.9 \times 10^6$ at $p=5$, $L=14$, with two strands), so the exact symmetries are used to block-diagonalise the Hamiltonian.
The wrapped invertible line $Y_{(p-1,1)} = \eta$ acts on the height basis as the reflection $a \mapsto p{-}a$, i.e. as a permutation of basis states, so projecting on its $\eta = \pm1$ eigenspaces halves the problem at no cost. At $p=4$, inside $\eta = +1$ the line $Y_{(2,1)}$ satisfies $Y_{(2,1)}^2 = \mathbf{1} + \eta = 2$, hence $Y_{(2,1)}/\sqrt2$ is a second involution that halves the block again. Each block is diagonalised separately (dense for small blocks, Lanczos otherwise) and the charges $(Y_{(2,1)}, \eta)$ are read back on the eigenvectors, which yields a fully sector-resolved spectrum at a fraction of the cost. The midpoint reflection combined with complex conjugation supplies an antiunitary consistency condition, rather than an additional sector label. For larger sizes ($L = 24, 32$ at $p=5$) we use DMRG \cite{ref40,ref41} on the same Hamiltonian, written as an MPO on the ring, isolating the charge sectors by adding to $H$ a multiple of the wrapped lines aligned with the target charges.

\subsection{Observables}\label{sec:3}

Before turning to the two flows we define the quantities that will be used to characterise an interface, and how each is measured on the ring of Sec.~\ref{sec:lattice}. Throughout, $L$ denotes the number of physical sites of the ring, which is also its linear size in the scaling limit.

\paragraph{Scaling dimensions.} In radial quantisation about one strand, the eigenstates of $H$ are the defect operators of the interface and their energies give the scaling dimensions \cite{ref31,ref32,ref33}
\begin{equation}
\Delta = \frac{(E-E_0)\,L}{2\pi v},
\label{eq:26}
\end{equation}
with $E_0$ the ground-state energy of the ring and $v$ the sound velocity of Sec.~\ref{sec:2.1}. For a topological line $D$ the spectrum of the two-strand ring is $\mathcal{H}_{D\times D} = \bigoplus_k N^{k}_{DD}\,\mathcal{H}_k$ (\S\ref{sec:lassos}), whose content is fixed by the fusion rules; for a non-topological conformal interface it is not known a priori and is one of the outputs of the computation. At a topological endpoint $D\times D$ contains the identity, so the ground state is the vacuum and \eqref{eq:26} measures dimensions from it.

\paragraph{Charge sectors.} The wrapped lines $Y_{(2,1)}$ and $\eta = Y_{(p-1,1)}$ commute with $H$ along the whole flow (\S\ref{sec:2.2}), so every level carries a pair of quantum numbers. Their eigenvalues on a state of the $(r',s')$ type are the Verlinde ratios $S_{(r,1),(r',s')}/S_{(1,1),(r',s')}$, where for a defect field $\phi_{(r',s)}\otimes\bar\phi_{(r'',s)}$ the \emph{left} label $(r',s)$ is the relevant one; the values we need are listed with the sectors in Secs.~\ref{sec:4} and \ref{sec:5}. At $p=4$, $Y_{(2,1)}\in\{+\sqrt{2},0,-\sqrt{2}\}$ and $\eta=\pm 1$; at $p=5$, $Y_{(2,1)} = Y_{\tau_{\mathrm{Fib}}}\,\eta$ with $Y_{\tau_{\mathrm{Fib}}}\in\{\varphi,-\varphi^{-1}\}$ on the Fibonacci-neutral and Fibonacci-charged states respectively, and $\eta=\pm 1$.

\paragraph{The $g$-function.} The $g$-function of an interface is its ``ground-state degeneracy'' $g=\langle 0|D|0\rangle$ \cite{ref42}, the overlap of the interface with the vacuum on the cylinder; it equals the quantum dimension for a topological line ($g_\eta = 1$, $g_{D_{(2,1)}} = \sqrt{2}$ at $p=4$ and $\varphi$ at $p=5$) and decreases along any RG flow of the interface \cite{ref42,ref43,ref44}. We extract it from the entanglement entropy, using the fact that the boundary or impurity entropy appears as an additive constant in the entanglement entropy of an interval that contains the defect \cite{ref45,ref46,ref47,ref48}. Let $S_D(\ell)$ be the von Neumann entropy of an interval of $\ell = 2k+1$ bonds centred on a strand (the strand bond included) in the ground state of the ring, and $S_{\mathrm{bulk}}(\ell)$ the same quantity for an interval of $\ell$ bonds in the pure chain. For an interval containing the impurity in its interior the $\ell$-dependence cancels in the difference and
\begin{equation}
\ln g \;=\; S_D(\ell) - S_{\mathrm{bulk}}(\ell)
\label{eq:27}
\end{equation}
up to corrections that vanish as $\ell,L\to\infty$. The method is calibrated on the two exactly known topological cases at $p=4$ (\S\ref{sec:4}).

\paragraph{The transmission coefficient.} A conformal interface is characterised, beyond $g$, by how much energy it lets through. Following Quella, Runkel and Watts \cite{ref7} (see also \cite{ref6,ref49}), the transmission coefficient $\mathcal{T}$ is defined by the two-point function of the stress tensor across the interface, $\langle T(z)\,T(w)\rangle$ with $z$ and $w$ on opposite sides, normalised by its bulk value: $\mathcal{T}=1$ for a topological line (which is transparent) and $\mathcal{T}=0$ for a factorising interface (a pair of boundaries). On the lattice we use the energy current
\begin{equation}
J_j = i\,[h_j, h_{j+1}],
\label{eq:28}
\end{equation}
the time derivative of the local energy; being the divergence of a conserved density, its scaling limit is proportional to $T-\bar T$ with no admixture of the identity or of scalar primaries, and its bulk two-point function decays as $1/x^4$. Each $J_j$ acts on a block of four consecutive sites. The measurement is made on the two-strand ring as follows. We take the ground state of the ring and evaluate the connected correlator $\langle J_iJ_j\rangle_c = \langle J_iJ_j\rangle - \langle J_i\rangle\langle J_j\rangle$ for every pair of blocks with $i$ in one half of the ring and $j$ in the other, so that the pair straddles the strand at site $L-1$; blocks that overlap either strand are excluded. The same correlator is evaluated in the ground state of the pure ring of the same length, as a function of the separation $d$ of the two blocks, and averaged over the position of the pair. The estimator is the ratio
\begin{equation}
\mathcal{T}(d) = \frac{\langle J_iJ_j\rangle_c^{\text{ring with strands}}}{\langle J_iJ_j\rangle_c^{\text{pure ring}}}\,, \qquad d = \text{separation of the blocks } i,j,
\label{eq:29}
\end{equation}
which divides out the non-universal normalisation of $J$, the $1/d^4$ decay and the curvature of the ring. Several pairs share the same $d$ and differ in where the two blocks sit relative to the two strands; in the continuum the ratio depends only on $d$, on the lattice the pairs scatter at the percent level. We therefore quote, for each $d$, the mean over all admissible pairs together with the standard error of that mean, $\sigma/\sqrt{n_{\mathrm{pairs}}}$. Two systematic effects have to be kept in mind. At small $d$ one of the two blocks is necessarily close to the strand and the ratio contains the reflected image contribution; $\mathcal{T}(d)$ starts well above its asymptotic value and decreases with $d$ until it reaches a plateau, which is why the number quoted for a given $L$ is the antipodal value $d=L/2$, where both blocks are as far from both strands as the ring allows. In the DMRG calculations the ring is represented by an open matrix-product state whose two ends are joined by the long-range terms of the MPO; the truncation error is largest across that cut, and a block lying on it gives a connected correlator that is low by 1--2\% at $\chi=200$. Pairs with a block within two sites of the cut are therefore discarded (the cut sits at the centre of one half-ring, away from the strands). Finally the antipodal values are extrapolated linearly in $1/L$.

\paragraph{Lattice $\beta$-function.} The KRW flow can be read off directly on the lattice, by the finite-size scaling method introduced by L\"uscher, Weisz and Wolff for asymptotically free theories \cite{ref51}: a running coupling is \emph{defined} by a finite-volume observable (there the mass gap of the $O(3)$ $\sigma$-model in a box of size $L$, building on the small-volume expansion of \cite{ref54}), and its change from $L$ to $sL$, the step-scaling function, is the integrated $\beta$-function of that scheme. Developed into the Schr\"odinger-functional scheme \cite{ref55,ref56}, this is how the running coupling of QCD is computed non-perturbatively \cite{ref57}. In the statistical-mechanics literature the same idea, applied to an exponent-like finite-size observable, is the phenomenological renormalisation group of Nightingale and of Roomany and Wyld \cite{ref52,ref53}. One needs a dimensionless running coupling, defined by a finite-volume quantity, that vanishes at the UV defect, is monotonic along the flow and takes a definite value at each fixed point. We take it to be a finite-volume Casimir energy of the \emph{single-strand} ring (one $D_{(2,1)}$ strand, spectrum $\mathcal{H}_D$ of \S\ref{sec:lassos}), which is precisely the type of coupling used in \cite{ref51}: the $L$-dependent part of the energy of one fixed state of the ring. At the UV point the two lowest states of $\mathcal{H}_{D_{(2,1)}}$ are the mirror pair $(1,2)\otimes\overline{(2,2)}$ and $(2,2)\otimes\overline{(1,2)}$, degenerate at
\begin{equation}
\Delta_{k=1} = h_{1,2}+h_{2,2} = \frac{(p-1)^2+2}{4p(p+1)} \qquad (\tfrac{11}{80} \text{ at } p=4,\ \tfrac{3}{20} \text{ at } p=5),
\label{eq:31}
\end{equation}
exchanged by the swap of the two sides of the defect and distinguished by their left labels, hence by their $(Y_{(2,1)},\eta)$ sectors. We track the member with left label $(1,2)$, on which $Y_{(2,1)}$ takes the Verlinde value $S_{(2,1),(1,2)}/S_{(1,1),(1,2)}=-2\cos(\pi/p)$, so that it sits in the sector $(Y_{(2,1)},\eta)=(-\sqrt2,+1)$ at $p=4$ and in $(-\varphi,-1)$, the sector $d{-}$ of \S\ref{sec:5.3}, at $p=5$, and call it the $k=1$ state (in the loop model of Section~\ref{sec:loop} it is the ground state of the sector whose non-contractible loops carry the weight $-2\cos(\pi/p)$). At integer $p$ it remains the lowest state of its charge sector along the whole flow, so the conserved charges identify it at every $K$; its mirror partner is the ground state of the ring and barely moves. Let $E^{(k=1)}(L,K)$ be its energy and $\delta E(L,K)=E^{(k=1)}(L,K)-E^{(k=1)}(L,0)$. Writing $\ell$ for the midpoint of two consecutive ring sizes $\ell\mp1$ (the ring is computed for even $L$, so $\ell=7,9,11,\dots$), the running coupling attached to the size $\ell$ is the symmetric size difference
\begin{equation}
u_C(\ell,K)=\frac{\ell^2-1}{4\pi v}\bigl[\delta E(\ell-1,K)-\delta E(\ell+1,K)\bigr],
\label{eq:48}
\end{equation}
a lattice version of $-\ell^2\partial_\ell\delta E/2\pi v$ evaluated at $\ell$. Since $E(L,K)=e_{\mathrm{bulk}}L+e_{\mathrm{def}}(K)+\frac{2\pi v}{L}(x-\frac{c}{12})+\ldots$ \cite{ref32,ref33,ref31}, subtracting $K=0$ removes the bulk energy and the size difference removes the non-universal defect energy, so $u_C\to x(K)-x(0)$, the change of the scaling dimension of the tracked state. It vanishes at $D_{(2,1)}$ by construction and takes a universal value at a fixed point: on the $\eta$ line of $p=4$ the state becomes $(\tfrac1{10},\tfrac35)$ of the $\mathbb{Z}_2$-twisted sector, so $u_C^*=0.7-\tfrac{11}{80}=0.5625$, while at a conformal interface its value is one of the outputs. Since $u_C$ is dimensionless, its logarithmic size derivative at fixed $K$,
\begin{equation}
B_C(K;\ell_1,\ell_2) \;=\; \frac{du_C}{d\ln \ell} \;\simeq\; \frac{u_C(\ell_2,K)-u_C(\ell_1,K)}{\ln(\ell_2/\ell_1)},
\label{eq:37}
\end{equation}
is a lattice $\beta$-function for the coordinate $u_C$: it vanishes where $u_C$ is scale-invariant (a fixed point), and its slope at a zero is minus the RG eigenvalue of the perturbation that moves $u_C$ away from it, so that a zero with negative slope is IR-attractive. In the UV, $u_C\propto K\ell^{1-\Delta_\phi}$ to first order, so the curve leaves the origin with slope $1-\Delta_\phi$. Plotting $B_C$ against the running coordinate $A_C=\tfrac12[u_C(\ell_1)+u_C(\ell_2)]$ for successive size pairs collapses the data onto a single curve if scaling holds.

\paragraph{Improved points.} A lattice Hamiltonian of the family \eqref{eq:22} sits at a fixed point only up to irrelevant operators, whose amplitudes depend on $(K,\lambda)$ and produce $1/L$ corrections to every quantity. We call an \emph{improved point} a choice of $(K,\lambda)$ at which the leading corrections are minimised, in the spirit of Symanzik improvement \cite{ref50}, as judged by the flatness of the finite-size spectrum in $1/L$. Different improved points are different lattice representatives of the same continuum interface.

\paragraph{Numerical methods.} Exact diagonalisation (ED) uses the symmetry-reduced blocks of \S\ref{sec:2.5} for $L\le 14$ (up to 16 for a few quantities). Density-matrix renormalisation group (DMRG) calculations \cite{ref40,ref41} treat the same ring at $L=24,32,48$ ($p=4$) and $L=24,32,48$ ($p=5$) with matrix-product states of bond dimension up to $\chi=300$; the vacuum and the first excited state of the neutral sector are obtained by adding to $H$ a multiple of the wrapped lines that penalises the other charge sectors, and excited states by orthogonalisation against the states already found. The entropies $S_D, S_{\mathrm{bulk}}$ and the current correlators are evaluated directly on the converged states.

\section{\texorpdfstring{Results at $p=4$: $D_{(2,1)}\to\eta$}{Results at p = 4: D(2,1) -> eta}}\label{sec:4}

For the tricritical Ising model the category is Ising, $D_{(2,1)}=\sigma'$ has $g=\sqrt{2}$, and the only candidate endpoints of a flow that preserves the $(r,1)$ symmetry and lowers $g$ are the invertible line $\eta$ ($g=1$) and the identity ($g=1$ as well). 
KRW do not determine the endpoint at $p=4$: the last model in which their TCSA shows convincing evidence of the defect $C$ is $M(5,6)$, while in $M(4,5)$ the signature of the fixed point is no longer seen and, in their words, the position of the conformal fixed point may have moved a large distance or may require a higher truncation level (\cite{ref1}, \S6.2.4 and \S7). 
They do note that at $p=4$ a factorising defect, $D_{(2,1)}F^{1|1}$, is an additional candidate with $g$ below that of $D_{(1,2)}$ (\cite{ref1}, \S4). The invertible line $\eta$ was not among the candidates considered by KRW. It is, however, singled out by the symmetry argument of Section~\ref{sec:cft} combined with the $g$-theorem: a flow that preserves the over-crossing lasso and lowers $g$ below $\sqrt2$ can only end on a defect that commutes with the $(r,1)$ lines and has $g\le\sqrt2$, which leaves $\eta$, the identity, and the factorising defects, and the lattice decides between them. The result of this section is that the endpoint is $\eta$: the KRW flow of the tricritical Ising model ends on a topological, invertible line rather than on a genuinely conformal defect. (See \cite{ref12} for the conformal defects of the tricritical Ising model.) The bulk chain used here, the dilute $A_3$ model at the integrable critical point of \cite{ref14}, belongs to the same class of Ising-fusion-category chains (labels $\mathbf{1},\sigma,\psi$, i.e.\ $\mathbf{1},\wl,\eta$ in our notation, with the fusion constraint on neighbours) as the two-parameter model of Roychowdhury and Wang \cite{ref18}. The integrable weights \eqref{eq:9} place us directly at that point with no tuning. 

\paragraph{Calibration of the $g$-function.} We first apply \eqref{eq:27} to the two undeformed topological rings, with strands $\eta$ and $\sigma'$ respectively:
\[
\eta = D_{(3,1)}:\ \ln g = -0.01\ (g=0.99), \qquad \sigma' = D_{(2,1)}:\ \ln g = +0.335\ (g=1.40),
\]
in agreement with $g_\eta = 1$ and $g_{\sigma'} = \sqrt{2} = 1.414$. The residual $\pm 0.01$ sets the accuracy of the method at these sizes.

\paragraph{Endpoint of the flow.} Driving $D_{(2,1)}$ with the clean field $\Phi_{\mathrm{cl}}$ and following $\ln g$ as a function of $g$, the entropy drops from $\ln\sqrt{2}$ and saturates at $\ln g = -0.04$ ($g=1.00$) at strong coupling, stable against further increase of $K$: the flow ends on an interface with $g=1$, and the sector-resolved spectrum below identifies it as $\eta$. 

\paragraph{Improved point.} The endpoint $\eta$ is invertible and topological. It corresponds to a conformal phase of \eqref{eq:22}.  We can tune away the leading finite-size effect by searching for a point with good conformal tower. The couplings are
\begin{equation}
{\,(K,\lambda)_\eta = (4.82,\,0.17).\,}
\label{eq:30}
\end{equation}

\begin{figure}[ht]\centering
\includegraphics[width=0.9\textwidth]{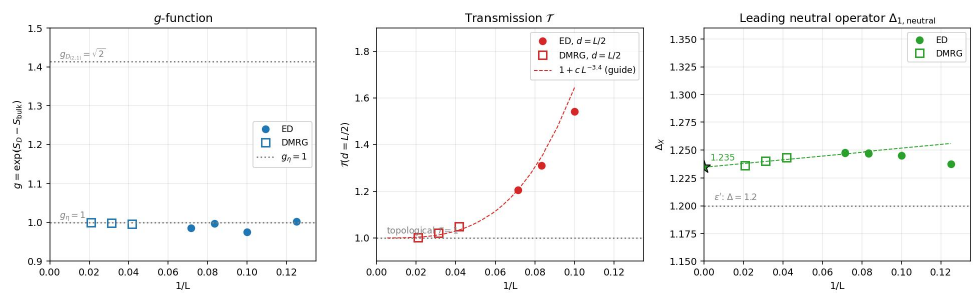}
\caption{$p=4$ at the improved point $(K,\lambda) = (4.82,0.17)$: finite-size dependence of the $g$-function (left), the transmission $\mathcal{T}$ at antipodal separation $d=L/2$ (centre) and the dimension of the leading neutral irrelevant operator (right). Filled circles: ED at $L=8,\dots,14$; open squares: DMRG at $L=24,32,48$. Dotted grey lines: reference values ($g_\eta = 1$ and $g_{D_{(2,1)}} = \sqrt{2}$; the topological $\mathcal{T}=1$; $\Delta_{\varepsilon'} = 1.2$). The dashed curve in the centre panel is a power-law guide $1 + cL^{-2.7}$ through the points; the dashed line and star on the right are the linear-in-$1/L$ extrapolation of the $L\ge 12$ points.}\label{fig:6}
\end{figure}

\begin{figure}[ht]\centering
\includegraphics[width=0.9\textwidth]{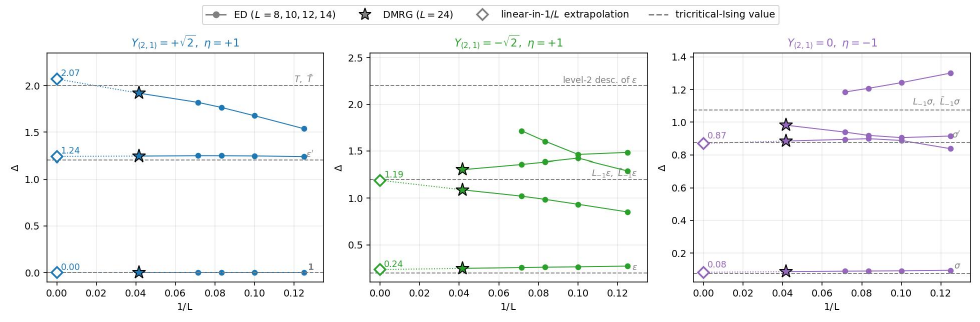}
\caption{$p=4$ sector-resolved spectrum at the improved point $(K,\lambda) = (4.82,0.17)$, $\Delta$ versus $1/L$, one panel per $(Y_{(2,1)},\eta)$ sector, showing the three (four) lowest levels of each sector from ED at $L=8,10,12,14$ (circles) and the three lowest from DMRG at $L=24$ (stars); lines connect the same level across sizes. Grey dashed lines: the diagonal tricritical-Ising spectrum expected on the $\eta$ line, primaries and their first descendants ($T,\bar T$ at 2 above the identity; the $L_{-1},\bar L_{-1}$ pairs at 1.2 above $\varepsilon$ and at 1.075 above $\sigma$; the level-2 descendants of $\varepsilon$ at 2.2). Open diamonds at $1/L=0$: linear-in-$1/L$ extrapolations of the leading levels through the three largest sizes. The primaries converge to within a few percent; the descendant pairs are still split at these sizes and approach their common value from both sides. Every level is consistent with the invertible-$\eta$ spectrum, confirming $D_{(2,1)}\to\eta$.}\label{fig:7}
\end{figure}

\paragraph{$g$-function, transmission and the leading irrelevant dimension.} Figure~\ref{fig:6} shows the three observables of Sec.~\ref{sec:3} at the improved point as functions of $1/L$, in the same format as for $C_5$ below, from ED at $L=8$--$14$ and DMRG at $L=24$, 32 and 48. The DMRG uses the commuting lines to restrict the search to the neutral sector: the ground state and first excited state of $H' = H - 1.0\,Y_{(2,1)} - 0.8\,\eta$, which pushes every non-neutral sector up by at least 1.4, are the vacuum and the leading neutral excitation of $H$ (their charges come out as $\langle Y_{(2,1)}\rangle = 1.4142$, $\langle\eta\rangle = 1.0000$). At $p=4$ the local Hilbert space dimension is 3, and the Hamiltonian MPO has bond dimension 20, so $\chi=160$ (truncation error $2\times 10^{-9}$) is enough. (i) The $g$-function from \eqref{eq:27} is flat at $g=0.99(2)$ for $L=8$--$14$ (individual values 1.002, 0.975, 0.998, 0.985) and $g = 0.9965, 0.9980, 0.9993$ at $L=24,32,48$, i.e. at
\noindent the invertible value $g_\eta = 1$ within the $\pm 0.01$ accuracy of the calibration, and far from $g_{D_{(2,1)}} = \sqrt{2}$. (ii) The transmission \eqref{eq:29} at $d=L/2$ is 1.54, 1.31, 1.21 for $L=10,12,14$ and 1.049, 1.021, 1.002 at $L=24,32,48$, above the topological value 1 and decreasing towards it approximately as a power law, $\mathcal{T}-1\propto L^{-2.7}$. 
A value above 1 is not possible for a conformal interface and is a short-distance artefact.
We check that the estimator itself is exact for a topological line: on the undeformed $\eta$ and $\sigma'$ rings at $L=12$ it gives $\mathcal{T}=1.000$ at every separation. The approach to 1 from above is therefore consistent with a topological endpoint, in contrast with the $p=5$ case where $\mathcal{T}$ approaches a value clearly below 1 from above (\S\ref{sec:5.2}). (iii) The leading irrelevant defect field in the neutral sector ($Y_{(2,1)} = +\sqrt{2}$, $\eta=+1$) has $\Delta = 1.238, 1.245, 1.247, 1.248$ for $L=8$--$14$ and $1.244, 1.241, 1.236$ at $L=24,32,48$, extrapolating to 1.23: it is the image of the driving field along the flow and is identified with the bulk field $\varepsilon'$ of dimension 1.2, the residual $\sim 3\%$ being of the same size as for the other primaries in Fig.~\ref{fig:7}.

\paragraph{Sector-resolved spectrum.} At the IR endpoint each strand has become $\eta$, and $\eta\times\eta = \mathbf{1}$, so the ring spectrum must be the untwisted, diagonal spectrum of the tricritical Ising model itself, with the Verlinde eigenvalues
\[
\begin{gathered} Y_{(2,1)} = +\sqrt{2},\ \eta = +1:\ \{\mathbf{1},\varepsilon'\}, \qquad
Y_{(2,1)} = -\sqrt{2},\ \eta = +1:\ \{\varepsilon,\varepsilon''\}, \\
Y_{(2,1)} = 0,\ \eta = -1:\ \{\sigma,\sigma'\}, \end{gathered}
\]
where $\varepsilon,\varepsilon',\varepsilon'',\sigma,\sigma'$ are the primaries of dimension $\Delta = 0.2, 1.2, 3, 0.075, 0.875$. Figure~\ref{fig:7} shows the three sectors at the improved point for $L=8,\dots,14$ (ED) and $L=24$ (DMRG, three states per sector, obtained as the lowest states of $H$ shifted by the wrapped lines aligned with the sector charges, as in \S\ref{sec:5.3}). Linear-in-$1/L$ extrapolation of the leading levels through the three largest sizes gives
\[
\Delta \to 0.08\ (\sigma,\ 0.075), \qquad 0.24\ (\varepsilon,\ 0.2), \qquad 0.87\ (\sigma',\ 0.875), \qquad 1.24\ (\varepsilon',\ 1.2),
\]
and the descendants behave as they should. Together with $g\to 1$ this establishes $D_{(2,1)}\to\eta$ or $D_{(2,1)}\to \mathbf{1}$ at $p=4$. The single strand spectrum will distinguish the two cases.

\paragraph{Lattice $\beta$-function.} We now apply the step-scaling construction of \S\ref{sec:3}: the Casimir coupling $u_C$ of \eqref{eq:48}, built from the $k=1$ state of the single-strand ring, which at $p=4$ is the lowest state of the sector $(Y_{(2,1)},\eta)=(-\sqrt2,+1)$ (its mirror partner, the lowest state of $(0,-1)$, is the ground state of the ring). On the $\eta$ line the tracked state becomes $(\tfrac1{10},\tfrac35)$ of the $\mathbb{Z}_2$-twisted sector, so the fixed-point value is $u_C^*=0.7-\tfrac{11}{80}=0.5625$. We follow the ray through the improved point $(K,\lambda)_\eta = (4.82,0.17)$, i.e. $H = H_0 + K\,\Phi_{\mathrm{cl}} + \lambda\,(n_{\mathrm{L}}+n_{\mathrm{R}})$ with $\lambda = 0.035\,K$ (the counterterm is kept in fixed proportion to the clean coupling), for couplings $0\le K\le 3$ in steps of $0.1$ ($0.05$ for $K\le0.2$), with $E^{(k=1)}(L,K)$ from the symmetry-reduced ED of \S\ref{sec:2.5} at $L=6,8,\dots,16$, i.e.\ $\ell=7,9,\dots,15$ and the size pairs $(7,9),\dots,(13,15)$.

\begin{figure}[ht]\centering
\includegraphics[width=\textwidth]{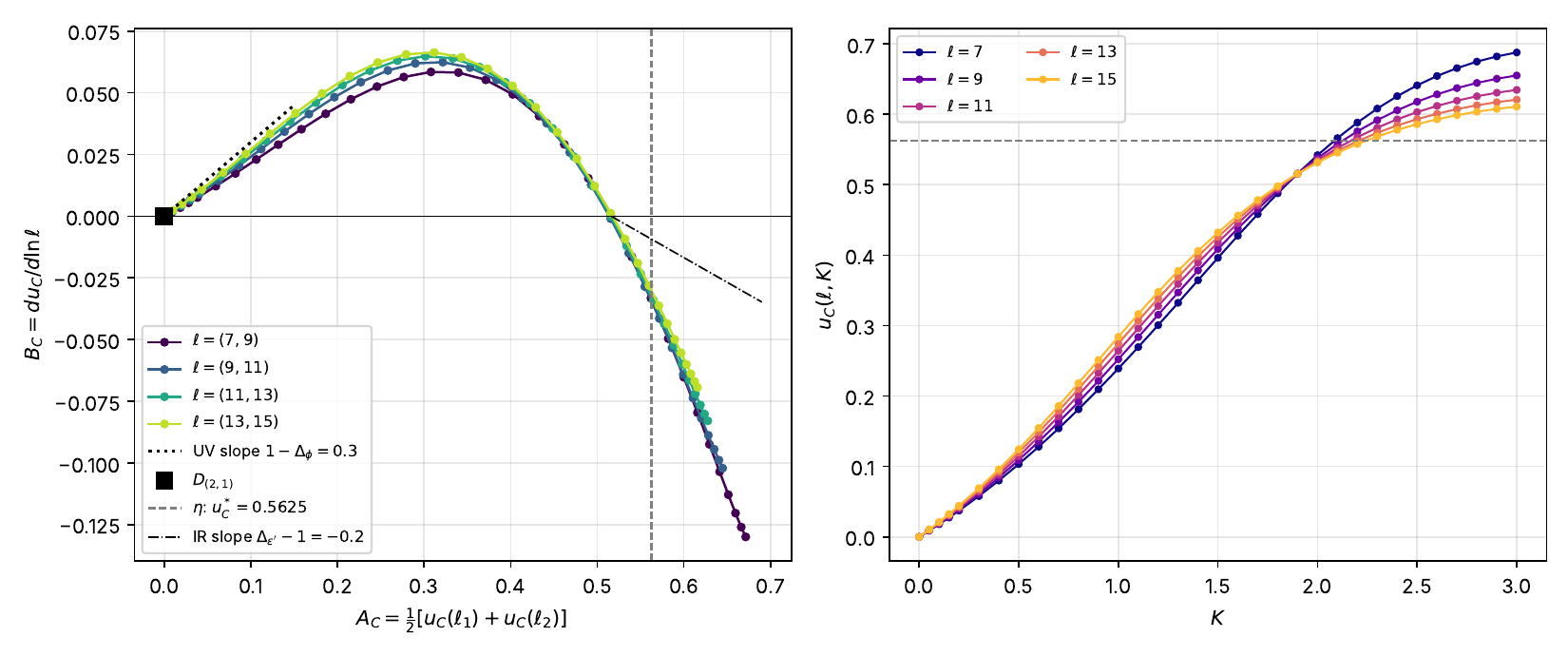}
\caption{Lattice $\beta$-function at $p=4$ along the ray through the improved point, $H = H_0 + K\,\Phi_{\mathrm{cl}} + \lambda\,(n_{\mathrm{L}}+n_{\mathrm{R}})$ with $\lambda = 0.035\,K$, from the $k=1$ state of the single-strand ring (ED, $L=6,\dots,16$). \textbf{Left:} $B_C = du_C/d\ln\ell$ of \eqref{eq:37} versus the running coordinate $A_C = \tfrac{1}{2}[u_C(\ell_1)+u_C(\ell_2)]$ for the size pairs $\ell=(7,9),\dots,(13,15)$, for $0\le K\le3$. The black square is $D_{(2,1)}$ ($u_C=0$), the dotted line the UV slope $1-\Delta_\phi = 0.3$, the vertical dashed line the $\eta$ value $u_C^* = 0.5625$, and the dash-dotted line the universal IR slope $\Delta_{\varepsilon'}-1=-0.2$ through the zero. \textbf{Right:} the raw coupling $u_C(\ell,K)$ versus $K$ for $\ell=7,\dots,15$; the dashed line is $u_C^*$.}\label{fig:8}
\end{figure}

Figure~\ref{fig:8} shows the result. The collapse of the successive size pairs is excellent---from $(7,9)$ to $(13,15)$ the curves lie on top of one another over the whole range---and the curve has the expected structure: it leaves $D_{(2,1)}$ with positive slope, rises to a maximum $B_C\simeq 0.065$ near $K\simeq 1.2$ ($A_C\simeq0.3$), and crosses zero with negative slope at
\begin{align*}
K^* &= 1.900,\ 1.892,\ 1.900,\ 1.912, \\
A_C^* &= 0.516,\ 0.514,\ 0.515,\ 0.518 \qquad \text{for } \ell=(7,9),\ (9,11),\ (11,13),\ (13,15),
\end{align*}
within 8\% of the $\eta$ value $u_C^*=0.5625$ and drifting slowly towards it; the right panel shows the same thing in the raw coupling, which rises through $u_C^*$ at $K\simeq2.1$, overshoots to $0.62$--$0.69$ at $K=3$ with the overshoot shrinking with $\ell$. 
The UV slope $B_C/A_C$ at the smallest couplings is $0.19, 0.21, 0.22, 0.23$ for the four pairs and extrapolates linearly in $1/\ell$ to $0.28$, consistent with the predicted $1-\Delta_\phi=0.3$ (the first-order coefficient of $u_C$ carries a lattice correction that decreases with $\ell$). The slope at the zero, $-0.59$ to $-0.63$, is much steeper than the universal $\Delta_{\varepsilon'}-1=-0.2$ of the leading irrelevant operator on $\eta$: as at $p=5$ below, the corrections to scaling at these sizes are dominated by more irrelevant operators, the amplitude of $\varepsilon'$ being what the improved-point tuning suppresses, and the slow $L^{-0.2}$ tail would only become visible at much larger sizes. The $\beta$-function thus reproduces, from the single-strand setup alone, the flow from $D_{(2,1)}$ to $\eta$ with the correct UV exponent and, to within the finite-size accuracy, the correct IR value of the Casimir coupling.

\begin{figure}[ht]\centering
\includegraphics[width=0.9\textwidth]{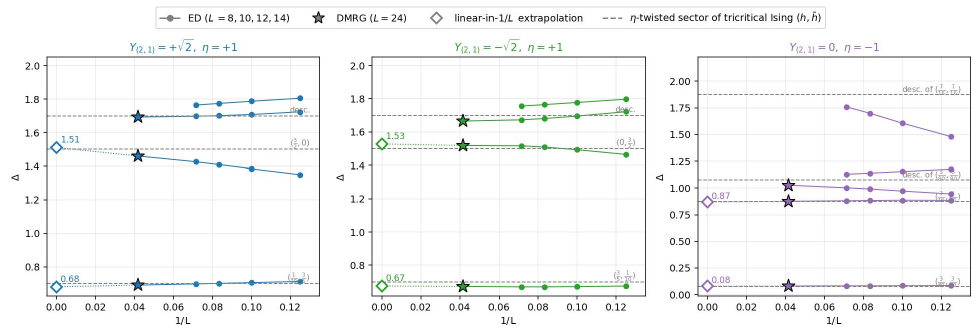}
\caption{$p=4$, ring with a \emph{single} $D_{(2,1)}$ strand at the improved point $(K,\lambda) = (4.82, 0.17)$: sector-resolved spectrum versus $1/L$, with $\Delta = (E - E_{\mathrm{bulk}} - e_D)L/2\pi v$ measured from the bulk ground state plus the defect energy of one strand. Circles: ED at $L = 8, \ldots, 14$; stars: DMRG at $L = 24$ (three states per sector); open diamonds at $1/L = 0$: linear extrapolations through the three largest sizes. Grey dashed lines: the $\eta$-twisted sector of the tricritical Ising model, primaries $(h, \bar h)$ and their first descendants. There is no state at $\Delta = 0$: the endpoint is $\eta$, not the identity line.}\label{fig:9}
\end{figure}

\paragraph{Single-strand ring: $\eta$, not the identity.} The two-strand spectrum above is the untwisted spectrum $\mathcal{H}_{D\times\bar D}$, and it would be the same, namely the diagonal tricritical-Ising spectrum, for either candidate endpoint with $g=1$: the invertible line $\eta$ ($\eta\times\eta = \mathbf{1}$) and the trivial line ($\mathbf{1}\times\mathbf{1} = \mathbf{1}$). A ring with a \emph{single} strand distinguishes them. In radial quantisation its spectrum is the twisted Hilbert space $\mathcal{H}_D$ of the endpoint line: for the identity line this is again the untwisted spectrum, with a vacuum at $\Delta=0$; for $\eta$ it is the $\mathbb{Z}_2$-twisted sector of the tricritical Ising model, $\mathcal{H}_\eta = \bigoplus_{(r,s)} V_{(r,s)}\otimes\bar V_{(4-r,s)}$, which has no vacuum and consists of the primaries $(\tfrac{3}{80},\tfrac{3}{80})$, $(\tfrac{1}{10},\tfrac{3}{5})$, $(\tfrac{3}{5},\tfrac{1}{10})$, $(\tfrac{7}{16},\tfrac{7}{16})$, $(0,\tfrac{3}{2})$, $(\tfrac{3}{2},0)$ of dimensions 0.075, 0.7, 0.7, 0.875, 1.5, 1.5 (the second, third, fifth and sixth carrying spin $\pm\tfrac{1}{2}$ and $\pm\tfrac{3}{2}$) and their descendants. We therefore repeat the computation on a ring with one $D_{(2,1)}$ strand at the same improved point, for $L=8$--$14$ by ED and $L=24$ by DMRG, measuring energies from $E_{\mathrm{bulk}}(L) + e_D$, where $e_D = \tfrac{1}{2}[E_0^{\text{two strands}}(L) - E_{\mathrm{bulk}}(L)]$ is the non-universal energy of one deformed strand extracted from the two-strand ring at the same $L$ ($e_D = -3.5921$ at $L = 14$ and $-3.5915$ at $L = 24$--$48$, so it is well converged). The wrapped lines still commute with $H$ and label the levels. The result (Fig.~\ref{fig:9}) is unambiguous: there is no state at $\Delta = 0$, the ground state is $(\tfrac{3}{80}, \tfrac{3}{80})$ at $\Delta = 0.081 \to 0.08$, and the remaining levels reproduce the twisted spectrum---the spin-$\tfrac12$ pair at $0.69/0.67 \to 0.68/0.67$, split by the $\sigma'$-line into $Y_{(2,1)} = \pm\sqrt2$ according to its left label (\S\ref{sec:lassos}); $(\tfrac{7}{16}, \tfrac{7}{16})$ at $0.877 \to 0.87$; the spin-$\tfrac32$ pair at $1.46/1.52 \to 1.51/1.53$; and descendants at $1.69/1.67$ and $1.03$. The single-strand ring thus confirms that the IR endpoint of the $p = 4$ flow is the invertible line $\eta$ and not the trivial line.

\section{\texorpdfstring{Results at $p=5$: $D_{(2,1)}\to C_5$}{Results at p = 5: D(2,1) -> C5}}\label{sec:5}

At $p = 5$ the category is $\mathrm{Fib} \boxtimes \mathbb{Z}_2$ and $D_{(2,1)} = \tau_{\mathrm{Fib}} \otimes \eta$ has $g = \varphi = 1.618$. The invertible line $\eta$ is still available as an endpoint with $g = 1$, but the flow does not reach it: it stops at a conformal interface $C_5$ with
\[
g_{C_5} = 1.503(2) \quad (\ln g = 0.408), \qquad \mathcal{T} = 0.665(10),
\]
to be compared with $\ln g = \ln\varphi = 0.481$ and $\mathcal{T} = 1$ for $D_{(2,1)}$. This section documents the location of $C_5$, its three defining numbers with their $1/L$ dependence, its sector-resolved spectrum, and the lattice $\beta$-function of the flow.

\subsection{Location of the fixed point}\label{sec:5.1}

We use two parametrisations of the family \eqref{eq:22}, with $\Phi_{\mathrm{cl}} = \varphi(e_{\mathrm{L}} + e_{\mathrm{R}}) - (n_{\mathrm{L}} + n_{\mathrm{R}})$ the clean field \eqref{eq:21} at $p = 5$. The first is the \emph{clean ray} $\lambda = 0$,
\begin{equation}
H = H_0 + K\,\Phi_{\mathrm{cl}},
\label{eq:32}
\end{equation}
on which a single coupling $K$ runs from the UV defect ($K = 0$) through $C_5$. On this ray the fixed point is located by the lattice $\beta$-function of the Casimir coupling (\S\ref{sec:3}), which for the clean ray is computed in the loop-model formulation of Section~\ref{sec:loop} (identical to the RSOS chain at $p=5$, \S\ref{sec:6.1}): its zeros for the successive size pairs converge with geometrically decreasing steps (\S\ref{sec:6.3}) to
\begin{equation}
\boxed{\; K^*_{C_5} = 0.95(5) \quad (\lambda = 0) \;}
\label{eq:33}
\end{equation}
The second is a two-parameter \emph{tower-improved} point,
\begin{equation}
(K, \lambda) = (0.84,\ 0.34),
\label{eq:34}
\end{equation}
the same clean field with an irrelevant counterterm in the channel-blind occupation, which gives the flattest finite-size spectrum and is used for the spectra of \S\ref{sec:5.3}. The two are distinct lattice representatives of the same continuum interface, differing only in irrelevant content, and they give consistent values of $g$, $\mathcal{T}$ and $\Delta_X$.
In the following, the $g$-function, the transmission and the leading irrelevant dimension of \S\ref{sec:5.2}, the sector-resolved spectrum and its $1/L$ extrapolation of \S\ref{sec:5.3}, and the Potts-interface spectrum of \S\ref{sec:5.5}---is computed at the tower-improved point \eqref{eq:34}, where the corrections to scaling are smallest, with ED for $L\le16$ and DMRG at $L=24$, $32$, $48$. 
The lattice $\beta$-function of \S\ref{sec:5.4} requires a one-parameter family that starts at the ultraviolet defect $D_{(2,1)}$ at $K=0$ and passes through the fixed point; it is measured along the ray through the tower-improved point, $\lambda=0.405\,K$, as at $p=4$; 
the clean ray \eqref{eq:32}, whose step-scaling curves collapse better and which fixes the location \eqref{eq:33} of $C_5$, is followed in the loop-model formulation of Sections~\ref{sec:loop} and \ref{sec:nonint}. The quantities that can be compared between the two points---the conserved charges and sector assignments, the leading irrelevant dimension $\Delta_X$ and the fixed-point value of the Casimir coupling---agree within the quoted uncertainties.

\subsection{$g$-function, transmission and the leading irrelevant dimension}\label{sec:5.2}

Figure~\ref{fig:11} and Table~\ref{tab:4} collect the three defining numbers of $C_5$ and their finite-size flow, combining ED ($L \le 16$) with DMRG ($L = 24, 32$ at $\chi = 200$ and $L = 48$ at $\chi = 300$).

(i) The $g$-function from \eqref{eq:27} is already flat by $L = 12$: $\ln g = 0.4085,$ $0.4089,$ $0.4084,$ $0.4076$ at $L = 14,$ $24,$ $32,$ $48$, i.e. $g = 1.503(2)$ (the residual drift between $L = 32$ and $48$ is $10^{-3}$ in $\ln g$ and is within the $\chi$-dependence of the entropy)\linebreak[0], well below the topological value $\varphi$ and above the invertible value 1.

\begin{figure}[ht]\centering
\includegraphics[width=0.9\textwidth]{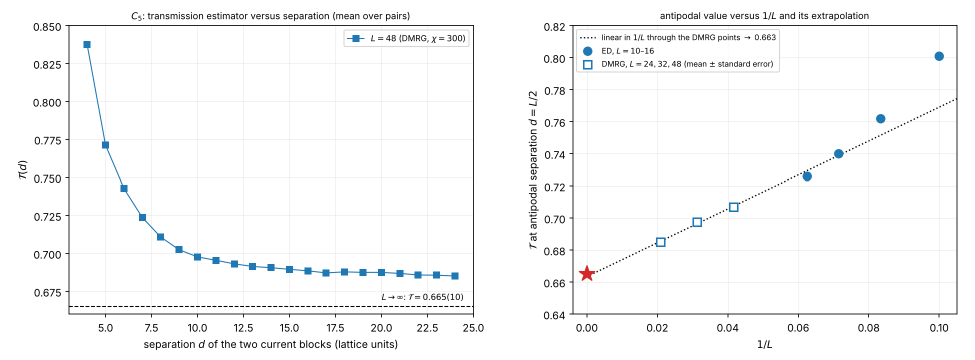}
\caption{Measurement of the transmission coefficient of $C_5$ at the tower-improved point. Left: the estimator \eqref{eq:29}, i.e. the connected current correlator across the strand divided by the bulk correlator at the same separation, versus the separation $d$ of the two four-site current blocks, for $L = 48$ (DMRG, $\chi=300$). Each point is the mean over all admissible pairs of blocks at that $d$ (both blocks away from the strands and at least three sites from the cut of the matrix-product state); the standard errors of the means, $\lesssim 2\times10^{-3}$, are smaller than the symbols. The dashed line is the $L \to \infty$ value. Right: the antipodal values $d = L/2$ against $1/L$ with their standard errors and the linear extrapolation through the three DMRG points (dotted). The star marks the quoted value $\mathcal{T} = 0.665(10)$.}\label{fig:10}
\end{figure}

\begin{figure}[ht]\centering
\includegraphics[width=0.9\textwidth]{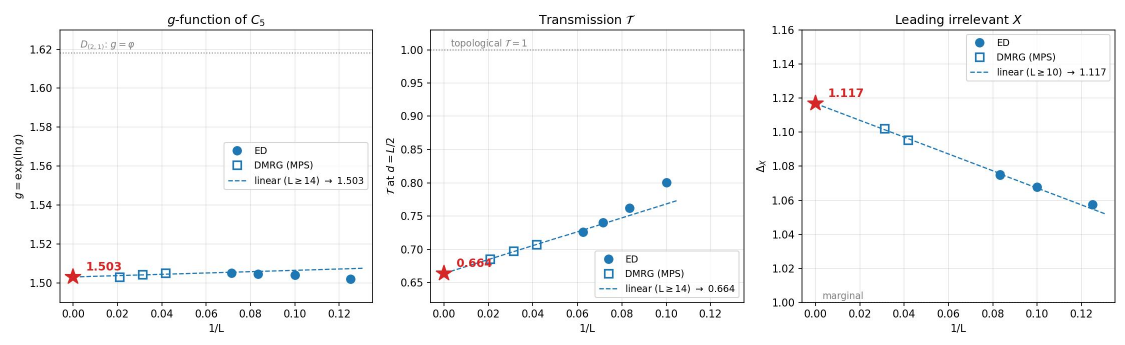}
\caption{Finite-size flow of the three $C_5$ observables versus $1/L$: the $g$-function (left), the transmission $\mathcal{T}$ at antipodal separation $d = L/2$ (centre), and the dimension $\Delta_X$ of the leading irrelevant operator (right). Filled circles: ED; open squares: DMRG at $L = 24, 32$ ($\chi = 200$) and $L = 48$ ($\chi = 300$; $g$ and $\mathcal{T}$ only); dashed line and star: linear-in-$1/L$ extrapolation. Dotted grey lines mark the reference values ($g = \varphi$ and the topological $\mathcal{T} = 1$).}\label{fig:11}
\end{figure}

(ii) Figure~\ref{fig:10} shows how the transmission is measured. The estimator \eqref{eq:29} is averaged over all admissible pairs of current blocks at the same separation $d$ (left panel, $L=48$); the ratio depends on $d$ only for $d\lesssim 8$, where the image contribution of the block nearest to the strand dominates, and then settles on a size-dependent plateau. The number quoted for each $L$ is the antipodal value $d=L/2$: $0.801, 0.762, 0.740, 0.726$ from ED ($L=10$--$16$) and $0.707, 0.6975(10), 0.685(2)$ from DMRG ($L=24,32,48$). These lie on a straight line in $1/L$ (right panel) and extrapolate to $\mathcal{T}=0.663$ through the DMRG points ($0.664$ through $L\ge14$; a $1/L+1/L^2$ fit through $L\ge12$ gives $0.677$). We quote $\mathcal{T} = 0.665(10)$, the uncertainty covering the spread between fit forms. This is clearly below 1: $C_5$ is genuinely partially reflecting, not a topological line in disguise.

(iii) The leading irrelevant defect field of $C_5$, which we call $X$, is the lowest excitation in the fully neutral sector ($Y_{\tau_{\mathrm{Fib}}} = \varphi$, $\eta = +1$; the sector of the identity). It is the image of the driving
field along the flow, and its dimension controls the approach to the fixed point. Its gap rises past marginality with $L$ and extrapolates to
\begin{equation}
\Delta_X = 1.12(1), \qquad \varepsilon_X \equiv \Delta_X - 1 = 0.12(1).
\label{eq:35}
\end{equation}
The DMRG points at $L = 24, 32, 48$ are essential here: they anchor the extrapolations far beyond the ED reach and confirm that the small-$L$ trends are not finite-size artefacts.

\subsection{Sector-resolved spectrum and its $1/L$ dependence}\label{sec:5.3}

\begin{figure}[ht]\centering
\includegraphics[width=0.9\textwidth]{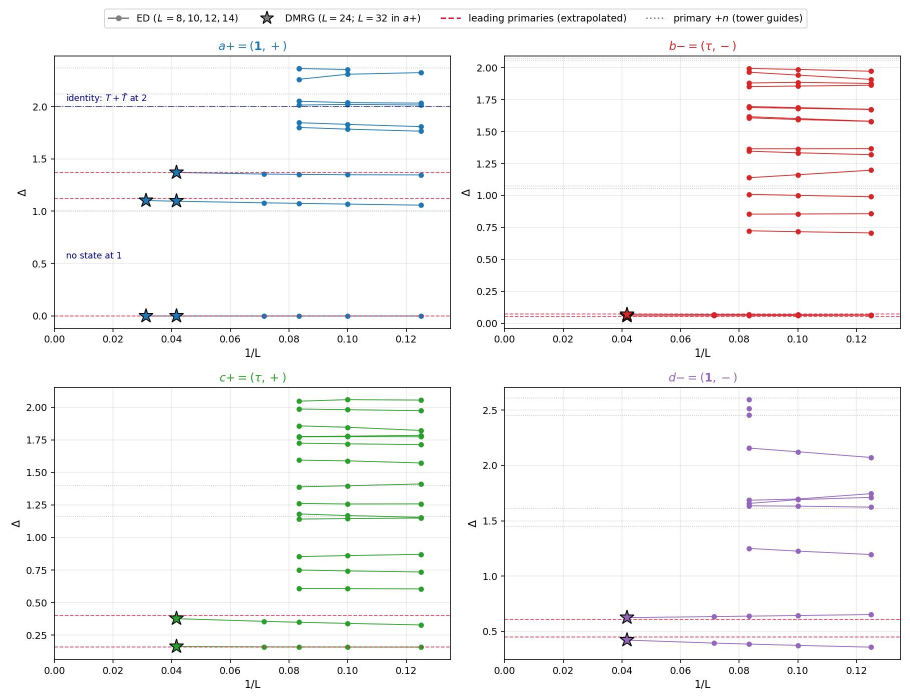}
\caption{$C_5$ sector-resolved spectrum versus $1/L$ at the tower-improved point $(K, \lambda) = (0.84, 0.34)$, one panel and one colour per charge sector. Circles: symmetry-reduced ED at $L = 8, 10, 12$ (up to 14--16 levels per sector) and the leading two levels at $L = 14$; stars: DMRG at $L = 24$ (ground state, $X$ and $X'$ in $a{+}$; the two lowest levels in each of $b{-}$, $c{+}$, $d{-}$) and $L = 32$ (ground state and $X$). Lines connect the same level across all system sizes, ED and DMRG alike. Red dashed lines mark the extrapolated leading primaries of each sector (Table~\ref{tab:5}), grey dotted lines their integer-shifted descendants $\Delta + n$. The levels are essentially flat in $1/L$, and the pattern is consistent with conformal towers: in the neutral sector $a{+}$ the identity module has no state at $\Delta = 1$ and a pair of states at $\Delta \simeq 2$ ($T$ and $\bar T$), while $X$ and $X'$ each come with a descendant one unit higher; in the charged sectors the leading primaries at $0.055/0.073$ ($b{-}$), $0.16/0.40$ ($c{+}$) and $0.45/0.61$ ($d{-}$) are followed by clusters of states near $\Delta + 1$ (e.g. the pair at $\Delta \simeq 1.15$ above the $c{+}$ primary at $0.16$, as expected for $L_{-1}$ and $\bar L_{-1}$ descendants of a primary with $h, \bar h \neq 0$). The finite-size drifts of the subleading $c{+}$ and $d{-}$ primaries ($0.35 \to 0.40$, $0.39 \to 0.45$) are the largest in the spectrum and are the main source of uncertainty in the identification of the higher levels.}\label{fig:12}
\end{figure}

The four charge sectors of the ring are labelled by the Fibonacci charge ($\mathbf{1}$ or $\tau_{\mathrm{Fib}}$, i.e. $Y_{\tau_{\mathrm{Fib}}} = \varphi$ or $-\varphi^{-1}$) and the $\mathbb{Z}_2$ parity $\eta = \pm1$; we use the shorthand
\[
a{+} = (\mathbf{1}, +), \qquad b{-} = (\tau_{\mathrm{Fib}}, -), \qquad c{+} = (\tau_{\mathrm{Fib}}, +), \qquad d{-} = (\mathbf{1}, -),
\]
which are the four combinations that occur (the wrapped $Y_{(2,1)} = Y_{\tau_{\mathrm{Fib}}}\,\eta$ takes the values $\varphi$, $\varphi^{-1}$, $-\varphi^{-1}$, $-\varphi$ on them, respectively). The identity and $X$ live in $a{+}$; the lowest state of the whole ring above the vacuum lies in $b{-}$. Table~\ref{tab:5} lists the leading levels of each sector at the tower-improved point for $L = 8, 10, 12$, the DMRG values at $L = 24, 32$, and $1/L$ extrapolations, and Fig.~\ref{fig:12} shows all levels obtained from the symmetry-reduced diagonalisation together with integer-shifted guide lines, as a check that the finite-size spectrum organises into conformal towers.

\begin{table}[ht]
\caption{$g$-function and transmission of $C_5$ at the tower-improved point $(K, \lambda) = (0.84, 0.34)$ versus system size. ED for $L \le 16$; DMRG at $L = 24, 32$ ($\chi = 200$) and $L = 48$ ($\chi = 300$). $\ln g$ is the entropy difference \eqref{eq:27} for the interval of $2k{+}1$ bonds centred on the strand, $\mathcal{T}$ the current-correlator ratio \eqref{eq:29} at antipodal separation $d = L/2$, averaged over the admissible pairs (blocks near the MPS cut excluded), with the standard error of the mean in parentheses where more than one pair contributes. The values below the table are the linear-in-$1/L$ extrapolations through $L \ge 14$ (Fig.~\ref{fig:11}).}\label{tab:4}
\centering
\adjustbox{max width=\textwidth}{\begin{tabular}{lcccccccc}
\toprule
$L$ & 8 & 10 & 12 & 14 & 16 & 24 & 32 & 48 \\
\midrule
$\ln g$ & 0.4069 & 0.4081 & 0.4085 & 0.4089 & --- & 0.4089 & 0.4084 & 0.4076 \\
$g$ & 1.502 & 1.504 & 1.505 & 1.505 & --- & 1.505 & 1.504 & 1.503 \\
$\mathcal{T}(L/2)$ & --- & 0.801 & 0.762 & 0.740 & 0.726 & 0.707 & 0.6975(10) & 0.685(2) \\
\bottomrule
\multicolumn{9}{c}{$L \to \infty$: $g = 1.503(2)$, $\mathcal{T} = 0.665(10)$.} \\
\end{tabular}}
\end{table}

\subsection{Lattice $\beta$-function}\label{sec:5.4}

\paragraph{Construction.} We apply the step-scaling construction of \S\ref{sec:3} on the single-strand ring. At $p=5$ the $k=1$ state of \eqref{eq:31} is the lowest state of the sector $d{-}$ ($Y_{(2,1)}=-\varphi$, $\eta=-1$). The Casimir coupling $u_C$ of \eqref{eq:48} is formed from its energy at consecutive even sizes, and the $\beta$-function $B_C$ of \eqref{eq:37} from $u_C$ on successive size pairs. We follow the ray through the tower-improved point, $\lambda=0.405\,K$, so that $K=0.84$ is the point \eqref{eq:34}. The energies come from the symmetry-reduced ED of \S\ref{sec:2.5} ($\eta=-1$ block) for $L=6,\dots,16$ and from DMRG \cite{ref40,ref41} for $L=18,20,22$, where the $d{-}$ state is made the ground state by adding to $H$ the wrapped lines with the signs of its charges, $H'=H+0.6\,Y_{(2,1)}+0.5\,\eta$ (bond dimension $\chi=120$, truncation error $\lesssim 4\times10^{-8}$; at $L=12$ the DMRG energy agrees with ED to $4\times10^{-7}$). The size pairs are $\ell=(7,9),\dots,(13,15)$ from ED and $(15,17)$, $(17,19)$, $(19,21)$ with DMRG.

\begin{table}[ht]
\caption{$C_5$ sector-resolved spectrum at the tower-improved point $(K, \lambda) = (0.84, 0.34)$, $v = 5/3$. Sectors as defined in \S\ref{sec:5.3}. DMRG column: $L = 24$ for all sectors ($X$ also at $L = 32$, second value); the last column is the $1/L$ extrapolation.}\label{tab:5}
\centering
\adjustbox{max width=\textwidth}{\begin{tabular}{llccccc}
\toprule
sector & field & $L{=}8$ & $L{=}10$ & $L{=}12$ & (DMRG) & $L \to \infty$ \\
\midrule
$a{+}\ (\mathbf{1}, +)$ & $X$ & 1.058 & 1.068 & 1.075 & 1.095, 1.102 & 1.12(1) \\
 & $X'$ & 1.347 & 1.348 & 1.351 & --- & 1.37(2) \\
$b{-}\ (\tau_{\mathrm{Fib}}, -)$ & lead. & 0.063 & 0.061 & 0.060 & 0.058 & 0.055(5) \\
 & next & 0.066 & 0.068 & 0.068 & 0.071 & 0.073(5) \\
$c{+}\ (\tau_{\mathrm{Fib}}, +)$ & lead. & 0.157 & 0.158 & 0.159 & 0.161 & 0.16 \\
 & 2nd & 0.328 & 0.340 & 0.349 & 0.376 & 0.40(3) \\
$d{-}\ (\mathbf{1}, -)$ & lead. & 0.359 & 0.374 & 0.386 & 0.422 & 0.45(3) \\
 & next & 0.653 & 0.644 & 0.639 & 0.625 & 0.61(2) \\
\bottomrule
\end{tabular}}
\end{table}

\begin{figure}[ht]\centering
\includegraphics[width=0.75\textwidth]{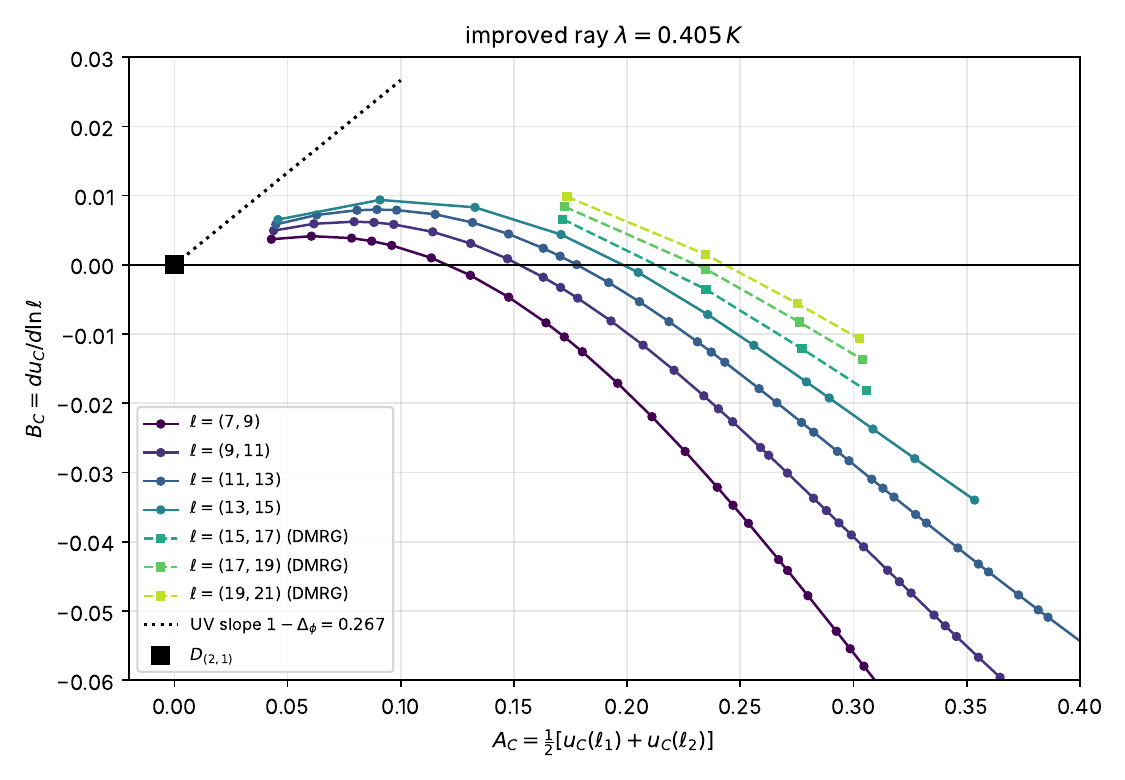}
\caption{Lattice $\beta$-function at $p = 5$ along the ray through the tower-improved point, $H = H_0 + K\Phi_{\mathrm{cl}} + \lambda\,(n_{\mathrm{L}}+n_{\mathrm{R}})$ with $\lambda = 0.405\,K$ (so that $K=0.84$ is the point \eqref{eq:34}), from the Casimir coupling $u_C$ of the $k=1$ state of the single-strand ring: $B_C = du_C/d\ln\ell$ of \eqref{eq:37} versus the running coordinate $A_C$, for $0 < K \le 1.6$ (the region containing the first fixed point). Circles: ED at $L=6,\dots,16$ (pairs $\ell=(7,9),\dots,(13,15)$); squares: DMRG at $L=18,20,22$ (pairs $(15,17)$, $(17,19)$, $(19,21)$, at $K=0.5,0.75,0.95,1.1$). The black square is the UV defect $D_{(2,1)}$ ($u_C = 0$) and the dotted line the predicted UV slope $1 - \Delta_\phi = 0.267$. The zero crossing with negative slope is the IR-stable interface $C_5$; it drifts to larger $K$ with increasing size (see text).}\label{fig:13}
\end{figure}

\paragraph{Result.} Figure~\ref{fig:13} shows $B_C$ against $A_C$. The curve has the structure of a flow to an IR-stable fixed point: it leaves $D_{(2,1)}$ with positive slope, turns over and crosses zero with negative slope; the strong-coupling regime beyond $K=1.6$, where the curve turns back up and whose nature is not yet understood, is left for future work. Along this ray, however, the successive size pairs do not collapse: the whole curve shifts upward from pair to pair and the zero drifts,
\begin{align*}
K^* &= 0.345,\ 0.441,\ 0.526,\ 0.600 \qquad (\text{ED, } \ell=(7,9),\dots,(13,15)), \\
K^* &= 0.663,\ 0.734,\ 0.793 \qquad\quad\ \ (\text{DMRG, } \ell=(15,17),\ (17,19),\ (19,21)),
\end{align*}
with $A_C^*$ rising from $0.12$ to $0.24$ and the slope at the zero between $-0.14$ and $-0.18$. Notice the leading irrelevant operator in the neutral channel has $\Delta_X-1=-0.12$, causing strong finite-size drift of the $B_C=0$ point.

\subsection{$C_5$ as an interface of the three-state Potts model}\label{sec:5.5}

The tetracritical Ising model $M(5,6)$ and the critical three-state Potts model are related by gauging a $\mathbb{Z}_2$ \cite{ref61,ref62,ref63}: the Potts model is the orbifold of $M(5,6)$ by the invertible line $\eta = D_{(4,1)}$, and conversely $M(5,6)$ is the orbifold of Potts by charge conjugation $C$. Under this correspondence the $S_3=\mathbb{Z}_3\rtimes\mathbb{Z}_2$ symmetry of Potts becomes the fusion category $\mathrm{Rep}(S_3)$ on the $M(5,6)$ side: the sign representation is $\eta$, and the two-dimensional representation, i.e.\ the sum $\omega+\omega^2$ of the two $\mathbb{Z}_3$ lines, is the non-invertible line $D_{(1,3)}$ of quantum dimension 2. The line $D_{(2,1)}$ commutes with $\eta$ and survives the gauging as the Fibonacci line $\tau_{\mathrm{Fib}}$ of the Potts model (quantum dimension $\varphi$, associated with the $h=2/5$ field) \cite{ref13}. Any $\eta$-symmetric interface of $M(5,6)$ therefore defines an interface of the Potts model, and since $C_5$ is reached from $D_{(2,1)}$ by an $\eta$-even deformation, the flow $D_{(2,1)}\to C_5$ is at the same time a flow $\tau_{\mathrm{Fib}}\to C_5^{\mathrm{Potts}}$ of Potts interfaces. In this subsection we work out the operator content of $C_5^{\mathrm{Potts}}$ and identify the symmetry under which it is a stable interface.

\paragraph{Orbifold rule and lattice implementation.}
Let $I$ be the gauging interface \cite{ref4,ref63}, with $M(5,6)$ on one side and Potts on the other; it has quantum dimension $\sqrt2$ and obeys $I\,I^\dagger=\mathbf{1}\oplus\eta$ on the $M(5,6)$ side and $I^\dagger I=\mathbf{1}\oplus C$ on the Potts side. The Potts interface associated with $C_5$ is $X=I^\dagger C_5 I$, and its interface operators live on a Potts ring with $X$ and $\bar X$ inserted antipodally. Fusing, $X\bar X=I^\dagger C_5\,(II^\dagger)\,\bar C_5\,I$, so on each of the two arcs between the $M(5,6)$ strands one sums over $\mathbf{1}$ and $\eta$; of the four resulting $\eta$-line configurations only two are inequivalent, because $C_5$ commutes with $\eta$ and an $\eta$ line on one arc can be slid through $C_5$ onto the other. The remaining $\eta$ loop of $II^\dagger$, wrapping the ring in the time direction, acts as the projector $\tfrac12(1+\eta)$. Hence
\begin{equation}
\mathcal{H}_{X\bar X}=\bigl[\mathcal{H}_{C_5\bar C_5}\oplus\mathcal{H}_{C_5\eta\bar C_5}\bigr]^{\eta\text{-even}},
\label{eq:41}
\end{equation}
the usual orbifold prescription: the $\eta$-even states of the untwisted interface Hilbert space, plus the $\eta$-even states of the $\eta$-twisted space in which an $\eta$ line is inserted between the two interfaces. The gauged theory carries the dual (``quantum'') $\mathbb{Z}_2$ of the gauging \cite{ref62,ref63}: gauging $\eta$ means summing over $\eta$ lines wrapping both cycles of the torus, the line along space being the projector $\tfrac12(1+\eta)$ and the line along time defining the twisted sector, and the dual symmetry is the operator that measures the latter, the holonomy of the $\mathbb{Z}_2$ gauge field around the spatial circle. It therefore acts as $+1$ on the untwisted summand of \eqref{eq:41} and as $-1$ on the $\eta$-twisted one; it is a symmetry because an $\eta$ line in the time direction is topological and cannot be removed by the dynamics, so the two summands never mix, and gauging it in turn returns $M(5,6)$. Since $M(5,6)$ is the orbifold of Potts by charge conjugation, this dual $\mathbb{Z}_2$ is $C$ itself (the same statement in the opposite direction is that the $C$-twisted sector of Potts supplies the $\eta$-odd fields of $M(5,6)$, such as $(2,2)$ with $h=1/40$; the Ising model, whose $\mathbb{Z}_2$ orbifold is again Ising with the disorder field $\mu$ of the twisted sector as the new odd spin, is the familiar prototype). The only convention is the overall sign of $C$ on the twisted sector, fixed by $C=+1$ on the untwisted vacuum. Note that $\eta\times C_5$ and $C_5$ give the same $X$, since $\eta$ is absorbed by $I$; the two are distinct interfaces of $M(5,6)$ (the twisted space below contains no identity operator) but indistinguishable from the Potts side. On the lattice both spaces in \eqref{eq:41} are two-strand rings of \S\ref{sec:2.5}: the untwisted one is the ring used throughout \S\ref{sec:5}, the twisted one is the same ring with a third, $\eta=(4,1)$ strand placed midway between the two $C_5$ strands (\S\ref{sec:2.2}; the $\eta$ strand carries no deformation). Two charges label the resulting states. The Fibonacci charge $q$ is the eigenvalue of the wrapped line $Y_{(2,1)}$ (equal to $Y_{(3,1)}$ on $\eta$-even states), $q=\mathbf{1}$ for eigenvalue $\varphi$ and $q=\tau_{\mathrm{Fib}}$ for $-\varphi^{-1}$. The charge under the dual $\mathbb{Z}_2$ of the gauging---charge conjugation $C$ of the Potts model---is, by the argument above, simply the ring a state comes from: $+1$ on the untwisted and $-1$ on the twisted ring. The $\mathbb{Z}_3$ charge is not diagonal in this basis, because the $\mathbb{Z}_3$ lines map the untwisted sector into the twisted one; on a $\mathbb{Z}_3$-symmetric interface it shows up instead as exact degeneracies between $C=+$ and $C=-$ states (the $\omega$ and $\omega^2$ components of a $\mathbb{Z}_3$ doublet are exchanged by $C$, so both $C$-eigencombinations must be present at the same energy).

\begin{table}[ht]
\centering
\adjustbox{max width=\textwidth}{\begin{tabular}{lcccccccccc}
\toprule
$L$ & $[\mathbf{1},+]$ & $[\tau_{\mathrm{Fib}},-]$ & $[\tau_{\mathrm{Fib}},-]$ & $[\tau_{\mathrm{Fib}},+]$ & $[\tau_{\mathrm{Fib}},+]$ & $[\mathbf{1},-]$ & $[\mathbf{1},-]$ & $[\tau_{\mathrm{Fib}},+]$ & $[\tau_{\mathrm{Fib}},-]$ & $[\mathbf{1},+]$ \\
\midrule
8  & 0 & 0.0545 & 0.0971 & 0.1572 & 0.3279 & 0.3792 & 0.6061 & 0.6051 & 0.7297 & 1.0576 \\
10 & 0 & 0.0549 & 0.0937 & 0.1579 & 0.3397 & 0.3943 & 0.6064 & 0.6068 & 0.7310 & 1.0678 \\
12 & 0 & 0.0553 & 0.0913 & 0.1586 & 0.3485 & 0.4060 & 0.6057 & 0.6087 & 0.7333 & 1.0749 \\
14 & 0 & 0.0557 & 0.0896 & 0.1593 & 0.3554 & 0.4153 & 0.6047 & 0.6105 & 0.7359 & 1.0802 \\
16 & 0 & 0.0561 & 0.0882 & 0.1598 & 0.3610 & 0.4230 & 0.6036 & 0.6121 & 0.7384 & 1.0844 \\
$24^*$ & 0 & 0.0575 & 0.0848 & 0.1614 & 0.3758 & 0.4440 & 0.5993 & -- & -- & 1.0953 \\
\bottomrule
\end{tabular}}
\caption{Interface operators of $C_5^{\mathrm{Potts}}$ at $(K,\lambda)=(0.84,0.34)$, labelled $[q,C]$ and listed by increasing dimension at $L=12$. The row marked ${}^*$ is DMRG at $L=24$ ($\chi=120$): the two leading states of each sector, from the untwisted ring for $C=+$ and from the $\eta$-twisted ring for $C=-$. The only Fibonacci-neutral, $C$-even operator below the marginal line is the identity; the first non-trivial one is the leading irrelevant operator $X$ (1.095 at $L=24$, extrapolating to 1.11), the same as on the $M(5,6)$ side (\S\ref{sec:5.3}).}
\label{tab:6}
\end{table}

\paragraph{Check at the topological point.}
At $K=\lambda=0$ the construction must reproduce the interface operators of the Potts $\tau_{\mathrm{Fib}}$ line, and it does. The merged $\eta$-even spectrum at $L=10$ is 0; $2/15$ with multiplicity four ($q=\tau_{\mathrm{Fib}}$, twice $C=+$ and twice $C=-$); $2/5$ twice ($q=\mathbf{1}$ and $q=\tau_{\mathrm{Fib}}$, both $C=-$); $11/15$ with multiplicity four ($q=\mathbf{1},\tau_{\mathrm{Fib}}$, $C=\pm$); 0.80 twice ($q=\tau_{\mathrm{Fib}}$, $C=+$). The fourfold levels are the $\mathbb{Z}_3$ doublets built on the Potts spin field $\sigma$ ($h=1/15$) and on the $h=2/3$ field, split evenly between $C=\pm$ exactly as the orbifold rule demands, while the $h=2/5$-type levels, which are $\mathbb{Z}_3$ neutral, appear without $C$ partners.

\begin{figure}[ht]\centering
\includegraphics[width=0.9\textwidth]{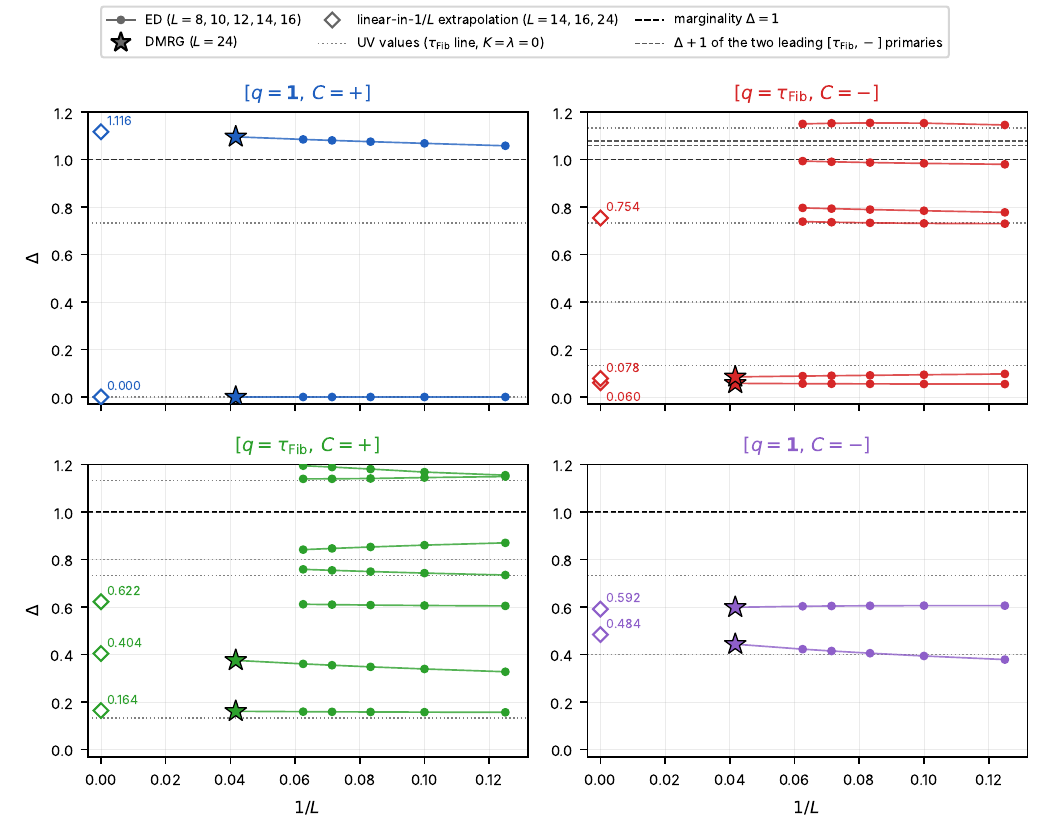}
\caption{Sector-resolved interface spectrum of $C_5^{\mathrm{Potts}}$ at $(K,\lambda)=(0.84,0.34)$ against $1/L$, one panel per $[q,C]$ sector, in the style of Fig.~\ref{fig:12}: circles are the ED levels at $L=8,10,12,14,16$ (untwisted ring for $C=+$, $\eta$-twisted ring for $C=-$), stars the two leading DMRG states per sector at $L=24$, lines connect the same level across sizes, open diamonds are linear-in-$1/L$ extrapolations of the leading levels through the three largest sizes ($L=14,16,24$; ED only, $L=12,14,16$, for the third $\tau_{\mathrm{Fib}}$ levels), dotted grey lines mark the values at the topological UV point ($\tau_{\mathrm{Fib}}$ line), and the dashed line is the marginal dimension. Levels above $\Delta=1.2$ are omitted. The only state below the marginal line in the fully neutral sector $[\mathbf{1},+]$ is the identity; the leading irrelevant operator extrapolates to $X\simeq1.11$, as on the $M(5,6)$ side. In $[\tau_{\mathrm{Fib}},-]$ the grey dashed lines mark $\Delta+1$ for the two leading primaries, $1.06$ and $1.08$: at the UV point the two $\Delta=2/15$ states have four level-one descendants at $17/15$ (left- and right-moving), and along the flow these become the levels at $0.99$, $1.15$ (shown) and $1.31$, $1.38$ (above the frame). On the conformal interface only a single Virasoro algebra acts, so only two of them remain $L_{-1}$ descendants; the pair at $0.99/1.15$ straddles the expected positions with a splitting that hardly moves between $L=8$ and $16$ ($0.98\to0.99$ and $1.145\to1.150$), as expected from mixing through the nearly marginal $X$ ($\Delta_X\simeq1.11$, corrections $\propto L^{-0.11}$), while their mean, $1.07$, agrees with the mean of the two expected descendant positions.}\label{fig:14}
\end{figure}

\paragraph{Spectrum of $C_5^{\mathrm{Potts}}$.}
Table~\ref{tab:6} and Fig.~\ref{fig:14} give the interface-operator spectrum at the improved point $(K,\lambda)=(0.84,0.34)$ for $L=8,10,12,14,16$ (ED) and $L=24$ (DMRG, two leading states per sector), sector by sector. The $L=24$ points are obtained as in \S\ref{sec:5.3}, with the $\eta$-twisted ring realised in the MPS by reflecting the heights across one site (the wrapped $(2,1)$ line is closed with the same reflection); the construction reproduces the ED twisted spectra at $L=8,12$ to all digits. The lowest levels are converged to two or three digits, and the linear extrapolations of the leading levels through $L=14,16,24$ are 0.060, 0.078, 0.754 in $[\tau_{\mathrm{Fib}},-]$, 0.164, 0.404, 0.622 in $[\tau_{\mathrm{Fib}},+]$, 0.484, 0.592 in $[\mathbf{1},-]$ and $X=1.116$ in $[\mathbf{1},+]$ (the third $\tau_{\mathrm{Fib}}$ levels from ED alone, $L=12,14,16$). The $\mathbb{Z}_3$ doublets of the UV point are gone: the four states at $2/15$ have split into two $C=-$ levels at 0.06, 0.08 and two $C=+$ levels at 0.16, 0.40, and no $C=+$ level has a degenerate $C=-$ partner anywhere in the spectrum. The interface $C_5^{\mathrm{Potts}}$ therefore breaks $\mathbb{Z}_3$. 
The deformation $\varphi_{(1,3)}\otimes\bar\varphi_{(3,3)}$ is charged under the lasso of $D_{(1,3)}$ (its eigenvalue is $S_{(1,3)(1,3)}/S_{(1,1)(1,3)}=-1=\omega+\omega^2$ rather than the neutral value 2), i.e.\ in Potts language it is one real component of a $\mathbb{Z}_3$ doublet of interface operators on $\tau_{\mathrm{Fib}}$, and turning it on selects one of three $\mathbb{Z}_3$-related images of the interface, in the same way that a fixed boundary condition of the Potts model selects one of the three colours. $C_5^{\mathrm{Potts}}$ preserves the $\mathbb{Z}_2=\langle C\rangle$ subgroup of $S_3$ (the dual of the $\eta$ symmetry of the lattice model) and the Fibonacci line $\tau_{\mathrm{Fib}}$.

\paragraph{Stability.}
An interface is stable under a symmetry if none of its relevant operators is neutral under that symmetry. Reading Table~\ref{tab:6}: the relevant operators of $C_5^{\mathrm{Potts}}$ are the $[\tau_{\mathrm{Fib}},-]$ levels at 0.06, 0.08, 0.75, 0.79, 0.99, the $[\tau_{\mathrm{Fib}},+]$ levels at 0.16, 0.40, 0.62, 0.74, 0.85, and the $[\mathbf{1},-]$ levels at 0.48, 0.59; the first $[\mathbf{1},+]$ operator is $X\simeq1.11$. Hence $C_5^{\mathrm{Potts}}$ is stable if and only if both the Fibonacci line $\tau_{\mathrm{Fib}}$ and charge conjugation $C$ are imposed. Imposing $C$ alone leaves the relevant Fibonacci-charged, $C$-even operators at 0.16, 0.40 and 0.62; imposing $\tau_{\mathrm{Fib}}$ alone leaves the Fibonacci-neutral, $C$-odd operators at 0.48 and 0.59. Since the interface breaks $\mathbb{Z}_3$, the full $S_3$ cannot be used, and $\mathbb{Z}_3$ plays no role in the stability count; in a Potts lattice realisation the interface would have to be protected by the non-invertible line $\tau_{\mathrm{Fib}}$ together with $C$, exactly as $C_5$ is protected by $Y_{(2,1)}$ together with $\eta$ on the $M(5,6)$ side. The $g$-function and the transmission are unchanged by the orbifold, since $\langle0|D|0\rangle$ and the current two-point function only involve the untwisted vacuum: $g_{C_5^{\mathrm{Potts}}}=1.503(2)$ and $\mathcal{T}=0.665(10)$.

\section{The loop-model formulation at continuous $p$}\label{sec:loop}

The RSOS construction fixes $p$ to an integer. Everything used above, however, lives in the dilute Temperley--Lieb algebra \cite{ref64} with loop weight $n=2\cos(\pi/p)=d_\wl$, and can be written in the loop (dilute $O(n)$) model \cite{ref65,ref66,ref67,ref29}, where $p$ is a continuous parameter. This section does so and follows the fixed point $C_p$ from $p=5$ down to $p=4$.

\subsection{Loop model with seams}\label{sec:6.1}

\begin{figure}[ht]\centering
\resizebox{0.9\textwidth}{!}{\input{figs/fig15_tikz.tex}}
\caption{The loop model in the oriented-strand representation. (a) Slot states. (b) The six face moves of \eqref{eq:43}, drawn with time upwards; the coefficients are the RSOS weights \eqref{eq:9}. (c) Orientation phases: a cup or cap carries $q^{\pm1/2}$ according to its orientation, so the two-slot singlet is $|c\rangle=q^{1/2}|{+}{-}\rangle+q^{-1/2}|{-}{+}\rangle$ (and $\langle\tilde c|$ likewise), and a contractible loop, summed over its two orientations, has weight $\langle\tilde c|c\rangle=q+q^{-1}=n$, which is the Temperley--Lieb relation $e^2=ne$.}\label{fig:15}
\end{figure}
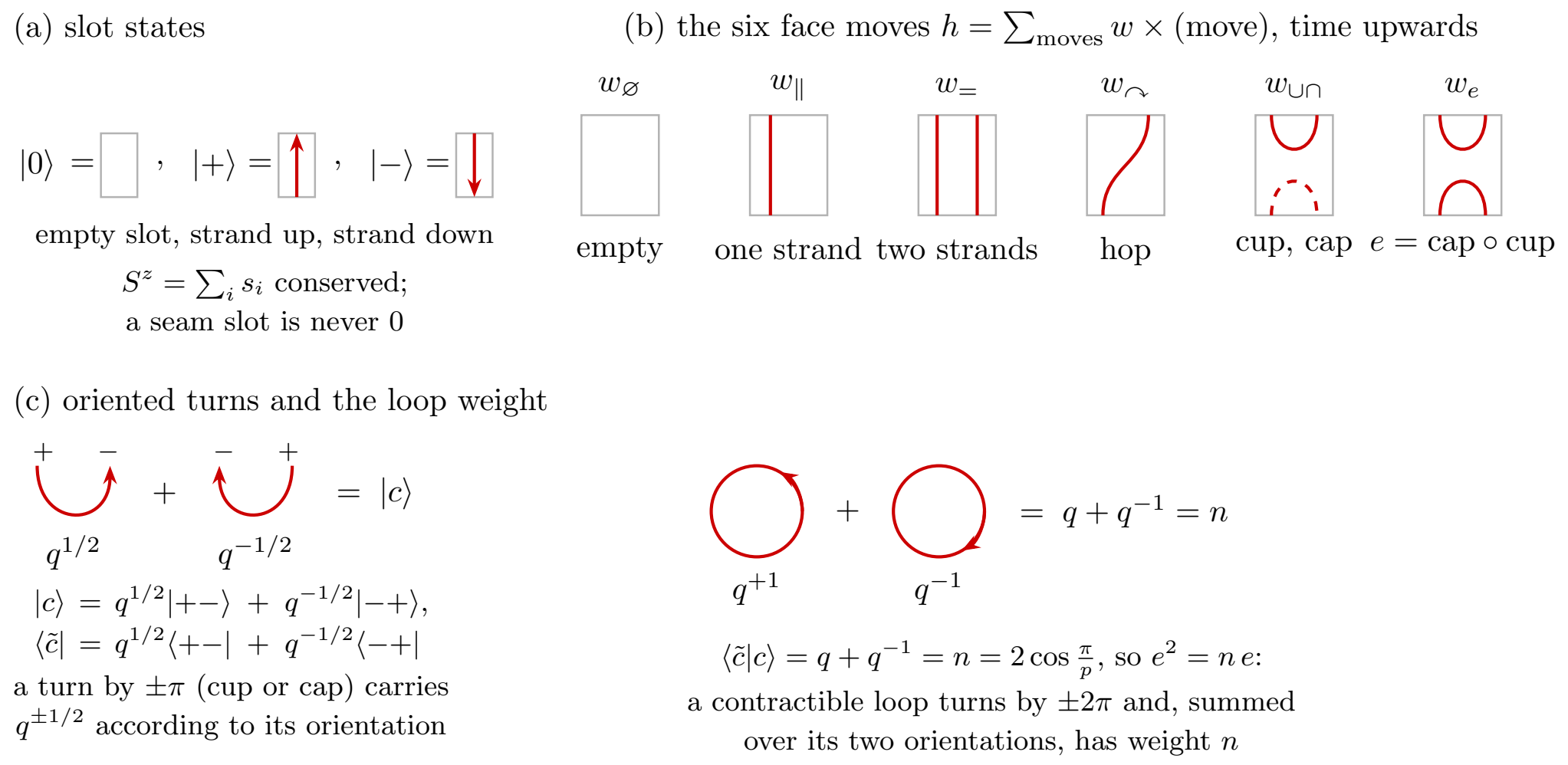

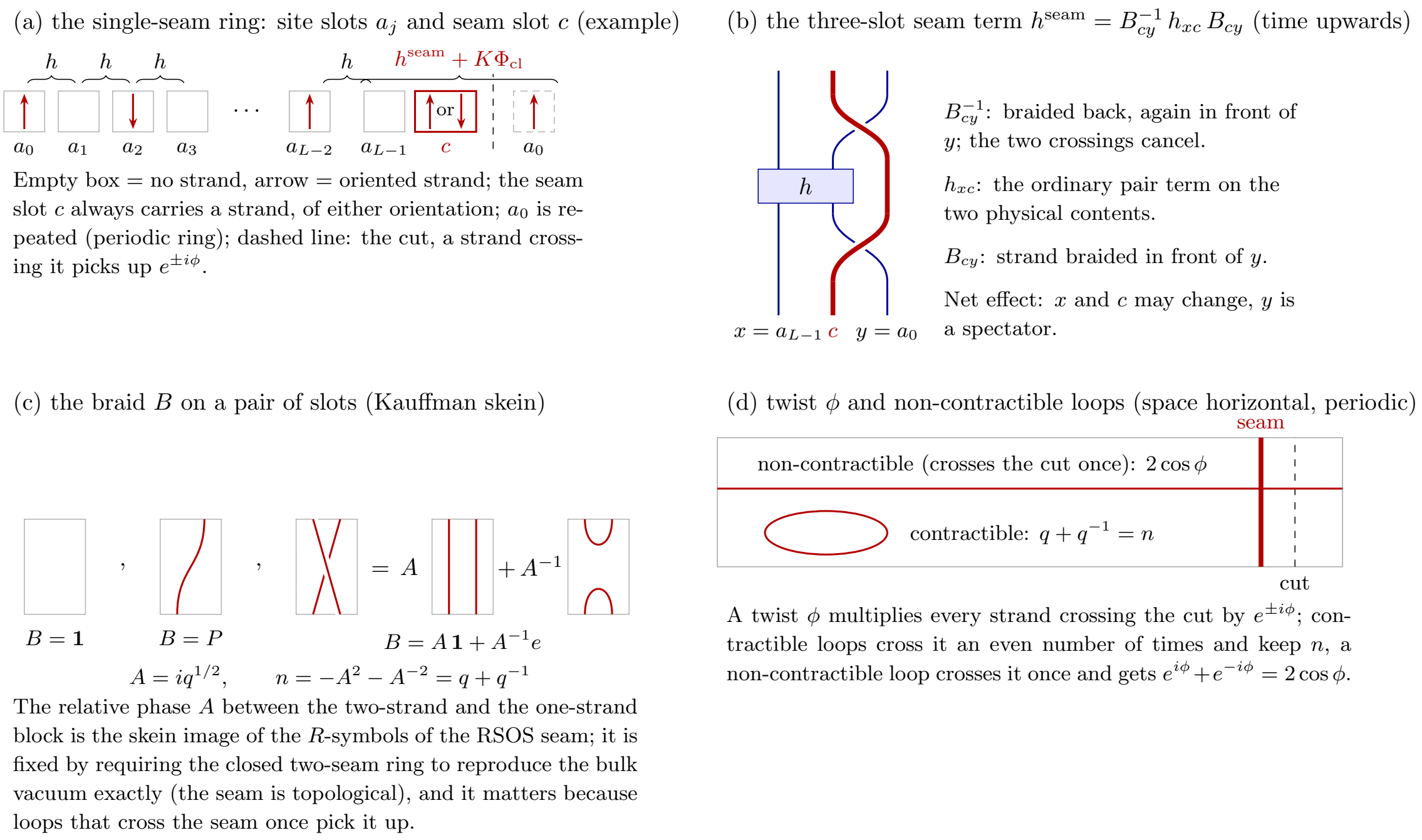
\begin{figure}[ht]\centering
\resizebox{\textwidth}{!}{\input{figs/fig16_tikz.tex}}
\caption{Seam and twist. (a) The single-seam ring: $L$ site slots and one always-occupied seam slot $c$, the $L-1$ ordinary pair terms, the three-slot seam term carrying the deformation, and the cut at which the twist is applied. (b) The seam term \eqref{eq:44}: the strand is braided in front of the content of $a_0$, the ordinary $h$ acts, and the strand is braided back with the inverse braid, again in front of $a_0$, so that the two crossings cancel. (c) The braid on a pair of slots, \eqref{eq:45}. (d) A twist $\phi$ gives non-contractible loops the weight $2\cos\phi$ while contractible loops keep $n$; the twist sector is the loop-model version of the $Y_{(2,1)}$ eigenvalue, \eqref{eq:47}.}\label{fig:16}
\end{figure}

\paragraph{Hilbert space.}
We use the oriented-strand (three-state vertex) representation of the dilute $O(n)$ model \cite{ref66,ref29,ref67}, Fig.~\ref{fig:15}(a). Each slot $i$ of the ring carries $s_i\in\{0,+,-\}$: either it is empty or it is crossed by a strand of the loop gas, oriented upwards ($+$) or downwards ($-$) in the time direction. A ring of $L$ sites with one seam has $M=L+1$ slots, the seam being an extra slot $c$ inserted between $a_{L-1}$ and $a_0$ (Fig.~\ref{fig:16}(a)) which is \emph{never empty}: it is the world-line of the $(2,1)$ strand. The total orientation $S^z=\sum_i s_i$, seam slot included, is conserved, and we work at $S^z=0$, the sector that contains every state with all strands closed into loops; at $L=12$ the seam ring has $3^{13}$ configurations of which $\dim=139\,152$ survive the two constraints. Because the seam slot is always occupied there is no separate seam label to track: what remains dynamical is the orientation $s_c=\pm$ of the seam strand and how it connects to the bulk strands, and that orientation changes under the dynamics, since the pair term can project the seam strand and its neighbour onto a singlet, annihilate the pair and recreate it. The seam strand carries the twist phase when it crosses the cut like any other strand.

\paragraph{Loop weight from orientations.}
Loops acquire their weight from the phases attached to their turns (Fig.~\ref{fig:15}(c)). A strand that turns by $\pm\pi$ (a cup or a cap) carries $q^{\pm1/2}$ according to its orientation, where $q=e^{i\pi/p}$; a closed contractible loop turns by $\pm2\pi$ in total and weighs $q^{\pm1}$, so that the sum over its two orientations gives $q+q^{-1}=2\cos(\pi/p)=n$. Concretely, with $|c\rangle=q^{1/2}|{+}{-}\rangle+q^{-1/2}|{-}{+}\rangle$ and $\langle\tilde c|=q^{1/2}\langle{+}{-}|+q^{-1/2}\langle{-}{+}|$, the Temperley--Lieb generator and the cup and cap on two adjacent slots are
\begin{equation}
e=|c\rangle\langle\tilde c|,\qquad \mathrm{cup}=|c\rangle\langle00|,\qquad \mathrm{cap}=|00\rangle\langle\tilde c|,\qquad e^2=\langle\tilde c|c\rangle\,e=n\,e,
\label{eq:42}
\end{equation}
so that $e$ is ``cap followed by cup''. Note that $\langle\tilde c|$ is \emph{not} the Hermitian conjugate of $|c\rangle$ (that would carry $q^{\mp1/2}$): for complex $q$ the representation is non-Hermitian, which is the price for a continuous $p$ (complex-symmetric defect Hamiltonians of the same kind arise for lattice defects of the non-unitary minimal models \cite{ref68}), and we diagonalise with left and right eigenvectors throughout. The eigenvalues are nevertheless real on all the \emph{Kac levels}, by which we mean the states that belong to the minimal model proper: those whose scaling limit lies in a representation $V_{(r,s)}\otimes\bar V_{(r',s')}$ with both labels inside the Kac table of $M(p,p+1)$, $1\le r\le p-1$, $1\le s\le p$, so that their dimension is $h_{r,s}+h_{r',s'}$ plus integers. These are the states that have a counterpart in the RSOS chain, with the same energies at integer $p$, varying smoothly in between. The loop model also contains \emph{non-Kac} levels, with labels outside the table or outside the $U_q(sl_2)$ highest-weight sector, which the RSOS restriction removes; they are discussed in \S\ref{sec:6.2}.

\paragraph{Heights and orientations.}
The RSOS chain of Section~\ref{sec:lattice} is related to this vertex chain by a map of configurations. A domain wall of the height model is a point where a loop crosses the chain, and the sign of the height step across it, $a_i-a_{i-1}=\pm1$, says on which side of the loop the height is larger. Since the inside of a closed domain wall differs from its outside by the same $\pm1$ all along the wall, this sign is constant around each loop: it is an orientation of the loop. The map $s_i=a_i-a_{i-1}\in\{0,+,-\}$ therefore sends a height configuration to an \emph{oriented} loop configuration, injectively (given $a_0$, the heights are the running sum of the steps), its image being restricted by $1\le a_i\le p-1$ and, on the ring, by $\sum_i s_i=0$; this is why the loop model is studied at $S^z=0$, and for the seam slot it is the statement that the RSOS seam label $c_j=a_j\pm1$ is the orientation $s_c=\pm$ of the seam strand. What the two descriptions do not share is the weight attached to an orientation. In the height representation the generators \eqref{eq:42} act, on a triple of heights with equal ends $l$, as
\begin{equation}
e\,|l\,a\,l\rangle=\frac{\sqrt{S_aS_{a'}}}{S_l}\,|l\,a'\,l\rangle,\qquad \mathrm{cup}\,|l\,l\,l\rangle=\sqrt{\frac{S_a}{S_l}}\,|l\,a\,l\rangle,\qquad \mathrm{cap}\,|l\,a\,l\rangle=\sqrt{\frac{S_a}{S_l}}\,|l\,l\,l\rangle ,
\label{eq:42b}
\end{equation}
with $a,a'$ adjacent to $l$ (these are the TL and cup/cap moves $w_e$, $w_{\cup\cap}$ of \eqref{eq:9} and the projector $P^{\mathbf 1}=e/d_\wl$ of Fig.~\ref{fig:proj}). A contractible loop with outside height $a$ and inside height $a\pm1$ thus weighs $S_{a\pm1}/S_a$, where the loop gas gives $q^{\pm1}$ for the corresponding orientation, and the two agree only after the sum over the two orientations, $\sum_{a'\,\mathrm{adj}\,a}S_{a'}/S_a=n=q+q^{-1}$ (the height-basis form of $\langle\tilde c|c\rangle=n$), not term by term; for a non-contractible loop they differ even then, which is where the twist introduced below enters. The height and vertex chains are therefore two representations of the same elements of the dilute Temperley--Lieb algebra rather than two gauges of one matrix. What coincides is the spectrum, sector by sector, through the standard RSOS--vertex correspondence \cite{ref27,ref69}: the RSOS sectors are reproduced by twists of the $S^z=0$ sector (point (iii) below), and the vertex representation simply continues to exist when $p$ is not an integer.

\paragraph{Hamiltonian.}
The pair term is the element of the dilute Temperley--Lieb algebra that defines the RSOS face operator of \S\ref{sec:2.1}, Fig.~\ref{fig:15}(b), now in the vertex representation:
\begin{equation}
h=w_\varnothing\,\mathbf{1}_\varnothing+w_{\parallel}\,\mathbf{1}_{\parallel}+w_{=}\,\mathbf{1}_{=}+w_{\curvearrowright}\,(\mathrm{hop})+w_{\cup\cap}\,(\mathrm{cup}+\mathrm{cap})+w_e\,e,\qquad H_0=-\sum_j h_{j,j+1},
\label{eq:43}
\end{equation}
where $\mathbf{1}_\varnothing,\mathbf{1}_{\parallel},\mathbf{1}_{=}$ project on the pair configurations with zero, one and two strands, ``hop'' moves a single strand to the other slot of the pair, and cup, cap and $e$ are the vertex generators \eqref{eq:42}. The six scalar coefficients are the same functions of $\lambda=\pi(p-1)/4p$ as in \eqref{eq:9}, now with $p$ real: the first four are unchanged, and $w_{\cup\cap}=\sin2\lambda/\mathcal N_0$, $w_e=-\sin\lambda/\mathcal N_0$, the height-dependent square roots of \eqref{eq:9} being the matrix elements \eqref{eq:42b} of the generators rather than part of the coefficients. In the vertex language $h$ is a $9\times9$ matrix acting on a pair of slots.

\paragraph{The seam.}
The pair term that straddles the seam is built exactly as in \S\ref{sec:2.2}: the strand is braided out of the way, the ordinary $h$ acts on the two physical contents, and the strand is braided back (Fig.~\ref{fig:16}(b)),
\begin{equation}
h^{\mathrm{seam}}_{xcy}=B_{cy}^{-1}\,h_{xc}\,B_{cy},\qquad x=a_{L-1},\ y=a_0,
\label{eq:44}
\end{equation}
a $27\times27$ matrix on the three slots $(x,c,y)$. The braid $B$ on a pair of slots is the Kauffman skein evaluation of a crossing \cite{ref70}, Fig.~\ref{fig:16}(c): it is 1 on two empty slots, the permutation $P$ when one slot is empty (the strand simply moves past a vacancy), and
\begin{equation}
B=A\,\mathbf{1}+A^{-1}e,\qquad A=i\,q^{1/2},\qquad n=-A^2-A^{-2}=q+q^{-1},
\label{eq:45}
\end{equation}
when both slots carry a strand. Equation \eqref{eq:45} is the $FRF^{-1}$ braiding \eqref{eq:12} of the RSOS seam written in the skein language, and the two blocks of $B$ are related by a definite relative phase: the one-strand block is $P$ with coefficient 1, the two-strand block carries $A$. This phase is not a convention. Loops that cross the seam once pick it up, and it is fixed by requiring that the two-seam ring (a $D_{(2,1)}$ strand and its conjugate) reproduce the bulk vacuum energy exactly, i.e.\ that the seam be topological; with the wrong phase the two-seam ring is off the vacuum by a finite amount at every $L$. The braid is unitary in the RSOS representation but not in the vertex one; $h^{\mathrm{seam}}$ is similar to $h$ and has its spectrum. One difference of bookkeeping between the two representations should be kept in mind. In the RSOS chain the seam term changes $a_j$ and $c_j$ only and leaves the neighbouring heights fixed (\S\ref{sec:2.2}); those heights are fusion labels, not the contents of slots. In the vertex chain there is no such spectator: \eqref{eq:44} acts on all three slots $(x,c,y)$, and the orientation, and even the occupation, of $y$ can change in one step---for instance when $y$ is empty, the braid moves the seam strand onto $y$, $h$ then creates a pair on $(x,c)$, and the inverse braid acts on the two occupied slots $(c,y)$. What is preserved is the total orientation $S^z$ and the occupation of the seam slot $c$ after the step.

\paragraph{The clean field.}
The deformation \eqref{eq:21} lives on the same three slots,
\begin{equation}
\Phi_{\mathrm{cl}}=n\,(e_{xc}+e_{cy})-(n_x+n_y),\qquad n_i=\mathbf{1}-|0\rangle\langle0|_i,
\label{eq:46}
\end{equation}
where $e_{xc}$ ($e_{cy}$) is the Temperley--Lieb generator \eqref{eq:42} on the strand and its left (right) neighbour and $n_i$ counts whether slot $i$ is occupied; this is $d_\wl(e_{\mathrm{L}}+e_{\mathrm{R}})-(n_{\mathrm{L}}+n_{\mathrm{R}})$ with $d_\wl=n$. The full Hamiltonian of the single-seam ring is then $H=H_0-h^{\mathrm{seam}}+K\,\Phi_{\mathrm{cl}}$ with $H_0=-\sum_{\text{ordinary pairs}}h$ containing the $L-1$ ordinary pair terms (Fig.~\ref{fig:16}(a)).

\paragraph{Twist sectors.}
In the RSOS calculation the symmetry sectors were obtained by diagonalising the wrapped $(2,1)$ line. The loop model has no such operator at non-integer $p$, but it does not need one: the sectors can be selected directly by a boundary condition, the \emph{twist}, in three steps.

\emph{(i) Conserved orientation and a boundary-condition parameter.} On the ring the total orientation $S^z=\sum_k s_k$ is conserved, since every face move creates or destroys strands in oppositely oriented pairs. In the height language $S^z$ is the net height change around the ring, so the periodic height configurations of the RSOS ring all have $S^z=0$, and we work in that sector throughout. Because $S^z$ is conserved we may also impose a twisted boundary condition: a cut is placed at one site of the ring, between the seam slot $c$ and $a_0$ (Fig.~\ref{fig:16}(a)), and every matrix element in which a strand of orientation $s=\pm1$ crosses the cut is multiplied by $e^{is\phi}$. The angle $\phi$ labels a choice of Hamiltonian, rather than an additional conserved operator; in practice only the wrap-around term \eqref{eq:44} and $\Phi_{\mathrm{cl}}$ act across the cut, so the twist enters only there. Concretely, the twist is not a projection but a twisted boundary condition, and each sector is a separate Hamiltonian $H(\phi)$: a matrix element that moves an up-strand from $c$ to $a_0$ is multiplied by $e^{i\phi}$ and one that moves it back by $e^{-i\phi}$, a down-strand gets the opposite phases, and a cup or cap created or annihilated across the cut gets the phase of the orientation that ends up on the far side; all other terms of $H$ are untouched. $H(\phi)$ is complex and non-Hermitian, and for each value of $\phi$ it is diagonalised separately in the $S^z=0$ sector with biorthogonal left and right eigenvectors. 

\emph{(ii) What the twist weighs.} Consider the loops traced out in space-time by the transfer matrix. A contractible loop crosses the cut an even number of times with opposite orientations, so the twist phases cancel and its weight stays $n=2\cos(\pi/p)$, which comes entirely from its turning phases. A loop that winds once around the ring has no net turning, so its turning phases give nothing, and without a twist it would have weight $1+1=2$; it crosses the cut exactly once, so with the twist its two orientations give $e^{i\phi}+e^{-i\phi}=2\cos\phi$ (Fig.~\ref{fig:16}(d)). The twist is therefore a knob that sets the weight of non-contractible loops independently of $n$; the choice $\phi=\pi/p$ gives them the same weight $n$ as contractible loops and is the vacuum sector. This is the standard twisted-boundary-condition construction of the loop-model sectors \cite{ref71,ref69}.

\emph{(iii) Which $\phi$ gives which RSOS sector.} The decomposition of the RSOS torus partition function into loop (Coulomb-gas) partition functions with twisted boundary conditions is standard: summing over the heights of the plateaus between loops that wind around the ring turns the sum over configurations into a sum over the eigenvalues $\lambda_r=2\cos(\pi r/p)$, $r=1,\dots,p-1$, of the adjacency matrix of the $A_{p-1}$ graph, i.e.\ into a sum of loop partition functions with non-contractible loop weight $\lambda_r$ \cite{ref27,refDFSZ,ref69}; in transfer-matrix language the RSOS levels are recovered from the corresponding twisted loop representations with $2\cos\phi=\lambda_r$ after the RSOS restriction. At integer $p$, this restriction selects the appropriate Kac subquotients and removes the extra states present in the unrestricted vertex representation; the full vertex spectra must not be identified with the RSOS spectrum. The label $r$ is the one of the wrapped $(2,1)$ line, since that line acts on the plateau heights as the adjacency matrix and is itself a non-contractible loop of the loop model. Including the sign $(-1)^{r+s}$ that the $\eta$ charge contributes in \eqref{eq:24}, the Kac state with holomorphic label $(r,s)$ has $Y_{(2,1)}=(-1)^{r+s}\,2\cos(\pi r/p)$, hence it lives in the twist sector
\begin{equation}
\phi=\frac{\pi r}{p}\quad(r+s\ \text{even}),\qquad \phi=\pi-\frac{\pi r}{p}\quad(r+s\ \text{odd}),
\label{eq:47}
\end{equation}
where the second case is the sign flip $r\to p-r$ produced by $\eta$: the sectors $\phi$ and $\pi-\phi$ are distinct in the oriented model because each strand carries one unit of phase. At integer $p$, after the RSOS restriction, these correspond to the sectors $a,b,c,d$ of \S\ref{sec:5.3}, $(Y_{(2,1)},\eta)=(\varphi,+),(\varphi^{-1},-),(-\varphi^{-1},+),(-\varphi,-)$ at $p=5$ corresponding to $\phi=\pi/5,\,2\pi/5,\,3\pi/5,\,4\pi/5$; the ``$k=1$'' sector of the next subsection, with non-contractible weight $-2\cos(\pi/p)$, is $\phi=\pi-\pi/p$. Choosing $\phi$ thus selects the twisted vertex Hamiltonian associated with a given RSOS charge sector, subject to the RSOS restriction at integer $p$; it is not itself a projector. The twisted Hamiltonians continue to be defined at non-integer $p$, where they organize the loop-model calculation without imposing a minimal-model quotient or requiring an RSOS wrapped-line operator.

\paragraph{Validation.}
The construction was validated at $p=5$ against the RSOS chain: the bulk ring reproduces the vacuum energy to $10^{-13}$ and the levels of all four sectors; the two-seam ring reproduces the identity at exactly the bulk vacuum energy (the seam is topological) and every level of the $D_{(2,1)}$ spectrum with its charges; the single-seam ring reproduces the single-strand spectrum, and the running coupling $u_C$ of \eqref{eq:48} along the clean ray, recomputed on the RSOS chain for $L\le14$, agrees with the loop model to all digits.
The loop model has more states than the minimal model: its sectors are Fock-type modules of the Coulomb gas \cite{ref66,ref69,ref72}, so besides the Kac levels they contain states that are null at integer $p$ (for instance, a level at $\Delta=0.20$ below the $(1,1)\otimes(2,1)$ state of the $a$ sector at $p=5$, and $\Delta=1$ current-like states in the bulk). The Hamiltonian is non-Hermitian for all $p$, so levels are tracked with left and right eigenvectors (biorthogonal overlaps).

\subsection{The running coupling at continuous $p$}\label{sec:6.2}

For continuous $p$ we use the same running coupling as at integer $p$, the Casimir coupling $u_C$ of \eqref{eq:48} built from the $k=1$ state of the single-seam ring, with the velocity $v(p)=4p/(3(p-1))$ (so $v(4)=16/9$ and $v(5)=5/3$ as before) and $\ell=7,9,11,13$ for $L=6,\dots,14$. 
At non-integer $p$ there are no exact charges to identify the state, so it is followed continuously in $K$ in steps of $0.05$ by biorthogonal overlap, never re-selected as the lowest level; at $K=0$ it is the ground state of the sector with non-contractible loop weight $-2\cos(\pi/p)$, i.e.\ the Kac state $(1,2)\otimes(2,2)$ with $\Delta_{k=1}$ of \eqref{eq:31}. Two properties make it suitable at continuous $p$: the construction involves no excited level that becomes null at $p=4$, and its energy stays real along the whole flow to $C_p$ (it crosses the lower non-Kac level of its sector near $K\simeq0.3$ without mixing, also at non-integer $p$) and only acquires an imaginary part at $K\gtrsim3.5$, far beyond the fixed point. The $\beta$-function $B_C$ of \eqref{eq:37} is then formed from $u_C$ on successive size pairs and plotted against $A_C$ exactly as in Sections~\ref{sec:4} and \ref{sec:5}. 

\section{Results at non-integer $p$: the family $C_p$ and its endpoint}\label{sec:nonint}

The motivation for following the flow at non-integer $p$ is the contrast between the two integer results: at $p=5$ the flow ends on a genuinely conformal, partially transmitting interface $C_5$, while at $p=4$ it ends on the invertible topological line $\eta$, which transmits everything. Two scenarios could connect them. In the first, $C_5$ belongs to a one-parameter family $C_p$ of stable fixed points that moves continuously with $p$ and simply arrives at $\eta$ when $p$ reaches $4$. In the second, the stable interface $C_p$ disappears before $p$ reaches $4$: it merges with another fixed point of the same ray---the familiar annihilation of a stable and an unstable fixed point, after which the pair continues into the complex plane \cite{ref75,ref76,ref77}---and the flow out of $D_{(2,1)}$, no longer stopped by $C_p$, runs on to a different stable fixed point, which at $p=4$ is $\eta$. In the second scenario $C_5$ and $\eta$ would be unrelated interfaces, the $p=4$ endpoint would be an accident of the annihilation happening before $p=4$, and one would expect the familiar signatures of a merger: a slow, walking flow near the would-be fixed point for $p$ just below the merger, and a $\beta$-function whose zero is approached with vanishing slope. The loop model lets us decide between the two, because it realises the same construction at every real $p$ in $[4,5]$ and gives the $\beta$-function of Section~\ref{sec:6.2} at each of them. The result, presented in this section, is the first scenario: the zero of the $\beta$-function moves smoothly from $C_5$ to $\eta$, the slope at the zero grows in magnitude rather than vanishing, and no second fixed point appears on the ray. The merger with $\eta$ at $p=4$ is then of a different kind, discussed in \S\ref{sec:6.4}.

\subsection{$C_p$ for $4 \le p \le 5$}\label{sec:6.3}

\begin{figure}[ht]\centering
\includegraphics[width=0.9\textwidth]{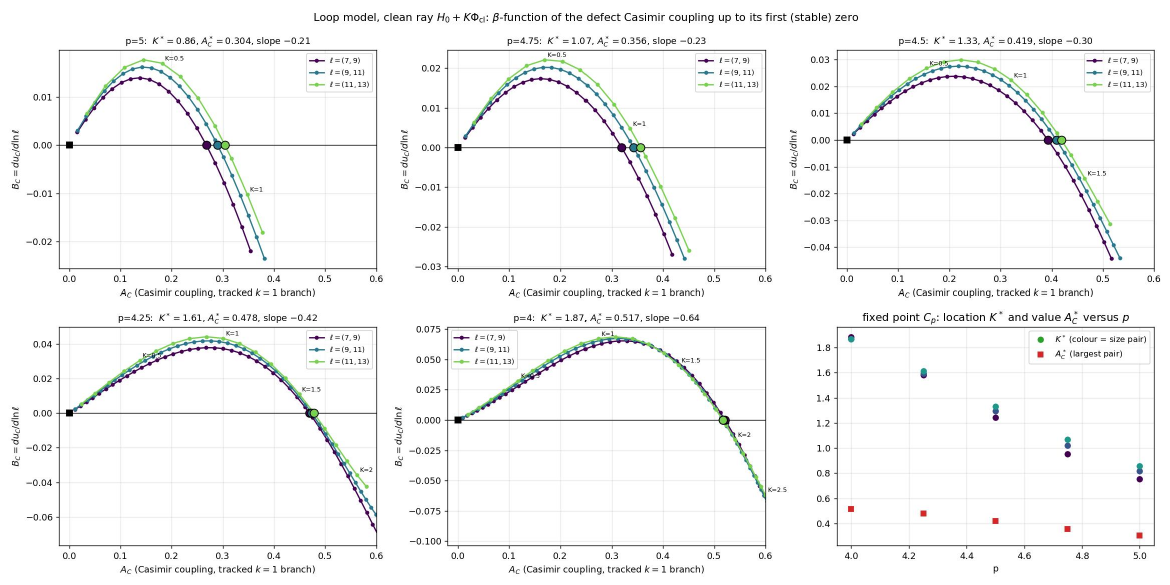}
\caption{Loop model on the clean ray $H_0 + K\Phi_{\mathrm{cl}}$: $\beta$-function $B_C$ of the defect Casimir coupling \eqref{eq:48} versus $A_C$ for $p = 5, 4.75, 4.5, 4.25, 4$ (ED, $L = 6$--$14$, size pairs $\ell = (7,9)$, $(9,11)$ and $(11,13)$), drawn up to $1.35\,K^*$; the black square is $D_{(2,1)}$, the filled circles the zero of each pair. The last panel collects $K^*$ (all pairs) and $A_C^*$ (largest pair) versus $p$.}\label{fig:17}
\end{figure}

\begin{figure}[ht]\centering
\includegraphics[width=0.9\textwidth]{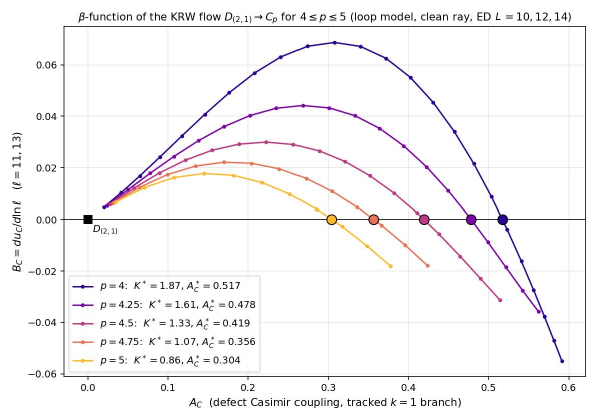}
\caption{The $\ell = (11,13)$ curves of Fig.~\ref{fig:17} ($L = 10, 12, 14$) overlaid: the stable fixed point $C_p$ moves continuously from $p=5$ to $p=4$.}\label{fig:18}
\end{figure}

Figures~\ref{fig:17} and \ref{fig:18} show $B_C$ for $p = 5, 4.75, 4.5, 4.25, 4$ from ED at $L = 6$--$14$ ($\ell = 7, 9, 11, 13$; the $L = 14$ sector has $1.17 \times 10^6$ states and was run in steps of $0.1$ up to just beyond the first zero), up to slightly beyond the first zero. At every $p$ the curve leaves $D_{(2,1)}$ with positive slope, turns over and crosses zero with negative slope, and the successive size pairs lie close to one another:

\begin{center}
\adjustbox{max width=\textwidth}{\begin{tabular}{lccccc}
\toprule
$p$ & 5 & 4.75 & 4.5 & 4.25 & 4 \\
\midrule
$K^*$ $(7,9)$ & 0.752 & 0.951 & 1.243 & 1.580 & 1.885 \\
$K^*$ $(9,11)$ & 0.816 & 1.022 & 1.296 & 1.595 & 1.867 \\
$K^*$ $(11,13)$ & 0.859 & 1.068 & 1.331 & 1.612 & 1.869 \\
$A_C^*$ $(9,11)$ & 0.290 & 0.342 & 0.409 & 0.474 & 0.517 \\
$A_C^*$ $(11,13)$ & 0.304 & 0.356 & 0.419 & 0.478 & 0.517 \\
slope $B_C'(A_C^*)$ $(11,13)$ & $-0.21$ & $-0.23$ & $-0.30$ & $-0.42$ & $-0.64$ \\
\bottomrule
\end{tabular}}
\end{center}

The stable fixed point $C_p$ therefore exists for every $p$ between 5 and 4: it moves smoothly to larger bare coupling ($K^*$ from $0.86$ to $1.87$) and to a larger Casimir value ($A_C^*$ from $0.30$ to $0.52$), and nothing happens to it along the way. The $p=5$ and $p=4$ columns are the clean-ray counterparts of the improved-ray $\beta$-functions of Sections~\ref{sec:4} and \ref{sec:5}. At $p=4$ the zero $K^*=1.87$, $A_C^*=0.517$ matches the improved-ray result $1.90$, $0.516$, the counterterm there being small. At $p=5$ the clean ray is the better behaved of the two: its three pairs lie close together, the steps of $K^*$ ($0.064$, $0.044$) decrease by a factor $0.7$ from pair to pair, which is the basis of the estimate $K^*_{C_5}=0.95(5)$ of \eqref{eq:33}, and the slope at the zero, $-0.21$, is close to the improved-ray value $-0.16(2)$; the improved ray, by contrast, drifts much more because of the $K$-proportional counterterm (\S\ref{sec:5.4}). The crossing value $A_C^*=0.27$--$0.30$ on the clean ray is also larger than the improved-ray values, the two being on their way to the common universal value with the slow $L^{-0.12}$ corrections.
Finite-size effects are stronger at large p. This is expected from the leading irrelevant dimension along the ray: the slope at the zero is $\Delta_X - 1$, and it grows in magnitude steadily from $-0.21$ at $p=5$ to $-0.64$ at $p=4$, so the $L^{\Delta_X - 1}$ corrections that make the $p=5$ crossing drift are strongly suppressed at $p=4$. 

\subsection{The endpoint at $p=4$: the line $D_{(1,4)}$}\label{sec:6.4}

The family $C_p$ ends, at $p=4$, on the invertible line $\eta$. In the Kac labelling of $M(p,p+1)$ this line is $\eta = D_{(p-1,1)} \cong D_{(1,p)}$, so at $p=4$ it is $D_{(1,4)}$. Unlike $(p-1,1)$, the label $(1,4)$ makes sense for every $p$: the generalised minimal model at $c = 1 - 6/p(p+1)$ \cite{ref73,ref74} has topological lines for all degenerate labels $(r,s)$, with fusion rules that factorise into an $r$-part and an $s$-part and quantum dimensions $d_{(r,s)} = [r]_q [s]_{q'}$, $q = e^{i\pi/p}$, $q' = e^{i\pi/(p+1)}$. The line $D_{(1,4)}$ therefore exists along the whole family, commutes with all the $(r,1)$ lines (it lies in the other factor of the fusion category and is mutually transparent with them), and has
\begin{equation}
d_{(1,4)} = \frac{\sin\big(4\pi/(p+1)\big)}{\sin\big(\pi/(p+1)\big)} = 1.73,\ 1.57,\ 1.40,\ 1.20,\ 1 \qquad \text{at } p = 5,\ 4.75,\ 4.5,\ 4.25,\ 4,
\label{eq:49}
\end{equation}
so it is invertible only at $p=4$; at $p=5$ the Kac identification makes it $D_{(1,4)} \cong D_{(4,2)}$, a non-invertible line of dimension $\sqrt{3}$. For $p>4$, however, $D_{(1,4)}$ cannot be the endpoint of the flow: as shown next, it carries a pair of relevant symmetric defect operators, so it is not infrared-stable, and the measured $C_p$ ($g_{C_5}=1.503$, $\mathcal T=0.665$ at $p=5$, against the topological values $\sqrt3$ and $1$) is a distinct, genuinely partially reflecting interface that merges with $D_{(1,4)}=\eta$ only at $p=4$.

\paragraph{Defect operators on $D_{(1,4)}$.} 
The operators on $D_{(1,4)}$ are the twisted sector of $D_{(1,4)} \otimes \bar{D}_{(1,4)}$, i.e.\ the sum over the channels $c \in (1,4) \times (1,4) = (1,1) + (1,3) + (1,5) + (1,7)$ of the $c$-twisted spaces; those neutral under the $(r,1)$ lines (chiral label $(1,s)$ with $s$ odd, \S\ref{sec:4}) are the $(1,s) \otimes \overline{(1,s')}$ with $s, s'$ odd:

\begin{center}
\adjustbox{max width=\textwidth}{\begin{tabular}{lllll}
\toprule
operator & $\Delta = h_{1,s} + h_{1,s'}$ & $p=4$ & $p=5$ & channel $c$ \\
\midrule
$(1,1) \otimes \overline{(1,3)}$, $(1,3) \otimes \overline{(1,1)}$ & $(p-1)/(p+1)$ & 0.60 & 0.667 & $(1,3)$ \\
$(1,3) \otimes \overline{(1,3)}$ & $2(p-1)/(p+1)$ & 1.20 & 1.33 & $(1,1), (1,3), (1,5)$ \\
$(1,1) \otimes \overline{(1,5)}$, $(1,5) \otimes \overline{(1,1)}$ & $(4p-2)/(p+1)$ & 2.8 & 3.0 & $(1,5)$ \\
\bottomrule
\end{tabular}}
\end{center}

For generic $p$ the line carries a pair of \emph{relevant} symmetric operators, the chiral defect fields $\phi^L = \phi_{(1,3)} \otimes \bar{\mathbf{1}}$ and $\phi^R = \mathbf{1} \otimes \bar{\phi}_{(1,3)}$ of weights $(h_{1,3}, 0)$ and $(0, h_{1,3})$, $\Delta = (p-1)/(p+1)$. They live in the $c = (1,3)$ channel, and that channel is exactly what the Kac quotient removes at $p=4$: there $(1,4) \cong (3,1)$ and $(3,1) \times (3,1) = (1,1)$ only, so the line becomes invertible and only the diagonal operators $(1,s) \otimes \overline{(1,s)}$ survive, the first being $(1,3) \otimes \overline{(1,3)}$ with $\Delta = 6/5$, the leading invariant operator on $\eta$ found in \S\ref{sec:4}. For $p>4$ the $(1,3)$ channel survives, so $D_{(1,4)}$ (at $p=5$, $D_{(1,4)} \cong D_{(4,2)}$) is unstable there; this is why it cannot be the endpoint of the flow for $p>4$, as anticipated above. 

The relevant fields $\phi^{L,R}$ of $D_{(1,4)}$ are null at $p=4$ and the line $D_{(1,4)}=\eta$ becomes infrared stable exactly at $p=4$. This is the analogue of the mechanism by which states of vanishing norm drop out of the $O(n)$ and Potts models at integer $n$ or $Q$ \cite{ref72,ref82}. 

\section{Summary and discussion}\label{sec:concl}

The dilute $A_{p-1}$ RSOS chain, read as an anyon chain of $\mathcal C_p$ with the $SU(2)_{p-2}$ fusion rules, realises the $(r,1)$ Verlinde defects of $M(p,p+1)$ exactly topologically: a defect is one more anyon threaded through the chain, coupled by the $F$- and $R$-moves of the category, and the wrapped lines commute with the Hamiltonian to machine precision. The quantum-dimension-weighted projector combination \eqref{eq:21} between the two fusion channels of the defect anyon and its neighbouring domain wall is the clean lattice representative of the KRW field $\phi_{(1,3)} \otimes \bar{\phi}_{(3,3)}$: its overlaps with the identity and the irrelevant primary $\phi_{(1,3)} \otimes \bar{\phi}_{(1,3)}$ vanish to numerical precision at the tested sizes, while descendants and higher fields can contribute irrelevant corrections. 

With the clean operator the KRW flow is reproduced from first principles. At $p=4$, it ends on the invertible line $\eta$, an endpoint that had not been proposed before: the $g$-function drops from $\sqrt{2}$ to $1.00$, every level of the two-strand sector-resolved spectrum flows to the diagonal tricritical-Ising value, a ring with a single strand reproduces the $\mathbb{Z}_2$-twisted spectrum with no vacuum, which rules out the trivial line as the endpoint. At $p=5$ it ends on the conformal interface $C_5$ with $g = 1.503(2)$ and $\mathcal{T} = 0.665(10)$, whose leading irrelevant field has $\Delta_X = 1.12(1)$.
Gauging $\eta$ maps $C_5$ to an interface of the three-state Potts model, $\tau_{\mathrm{Fib}} \to C_5^{\mathrm{Potts}}$, which breaks $\mathbb{Z}_3$ and is stable precisely when the Fibonacci line $\tau_{\mathrm{Fib}}$ and charge conjugation $C$ are both imposed (\S\ref{sec:5.5}). 
Followed at continuous $p$ in the loop model (Section~\ref{sec:loop}), the family $C_p$ moves smoothly from $C_5$ to $\eta$, and the merger at $p=4$ is not a fixed-point annihilation: $C_p$ is the endpoint of the symmetric $\phi_{(1,3)}$ deformation of the line $D_{(1,4)}$, whose relevant $(1,3)$-channel fields become null at $p=4$ and decouple, so that $\eta=D_{(1,4)}$ becomes stable exactly there (\S\ref{sec:6.4}). 

\paragraph{Comparison with conformal perturbation theory and with integrable defect flows.}
The transmission of $C$ has been computed in conformal perturbation theory by Makabe and Watts \cite{ref84}: for the diagonal KRW fixed point they find the reflection coefficient $R_C = 1 - \mathcal{T} = \tfrac{9}{8}\pi^2 y^4 + O(y^5)$ with $y = 1 - h_{1,3} = 2/(p+1)$, which gives $R = 0.284$ at $p=4$ (against $0.366$ from their non-perturbative work on tricritical Ising defects \cite{ref12}) and $R = \pi^2/72 = 0.137$ at $p=5$, to be compared with our $R = 1 - \mathcal{T} = 0.335(10)$. As for the $g$-function, the leading term underestimates the departure from the topological value at $y = 1/3$, by a factor $2.4$ here; the general leading-order formulas for the change of $g$, of the reflectivity and of the entanglement entropy along a defect flow driven by a spinless field are given in \cite{ref85}. On the lattice side, the chiral KRW flows---the axes of the $(\lambda_l, \lambda_r)$ plane---have been realised in the dense RSOS models by an inhomogeneous spectral parameter on one column, which keeps the model integrable and gives the running $g$-function from the thermodynamic Bethe ansatz \cite{ref78,ref24}; those flows end on topological lines, whose identity is now fixed in general by the functional-relation analysis of \cite{refAKRW}, and the present construction is the complementary one, in which the non-chiral deformation of an exactly topological seam reaches the non-topological interface $C$.

\paragraph{Comparison with open string field theory.}
The $g$-function of the KRW defect $C$ has also been computed by Kudrna, Maccaferri and Schnabl \cite{ref86} as a level-truncated solution of open string field theory in the doubled theory $M(p,p+1)\otimes\overline{M(p,p+1)}$. Their extrapolated values are
\[
g_C^{\mathrm{OSFT}} \simeq 1.08\text{--}1.10 \quad (p=4), \qquad g_C^{\mathrm{OSFT}} \simeq 1.489 \quad (p=5),
\]
with the $p=4$ number flagged there as requiring higher levels, and the leading conformal-perturbation-theory estimate quoted in the same work, $\ln g_C - \ln g_{D_{(2,1)}} = -\tfrac{\pi^2}{4} y^3 + O(y^4)$ with $y = \tfrac{2}{p+1}$, gives $1.477$ at $p=5$. These numbers are comparable with our lattice results, $g = 1.503(2)$ for $C_5$ and $g = 1.00$ for the $p=4$ endpoint.

\acknowledgments

This work was heavily assisted by Claude Fable 5 (Anthropic): the lattice constructions, the numerical codes, the computations reported here and the drafting of the text were developed in collaboration with it, with the author directing the questions and checking the results.
The work of the author is supported by the European Union (ERC, QFTinAdS, project number 101087025, PI Balt van Rees). Views and opinions expressed are, however, those of the authors only and do not necessarily reflect those of the European Union or the European Research Council Executive Agency.

\bibliographystyle{JHEP}
\bibliography{krw}

\end{document}

%% file: figs/fig1_tikz.tex
\begin{tikzpicture}[x=1cm,y=1cm,
  hbox/.style={draw=blue!60!black,fill=blue!8,thick,minimum size=0.42cm,inner sep=1pt,font=\small},
  occ/.style={circle,fill=red!85!black,draw=red!85!black,inner sep=0pt,minimum size=5pt},
  emp/.style={circle,draw=gray!80,thick,fill=white,inner sep=0pt,minimum size=5pt},
  font=\small]
  \def\dx{1.0}
  \foreach \j/\h in {0/2,1/2,2/3,3/3,4/2,5/3,6/3,7/4}{
    \node[hbox] (b\j) at (\j*\dx,0) {$\h$};
    \node[blue!60!black,font=\footnotesize] at (\j*\dx,-0.62) {$a_{\j}$};
  }
  \draw[gray!80] (-0.45,0) -- (b0.west);
  \draw[gray!80] (b7.east) -- (7*\dx+0.45,0);
  \foreach \j/\s in {0/emp,1/occ,2/emp,3/occ,4/occ,5/emp,6/occ}{
    \pgfmathtruncatemacro{\k}{\j+1}
    \draw[gray!80] (b\j.east) -- (b\k.west);
    \node[\s] at (\j*\dx+0.5*\dx,0) {};
  }
  \foreach \j in {1,3,4,6}{
    \node[red!85!black,font=\footnotesize] at (\j*\dx+0.5*\dx,0.42) {$\wl$};
  }
  \begin{scope}[shift={(8.2,0.2)}]
    \node[hbox] at (0,0) {$a$};
    \node[anchor=west,font=\footnotesize] at (0.3,0) {link $j$: height $a_j\in\{1,\dots,L\}$, $|a_j-a_{j+1}|\le 1$};
    \node[occ] at (0,-0.5) {};
    \node[anchor=west,font=\footnotesize] at (0.3,-0.5) {occupied site ($a_j\neq a_{j+1}$): carries a $\wl$};
    \node[emp] at (0,-1.0) {};
    \node[anchor=west,font=\footnotesize] at (0.3,-1.0) {empty site ($a_j=a_{j+1}$)};
  \end{scope}
  \begin{scope}[shift={(0,1.6)},y=0.42cm]
    \def\x0{-0.55}\def\xe{7*\dx+0.55}
    \fill[blue!8]
      (\x0,1) -- (\x0,2) -- (1.5*\dx,2) -- (1.5*\dx,3) -- (3.5*\dx,3) -- (3.5*\dx,2)
      -- (4.5*\dx,2) -- (4.5*\dx,3) -- (6.5*\dx,3) -- (6.5*\dx,4) -- (\xe,4) -- (\xe,1) -- cycle;
    \draw[blue!40!black,very thick]
      (\x0,2) -- (1.5*\dx,2)  (1.5*\dx,3) -- (3.5*\dx,3)  (3.5*\dx,2) -- (4.5*\dx,2)
      (4.5*\dx,3) -- (6.5*\dx,3)  (6.5*\dx,4) -- (\xe,4);
    \foreach \j in {1,3,4,6}{
      \draw[red!85!black,thick] (\j*\dx+0.5*\dx,1) -- (\j*\dx+0.5*\dx,4.4);
    }
    \draw[-{Stealth[length=4pt]},gray!80] (\x0,1) -- (\x0,4.9);
    \foreach \h in {1,2,3,4}{
      \node[gray,font=\footnotesize,anchor=east] at (\x0-0.1,\h) {$\h$};
    }
    \node[gray,font=\footnotesize,anchor=south] at (\x0+0.2,5.0) {height};
    \node[anchor=west,blue!40!black,font=\footnotesize] at (7.9,3.2) {height profile: domains};
    \node[anchor=west,red!85!black,font=\footnotesize] at (7.9,2.3) {domain wall $=$ a $\wl$ on the site};
  \end{scope}
\end{tikzpicture}

%% file: figs/fig2_tikz.tex
\begin{tikzpicture}[
  font=\small,
  hbox/.style={draw=blue!70!black, fill=blue!8, minimum width=0.42cm, minimum height=0.42cm, inner sep=0pt, font=\small},
  nbox/.style={hbox, draw=red!80!black, fill=red!10},
  wall/.style={circle, fill=red!80!black, inner sep=0pt, minimum size=0.13cm},
  empty/.style={circle, draw=gray, fill=white, inner sep=0pt, minimum size=0.12cm},
  mv/.style={-{Latex[length=1.6mm]}, semithick},
  pics/move/.style n args={9}{code={
    \node[font=\small] at (1.65,0.62) {#1};
    \node[hbox] (b1) at (0,0) {$#2$};
    \node[#5] at (0.34,0) {};
    \node[hbox] (b2) at (0.68,0) {$#3$};
    \node[#6] at (1.02,0) {};
    \node[hbox] (b3) at (1.36,0) {$#4$};
    \draw[mv] (1.68,0) -- (2.22,0);
    \node[hbox] (c1) at (2.56,0) {$#2$};
    \node[#8] at (2.90,0) {};
    \node[nbox] (c2) at (3.24,0) {$#7$};
    \node[#9] at (3.58,0) {};
    \node[hbox] (c3) at (3.92,0) {$#4$};
  }},
]
\pic at (0,0)   {move={empty: no wall, stays}{2}{2}{2}{empty}{empty}{2}{empty}{empty}};
\node[font=\small] at (1.96,-0.62) {$w_{\varnothing}=\frac{\sin 3\lambda}{\mathcal N_0}$};
\pic at (5.9,0) {move={one wall, stays}{1}{2}{2}{wall}{empty}{2}{wall}{empty}};
\node[font=\small] at (7.86,-0.62) {$w_{\scriptscriptstyle\parallel}=-\frac{\sin 2\lambda\cos 3\lambda}{\mathcal N_0}$};
\pic at (11.8,0){move={two walls, stay}{1}{2}{1}{wall}{wall}{2}{wall}{wall}};
\node[font=\small] at (13.76,-0.62) {$w_{=}=-\frac{\sin 5\lambda}{\mathcal N_0}$};
\pic at (0,-2.4)   {move={hop: wall moves across}{1}{2}{2}{wall}{empty}{1}{empty}{wall}};
\node[font=\small] at (1.96,-3.02) {$w_{\curvearrowright}=\frac{\sin 3\lambda}{\mathcal N_0}$};
\pic at (5.9,-2.4) {move={create / annihilate a pair ($l=r$)}{2}{2}{2}{empty}{empty}{3}{wall}{wall}};
\node[font=\small] at (7.86,-3.02) {$w_{\cup\cap}=\frac{\sin 2\lambda}{\mathcal N_0}\sqrt{S_a/S_l}$};
\pic at (11.8,-2.4){move={TL generator $e$ ($l=r$, both walls)}{2}{1}{2}{wall}{wall}{3}{wall}{wall}};
\node[font=\small] at (13.76,-3.08) {$w_{e}=-\frac{\sin\lambda}{\mathcal N_0}\,\frac{\sqrt{S_aS_{a'}}}{S_l}$};
\begin{scope}[yshift=-4.3cm, font=\footnotesize]
  \node[wall] (L1) at (0,0) {};
  \node[anchor=west, inner sep=2pt] (L2) at (0.12,0) {wall ($\wl$)};
  \node[empty, right=0.5cm of L2] (L3) {};
  \node[anchor=west, inner sep=2pt] (L4) at ([xshift=0.08cm]L3.east) {empty};
  \node[nbox, right=0.5cm of L4] (L5) {$a'$};
  \node[anchor=west, inner sep=2pt] (L6) at ([xshift=0.08cm]L5.east) {middle height after the move ($a'$)};
  \node[anchor=west, inner sep=2pt] (L7) at (-0.1,-0.55) {each move: $(l,a,r)\longrightarrow(l,a',r)$ with amplitude $W^{(l,r)}_{a\to a'}$; the spectators $l,r$ never change.};
\end{scope}
\end{tikzpicture}

%% file: figs/fig3_tikz.tex
\begin{tikzpicture}[
  x=1cm,y=1cm,
  hbox/.style={draw=blue!60!black,fill=blue!8,minimum width=0.72cm,minimum height=0.72cm,inner sep=1pt,font=\small},
  sbox/.style={draw=red!70!black,fill=red!10,minimum width=0.72cm,minimum height=0.72cm,inner sep=1pt,font=\small},
  dbox/.style={draw=blue!60!black,densely dashed,fill=blue!8,minimum width=0.72cm,minimum height=0.72cm,inner sep=1pt,font=\small},
  rowlab/.style={font=\small,align=right,anchor=east},
  note/.style={font=\small,text=blue!60!black,align=center},
  >=Stealth]
\newcommand{\hdot}[1]{%
  \fill[white] (#1) circle (0.11);
  \begin{scope}
    \clip (#1) circle (0.11);
    \fill[red!70!black] ($(#1)+(-0.11,-0.11)$) rectangle ($(#1)+(0,0.11)$);
  \end{scope}
  \draw[red!70!black,line width=0.6pt] (#1) circle (0.11);}
\newcommand{\fdot}[1]{\fill[red!70!black] (#1) circle (0.13);}

\def\dx{1.27}
\begin{scope}[yshift=0cm]
  \draw[gray] (-0.55,0) -- (5*\dx+0.55,0);
  \node[hbox] (a1) at (0,0) {$a_{j-2}$};
  \node[hbox] (b1) at (\dx,0) {$a_{j-1}$};
  \node[hbox] (c1) at (2*\dx,0) {$a_{j}$};
  \node[sbox] (d1) at (3*\dx,0) {$c_{j}$};
  \node[hbox] (e1) at (4*\dx,0) {$a_{j+1}$};
  \node[hbox] (f1) at (5*\dx,0) {$a_{j+2}$};
  \hdot{0.5*\dx,0}\hdot{1.5*\dx,0}\fdot{2.5*\dx,0}\hdot{3.5*\dx,0}\hdot{4.5*\dx,0}
  \node[font=\small,text=red!70!black] at (2.5*\dx,0.5) {$x$};
  \node[font=\small] at (3.5*\dx,0.5) {$y$ (site $j{+}1$)};
  \node[rowlab] at (-1.0,0) {(1) with the strand\\in place};
  \draw[decorate,decoration={brace,mirror,amplitude=4pt},blue!60!black]
    (\dx-0.38,-0.5) -- (4*\dx+0.38,-0.5);
  \node[note,anchor=north] at (2.5*\dx,-0.75)
    {$h_j$ needs the neighbours $(a_{j-1},\,a_j,\,a_{j+1})$ of the middle link $a_j$,\\
     but the strand $c_j$ sits between $a_j$ and $a_{j+1}$};
\end{scope}
\draw[->,thick] (2.5*\dx,-1.8) -- (2.5*\dx,-2.5);
\node[font=\small,anchor=west] at (2.5*\dx+0.15,-2.15)
  {braid the strand past site $j{+}1$:\ \ $B=F^{a\,y\,x}R^{yx}(F^{a\,x\,y})^{-1}$};
\begin{scope}[yshift=-3.5cm]
  \draw[gray] (-0.55,0) -- (5*\dx+0.55,0);
  \draw[draw=blue!60!black,fill=blue!8,rounded corners=4pt]
    (\dx-0.55,-0.5) rectangle (3*\dx+0.55,0.5);
  \node[hbox] (a2) at (0,0) {$a_{j-2}$};
  \node[hbox] (b2) at (\dx,0) {$a_{j-1}$};
  \node[hbox] (c2) at (2*\dx,0) {$a_{j}$};
  \node[dbox] (d2) at (3*\dx,0) {$b'$};
  \node[sbox] (e2) at (4*\dx,0) {$\tilde c$};
  \node[hbox] (f2) at (5*\dx,0) {$a_{j+2}$};
  \hdot{0.5*\dx,0}\hdot{1.5*\dx,0}\hdot{2.5*\dx,0}\fdot{3.5*\dx,0}\hdot{4.5*\dx,0}
  \node[font=\small] at (2.5*\dx,0.72) {$y$};
  \node[font=\small,text=red!70!black] at (3.5*\dx,0.72) {$x$};
  \node[rowlab] at (-1.0,0) {(2) strand moved\\out of the way};
  \node[font=\small,text=gray,anchor=west,align=left] at (5*\dx+0.5,0)
    {$b'\in a_j\otimes y$\\[2pt]$\tilde c=a_{j+1}$};
  \node[note,anchor=north] at (2.5*\dx,-0.75)
    {$h^{\mathrm{face}}$ acts on the triple $(a_{j-1},\,a_j,\,b')$ and changes $a_j$};
\end{scope}
\draw[->,thick] (2.5*\dx,-4.9) -- (2.5*\dx,-5.6);
\node[font=\small,anchor=west] at (2.5*\dx+0.15,-5.25)
  {braid back:\ \ $B^{-1}$ (evaluated with the new $a_j$)};
\begin{scope}[yshift=-6.4cm]
  \draw[gray] (-0.55,0) -- (5*\dx+0.55,0);
  \node[hbox] (a3) at (0,0) {$a_{j-2}$};
  \node[hbox] (b3) at (\dx,0) {$a_{j-1}$};
  \node[sbox] (c3) at (2*\dx,0) {$a'_{j}$};
  \node[sbox] (d3) at (3*\dx,0) {$c'_{j}$};
  \node[hbox] (e3) at (4*\dx,0) {$a_{j+1}$};
  \node[hbox] (f3) at (5*\dx,0) {$a_{j+2}$};
  \hdot{0.5*\dx,0}\hdot{1.5*\dx,0}\fdot{2.5*\dx,0}\hdot{3.5*\dx,0}\hdot{4.5*\dx,0}
  \node[rowlab] at (-1.0,0) {(3) strand back\\in place};
  \node[font=\small,anchor=north] at (2.5*\dx,-0.5)
    {$h_j^{\mathrm{seam}}=B^{-1}\,h_j^{\mathrm{face}}\,B$:\ \ $a_j\to a'_j$ and $c_j\to c'_j$ may change, $a_{j-1}$ and $a_{j+1}$ do not};
\end{scope}

\begin{scope}
  \def\xl{12.6}\def\xc{14.0}\def\xr{15.4}
  \def\yb{-7.1}\def\yt{0.5}
  \node[font=\small,anchor=south] at (\xc,\yt+0.35) {world-lines, time $\uparrow$};
  \fill[gray!12] (\xl-0.35,-4.3) rectangle (\xc+0.35,-3.0);
  \draw[gray!60] (\xl-0.35,-4.3) rectangle (\xc+0.35,-3.0);
  \node[font=\small,anchor=west] at (\xr+0.3,-3.65) {(2) $h^{\mathrm{face}}$ on the walls of sites $j$, $j{+}1$};
  \draw[red!70!black,thick] (\xl,\yb) -- (\xl,\yt);
  \draw[red!70!black,thick] (\xr,\yb) -- (\xr,-5.9);
  \draw[red!70!black,thick] (\xr,-5.9) .. controls (\xr,-5.3) and (\xc,-5.2) .. (\xc,-4.6);
  \draw[red!70!black,line width=2pt] (\xc,\yb) -- (\xc,-5.9);
  \draw[white,line width=5pt] (\xc,-5.9) .. controls (\xc,-5.3) and (\xr,-5.2) .. (\xr,-4.6);
  \draw[red!70!black,line width=2pt] (\xc,-5.9) .. controls (\xc,-5.3) and (\xr,-5.2) .. (\xr,-4.6);
  \draw[red!70!black,thick] (\xc,-4.6) -- (\xc,-2.7);
  \draw[red!70!black,line width=2pt] (\xr,-4.6) -- (\xr,-2.7);
  \draw[red!70!black,thick] (\xc,-2.7) .. controls (\xc,-2.1) and (\xr,-2.0) .. (\xr,-1.4);
  \draw[white,line width=5pt] (\xr,-2.7) .. controls (\xr,-2.1) and (\xc,-2.0) .. (\xc,-1.4);
  \draw[red!70!black,line width=2pt] (\xr,-2.7) .. controls (\xr,-2.1) and (\xc,-2.0) .. (\xc,-1.4);
  \draw[red!70!black,thick] (\xr,-1.4) -- (\xr,\yt);
  \draw[red!70!black,line width=2pt] (\xc,-1.4) -- (\xc,\yt);
  \node[font=\small,anchor=north] at (\xl,\yb-0.05) {site $j$};
  \node[font=\small,anchor=north,text=red!70!black] at (\xc,\yb-0.05) {$x$ ($c_j$)};
  \node[font=\small,anchor=north] at (\xr,\yb-0.05) {$y$};
  \node[font=\small,anchor=south] at (\xl,\yt-0.05) {site $j$};
  \node[font=\small,anchor=south,text=red!70!black] at (\xc,\yt-0.05) {$x$ ($c'_j$)};
  \node[font=\small,anchor=south] at (\xr,\yt-0.05) {$y$};
  \node[font=\small,anchor=west] at (\xr+0.3,-5.25) {$B$: strand $x$ crosses in front of $y$};
  \node[font=\small,anchor=west] at (\xr+0.3,-2.05) {$B^{-1}$: strand crosses back, again in front of $y$};
  \node[font=\small,anchor=west,text=gray] at (\xr+0.3,-6.5) {(1)};
  \node[font=\small,anchor=west,text=gray] at (\xr+0.3,-0.6) {(3)};
  \node[font=\footnotesize,anchor=north,align=center,text=gray] at (\xc+1.2,\yb-0.55) {thin lines: domain walls of sites $j$ and $j{+}1$ ($=y$), if occupied;\\ thick line: the $(2,1)$ strand $x$ (height $c_j$); grey box: the face operator};
\end{scope}
\end{tikzpicture}

%% file: figs/fig_latlasso_tikz.tex
\begin{tikzpicture}[x=1cm,y=1cm,>=Stealth, every node/.style={font=\small}]
\newcommand{\tree}{%
  \draw[thick,dotted] (-0.6,0) -- (0,0); \draw[thick] (0,0) -- (9.9,0); \draw[thick,dotted] (9.9,0) -- (10.5,0);
  \foreach \xv/\lab in {0.9/\ell_0,2.1/\ell_1,3.3/\ell_2,5.4/\ell_4,6.6/\ell_5,7.8/\ell_6,9.0/\ell_7}{
     \draw[thick] (\xv,0) -- (\xv,0.8); \fill (\xv,0) circle (1.4pt); \node[anchor=south,inner sep=1pt] at (\xv,0.8) {$\lab$};}
  \draw[red,very thick] (4.5,0) -- (4.5,0.8); \fill (4.5,0) circle (1.4pt); \node[red,anchor=south,inner sep=1pt] at (4.5,0.8) {$x'$};}
\begin{scope}
  \node[anchor=west] at (-1.3,1.0) {(a)};
  \tree
  \foreach \xv/\lab in {0.4/a_{7},1.5/a_0,2.7/a_1,3.9/a_2,4.95/c,6.0/a_4,8.4/b_6,9.5/b_7}{\node[anchor=north,inner sep=2pt] at (\xv,-0.02) {$\lab$};}
  \draw[blue,thick] (6.3,-1.1) -- (7.2,0); \fill[blue] (7.2,0) circle (1.4pt);
  \node[anchor=north,inner sep=2pt] at (6.8,-0.02) {$a_5$}; \node[anchor=north,inner sep=2pt] at (7.62,-0.02) {$b_5$};
  \draw[blue,thick,dotted] (5.6,-1.1) -- (6.3,-1.1); \node[blue,anchor=east] at (5.5,-1.1) {$x$};
\end{scope}
\node at (5.0,-2.4) {$\xrightarrow{\ \text{move the attachment point across the vertex of slot }5:\ F^{\,x\,a_4\,\ell_5}_{\,b_5\,b_4\,a_5}\ }$};
\begin{scope}[yshift=-4.3cm]
  \tree
  \foreach \xv/\lab in {0.4/a_{7},1.5/a_0,2.7/a_1,3.9/a_2,4.95/c,7.2/b_5,8.4/b_6,9.5/b_7}{\node[anchor=north,inner sep=2pt] at (\xv,-0.02) {$\lab$};}
  \draw[blue,thick] (5.1,-1.1) -- (6.0,0); \fill[blue] (6.0,0) circle (1.4pt);
  \node[anchor=north,inner sep=2pt] at (5.6,-0.02) {$a_4$}; \node[anchor=north,inner sep=2pt] at (6.42,-0.02) {$b_4$};
  \draw[blue,thick,dotted] (4.4,-1.1) -- (5.1,-1.1); \node[blue,anchor=east] at (4.3,-1.1) {$x$};
\end{scope}
\begin{scope}[yshift=-7.6cm]
  \node[anchor=west] at (-1.3,1.0) {(b)};
  \draw[thick] (-0.6,0) -- (3.4,0);
  \draw[thick] (1.2,0) -- (1.2,0.8); \node[anchor=south,inner sep=1pt] at (1.2,0.8) {$\ell_k$};
  \draw[blue,thick] (1.6,-1.1) -- (2.7,0); \node[blue,anchor=north east] at (1.6,-1.1) {$x$};
  \fill[blue] (2.7,0) circle (1.4pt); \fill (1.2,0) circle (1.4pt);
  \node[anchor=north,inner sep=2pt] at (0.3,-0.02) {$a_{k-1}$}; \node[anchor=north,inner sep=2pt] at (1.95,-0.02) {$a_k$}; \node[anchor=north,inner sep=2pt] at (3.1,-0.02) {$b_k$};
  \node at (5.0,0) {$=\ \displaystyle\sum_{b_{k-1}}\ F^{\,x\,a_{k-1}\,\ell_k}_{\,b_k\,b_{k-1}\,a_k}$};
  \begin{scope}[xshift=7.3cm]
    \draw[thick] (-0.6,0) -- (3.4,0);
    \draw[thick] (2.0,0) -- (2.0,0.8); \node[anchor=south,inner sep=1pt] at (2.0,0.8) {$\ell_k$};
    \draw[blue,thick] (-0.6,-1.1) -- (0.5,0); \node[blue,anchor=north east] at (-0.6,-1.1) {$x$};
    \fill[blue] (0.5,0) circle (1.4pt); \fill (2.0,0) circle (1.4pt);
    \node[anchor=north,inner sep=2pt] at (-0.2,-0.02) {$a_{k-1}$}; \node[anchor=north,inner sep=2pt] at (1.25,-0.02) {$b_{k-1}$}; \node[anchor=north,inner sep=2pt] at (2.8,-0.02) {$a_k$};
  \end{scope}
\end{scope}
\begin{scope}[yshift=-10.9cm]
  \node[anchor=west] at (-1.3,0.8) {(c)};
  \draw[red,very thick] (0,-0.8) -- (1.4,0.8); \node[red,anchor=north] at (0,-0.85) {$x'$};
  \draw[blue,thick] (1.4,-0.8) -- (0.85,-0.17); \draw[blue,thick] (0.55,0.17) -- (0,0.8); \node[blue,anchor=north] at (1.4,-0.85) {$x$};
  \node at (2.3,0) {$=\ -q^{-3/2}$};
  \begin{scope}[xshift=3.6cm]
    \draw[red,very thick] (0,-0.8) .. controls (0,-0.1) and (1.4,-0.1) .. (1.4,-0.8);
    \draw[blue,thick] (0,0.8) .. controls (0,0.1) and (1.4,0.1) .. (1.4,0.8);
    \node[anchor=west,font=\footnotesize] at (1.5,0) {$u_{\mathbf 1}$};
  \end{scope}
  \node at (6.3,0) {$+\ q^{1/2}$};
  \begin{scope}[xshift=7.4cm]
    \draw[red,very thick] (0,-0.8) -- (0.7,-0.3); \draw[blue,thick] (1.4,-0.8) -- (0.7,-0.3);
    \draw[thick,green!50!black] (0.7,-0.3) -- (0.7,0.3); \node[green!50!black,anchor=west,font=\footnotesize] at (0.75,0) {adj};
    \draw[red,very thick] (0.7,0.3) -- (1.4,0.8); \draw[blue,thick] (0.7,0.3) -- (0,0.8);
    \node[anchor=west,font=\footnotesize] at (1.5,0) {$u_{\mathrm{adj}}$};
  \end{scope}
\end{scope}
\end{tikzpicture}

%% file: figs/fig_twist_tikz.tex
\begin{tikzpicture}[x=1cm,y=1cm,every node/.style={font=\small},lab/.style={anchor=north,inner sep=1.5pt}]
\node[anchor=east] at (-1.0,-0.2) {before the move:};
\begin{scope}
\draw[red,very thick] (1.4,0) .. controls (1.4,0.9) and (0.3,0.9) .. (0.3,0.25);
\draw[red,very thick] (0.3,0.25) -- (0.3,0) -- (-0.3,-1.1);
\draw[thick] (0,0)--(0.15,0); \draw[thick] (0.45,0)--(4.2,0);
\node[red,anchor=north] at (-0.3,-1.15) {$x'$}; \fill (1.4,0) circle (1.6pt); \node[anchor=south west,inner sep=1pt] at (1.45,0.02) {$V$};
\draw[blue,thick] (2.4,-1.1)--(3.0,0); \node[blue,anchor=north] at (2.4,-1.15) {$x$}; \fill[blue] (3.0,0) circle (1.6pt);
\node[lab] at (0.85,-0.03) {$a_{k-1}$}; \node[lab] at (2.1,-0.03) {$a_k$}; \node[lab] at (3.7,-0.03) {$b_k$};
\node[anchor=north,align=center,font=\small] at (2.1,-1.5) {$x'$ from above\quad ($Y_x$)};
\end{scope}
\begin{scope}[xshift=7.4cm]
\draw[thick] (0,0)--(4.2,0);
\draw[red,very thick] (1.4,0) .. controls (1.4,0.9) and (0.3,0.9) .. (0.3,0.25);
\draw[red,very thick] (0.3,0.25) -- (0.3,0.12); \draw[red,very thick] (0.23,-0.12) -- (-0.3,-1.1);
\node[red,anchor=north] at (-0.3,-1.15) {$x'$}; \fill (1.4,0) circle (1.6pt); \node[anchor=south west,inner sep=1pt] at (1.45,0.02) {$V$};
\draw[blue,thick] (2.4,-1.1)--(3.0,0); \node[blue,anchor=north] at (2.4,-1.15) {$x$}; \fill[blue] (3.0,0) circle (1.6pt);
\node[lab] at (0.85,-0.03) {$a_{k-1}$}; \node[lab] at (2.1,-0.03) {$a_k$}; \node[lab] at (3.7,-0.03) {$b_k$};
\node[anchor=north,align=center,font=\small] at (2.1,-1.5) {$x'$ from below\quad ($Y'_x$): factor $\theta_{a_k}/(\theta_{a_{k-1}}\theta_{x'})$};
\end{scope}

\begin{scope}[yshift=-3.6cm]
\node[anchor=east] at (-1.0,-0.2) {after the move:};
\begin{scope}
\draw[red,very thick] (2.6,0) .. controls (2.6,0.9) and (1.75,0.9) .. (1.75,0.25);
\draw[red,very thick] (1.75,0.25) -- (1.75,0) -- (1.15,-1.1);
\draw[thick] (0,0)--(1.6,0); \draw[thick] (1.9,0)--(4.2,0);
\node[red,anchor=north] at (1.15,-1.15) {$x'$}; \fill (2.6,0) circle (1.6pt); \node[anchor=south west,inner sep=1pt] at (2.65,0.02) {$V$};
\draw[blue,thick] (0.5,-1.1)--(1.1,0); \node[blue,anchor=north] at (0.5,-1.15) {$x$}; \fill[blue] (1.1,0) circle (1.6pt);
\node[lab] at (0.5,-0.03) {$a_{k-1}$}; \node[lab] at (2.25,-0.03) {$b_{k-1}$}; \node[lab] at (3.5,-0.03) {$b_k$};
\node[anchor=north,align=center,font=\small] at (2.1,-1.5) {$x'$ from above\quad ($Y_x$)};
\end{scope}
\begin{scope}[xshift=7.4cm]
\draw[thick] (0,0)--(4.2,0);
\draw[red,very thick] (2.6,0) .. controls (2.6,0.9) and (1.75,0.9) .. (1.75,0.25);
\draw[red,very thick] (1.75,0.25) -- (1.75,0.12); \draw[red,very thick] (1.68,-0.12) -- (1.15,-1.1);
\node[red,anchor=north] at (1.15,-1.15) {$x'$}; \fill (2.6,0) circle (1.6pt); \node[anchor=south west,inner sep=1pt] at (2.65,0.02) {$V$};
\draw[blue,thick] (0.5,-1.1)--(1.1,0); \node[blue,anchor=north] at (0.5,-1.15) {$x$}; \fill[blue] (1.1,0) circle (1.6pt);
\node[lab] at (0.5,-0.03) {$a_{k-1}$}; \node[lab] at (2.25,-0.03) {$b_{k-1}$}; \node[lab] at (3.5,-0.03) {$b_k$};
\node[anchor=north,align=center,font=\small] at (2.1,-1.5) {$x'$ from below\quad ($Y'_x$): factor $\theta_{b_k}/(\theta_{b_{k-1}}\theta_{x'})$};
\end{scope}
\end{scope}
\end{tikzpicture}

%% file: figs/fig_proj_tikz.tex
\begin{tikzpicture}[x=1cm,y=1cm,every node/.style={font=\small},w/.style={red!70!black,thick}]
\begin{scope}
\node at (0.3,0.2) {$P^{\mathbf 1}\ =\ \dfrac{1}{d_\wl}$};
\begin{scope}[xshift=1.5cm]
\draw[w] (0,-0.8) .. controls (0,-0.15) and (1.0,-0.15) .. (1.0,-0.8);
\draw[w] (0,1.2) .. controls (0,0.55) and (1.0,0.55) .. (1.0,1.2);
\end{scope}
\node at (3.4,0.2) {$=\ \dfrac{e}{d_\wl}$,};
\node at (6.0,0.2) {$P^{\mathrm{adj}}\ =\ \dfrac{d_{\mathrm{adj}}}{\theta(\wl,\wl,\mathrm{adj})}$};
\begin{scope}[xshift=8.2cm]
\draw[w] (0,-0.8) .. controls (0,-0.3) and (0.5,-0.3) .. (0.5,-0.1);
\draw[w] (1.0,-0.8) .. controls (1.0,-0.3) and (0.5,-0.3) .. (0.5,-0.1);
\draw[green!50!black,thick] (0.5,-0.1) -- (0.5,0.5); \node[green!50!black,anchor=west,font=\footnotesize] at (0.55,0.2) {adj};
\draw[w] (0,1.2) .. controls (0,0.7) and (0.5,0.7) .. (0.5,0.5);
\draw[w] (1.0,1.2) .. controls (1.0,0.7) and (0.5,0.7) .. (0.5,0.5);
\fill (0.5,-0.1) circle (1.4pt); \fill (0.5,0.5) circle (1.4pt);
\end{scope}
\node[anchor=west] at (9.7,0.2) {$,\qquad P^{\mathbf 1}+P^{\mathrm{adj}}\ =$};
\begin{scope}[xshift=13.1cm]
\draw[w] (0,-0.8)--(0,1.2); \draw[w] (1.0,-0.8)--(1.0,1.2);
\end{scope}
\node[anchor=west] at (14.3,0.2) {$=\,n$};
\end{scope}
\end{tikzpicture}

%% file: figs/fig5_tikz.tex
\begin{tikzpicture}[
  font=\small,
  hbox/.style={draw=blue!70!black, fill=blue!8, minimum width=0.68cm, minimum height=0.66cm, inner sep=1pt},
  sbox/.style={draw=red!80!black, fill=red!8, minimum width=0.68cm, minimum height=0.66cm, inner sep=1pt},
  wall/.style={circle, fill=red!85!black, draw=red!85!black, inner sep=0pt, minimum size=0.17cm},
  empty/.style={circle, draw=gray!70!black, thick, fill=white, inner sep=0pt, minimum size=0.17cm},
  pics/half/.style={code={
    \draw[gray!70!black, thick, fill=white] (0,0) circle (0.085cm);
    \fill[red!85!black] (0,-0.085) arc (-90:-270:0.085cm) -- cycle;
    \draw[red!85!black] (0,-0.085) arc (-90:-270:0.085cm);
  }},
]
  \draw[gray!70] (-0.45,0) -- (7.95,0);
  \node[hbox] (am2) at (0,0) {$a_{j-2}$};
  \node[empty] at (0.75,0) {};
  \node[hbox] (am1) at (1.5,0) {$a_{j-1}$};
  \pic at (2.25,0) {half};
  \node[hbox] (a0) at (3,0) {$a_{j}$};
  \node[wall] at (3.75,0) {};
  \node[sbox] (c) at (4.5,0) {$c_{j}$};
  \pic at (5.25,0) {half};
  \node[hbox] (ap1) at (6,0) {$a_{j+1}$};
  \node[empty] at (6.75,0) {};
  \node[hbox] (ap2) at (7.5,0) {$a_{j+2}$};
  \node[anchor=south] at (2.3,0.4) {site $j$};
  \node[anchor=south, red!85!black] at (3.75,0.4) {$\wl$ (always)};
  \node[anchor=south] at (5.75,0.4) {site $j{+}1$};
  \node[anchor=north, red!85!black] at (3.75,-0.42) {strand link, $|c_j - a_j| = 1$};
  \draw[decorate, decoration={brace, amplitude=5pt}] (2.65,0.95) -- (6.35,0.95);
  \node[anchor=south, align=center] at (4.5,1.15)
    {right pair: (strand, site $j{+}1$)\\ $e_{\mathrm{R}},\ n_{\mathrm{R}}$; channel from ends $(a_j, a_{j+1})$};
  \draw[decorate, decoration={brace, mirror, amplitude=5pt}] (1.15,-1.0) -- (4.85,-1.0);
  \node[anchor=north, align=center] at (3,-1.2)
    {left pair: (site $j$, strand)\\ $e_{\mathrm{L}},\ n_{\mathrm{L}}$; channel from ends $(a_{j-1}, c_j)$};
  \node[wall] at (8.6,0.55) {};
  \node[anchor=west] at (8.85,0.55) {domain wall ($\wl$)};
  \pic at (8.6,0) {half};
  \node[anchor=west] at (8.85,0) {wall or empty, depending on the heights};
  \node[empty] at (8.6,-0.55) {};
  \node[anchor=west] at (8.85,-0.55) {empty};
\end{tikzpicture}

%% file: figs/fig15_tikz.tex
\begin{tikzpicture}[x=1cm,y=1cm,font=\small,
  strand/.style={draw=red!80!black,line width=1pt},
  sbox/.style={draw=blue!60!black,fill=blue!8,minimum size=0.55cm,inner sep=1pt},
  fbox/.style={draw=gray!60,line width=0.6pt},
  >={Stealth[length=2mm]}]
\node[anchor=west] at (0,0) {(a) slot states};
\node at (0.4,-1.5) {$|0\rangle$}; \node at (0.9,-1.5) {$=$}; \draw[fbox] (1.1,-1.85) rectangle (1.5,-1.15); \node at (1.75,-1.5) {,};
\node at (2.35,-1.5) {$|{+}\rangle$}; \node at (2.85,-1.5) {$=$}; \draw[fbox] (3.05,-1.85) rectangle (3.45,-1.15); \draw[strand,->] (3.25,-1.85) -- (3.25,-1.15); \node at (3.7,-1.5) {,};
\node at (4.3,-1.5) {$|{-}\rangle$}; \node at (4.8,-1.5) {$=$}; \draw[fbox] (5.0,-1.85) rectangle (5.4,-1.15); \draw[strand,->] (5.2,-1.15) -- (5.2,-1.85);
\node[anchor=north,align=center,font=\footnotesize] at (2.9,-2.0) {empty slot, strand up, strand down\\[3pt]
  $S^z=\sum_i s_i$ conserved;\\ a seam slot is never $0$};
\node[anchor=west] at (6.7,0) {(b) the six face moves $h=\sum_{\mathrm{moves}} w\times(\mathrm{move})$, time upwards};
\def\bw{0.85}\def\bh{1.1}
\foreach \i/\lab/\wt in {0/empty/{w_\varnothing},1/one strand/{w_\parallel},2/two strands/{w_{=}},3/hop/{w_{\curvearrowright}},4/{cup, cap}/{w_{\cup\cap}},5/{$e=\mathrm{cap}\circ\mathrm{cup}$}/{w_e}}{
  \coordinate (b\i) at ({6.8+1.85*\i},-1.5);
  \draw[fbox] ($(b\i)+(-\bw/2,-\bh/2)$) rectangle ($(b\i)+(\bw/2,\bh/2)$);
  \node at ($(b\i)+(0,0.85)$) {$\wt$};
  \node[anchor=north] at ($(b\i)+(0,-0.65)$) {\lab};
}
\draw[strand] ($(b1)+(-0.2,-\bh/2)$) -- ($(b1)+(-0.2,\bh/2)$);
\draw[strand] ($(b2)+(-0.22,-\bh/2)$) -- ($(b2)+(-0.22,\bh/2)$);
\draw[strand] ($(b2)+(0.22,-\bh/2)$) -- ($(b2)+(0.22,\bh/2)$);
\draw[strand] ($(b3)+(-0.25,-\bh/2)$) .. controls ($(b3)+(-0.25,0)$) and ($(b3)+(0.25,0)$) .. ($(b3)+(0.25,\bh/2)$);
\draw[strand] ($(b4)+(-0.25,\bh/2)$) .. controls ($(b4)+(-0.25,0.05)$) and ($(b4)+(0.25,0.05)$) .. ($(b4)+(0.25,\bh/2)$);
\draw[strand,dashed] ($(b4)+(-0.25,-\bh/2)$) .. controls ($(b4)+(-0.25,-0.05)$) and ($(b4)+(0.25,-0.05)$) .. ($(b4)+(0.25,-\bh/2)$);
\draw[strand] ($(b5)+(-0.25,\bh/2)$) .. controls ($(b5)+(-0.25,0.05)$) and ($(b5)+(0.25,0.05)$) .. ($(b5)+(0.25,\bh/2)$);
\draw[strand] ($(b5)+(-0.25,-\bh/2)$) .. controls ($(b5)+(-0.25,-0.05)$) and ($(b5)+(0.25,-0.05)$) .. ($(b5)+(0.25,-\bh/2)$);
\node[anchor=west] at (0,-4.1) {(c) oriented turns and the loop weight};
\begin{scope}[shift={(0.8,-5.3)}]
  \node at (0,-0.45) {$q^{1/2}$};
  \draw[strand,->] (-0.4,0.5) .. controls (-0.4,-0.2) and (0.4,-0.2) .. (0.4,0.5) ;
  \node[font=\scriptsize] at (-0.33,0.65) {$+$};
  \node[font=\scriptsize] at (0.38,0.65) {$-$};
  \node at (1.0,0.2) {$+$};
  \node at (2.0,-0.45) {$q^{-1/2}$};
  \draw[strand,->] (2.4,0.5) .. controls (2.4,-0.2) and (1.6,-0.2) .. (1.6,0.5);
  \node[font=\scriptsize] at (1.65,0.65) {$-$};
  \node[font=\scriptsize] at (2.36,0.65) {$+$};
  \node[anchor=west] at (2.75,0.2) {$=\;|c\rangle$};
\end{scope}
\node[anchor=west,align=center,font=\footnotesize] at (0.0,-6.95) {
  $\begin{array}{rcl}|c\rangle &=& q^{1/2}|{+}{-}\rangle \;+\; q^{-1/2}|{-}{+}\rangle,\\[1pt]
   \langle\tilde c| &=& q^{1/2}\langle{+}{-}| \;+\; q^{-1/2}\langle{-}{+}|\end{array}$\\[2pt]
  a turn by $\pm\pi$ (cup or cap) carries\\ $q^{\pm1/2}$ according to its orientation};
\begin{scope}[shift={(8.3,-5.3)}]
  \draw[strand] (0,0) circle (0.5);
  \draw[strand,->] (0.5,0) arc (0:60:0.5);
  \node at (0,-0.85) {$q^{+1}$};
  \node at (1.0,0) {$+$};
  \draw[strand] (2.0,0) circle (0.5);
  \draw[strand,->] (2.5,0) arc (0:-60:0.5);
  \node at (2.0,-0.85) {$q^{-1}$};
  \node[anchor=west] at (2.75,0) {$=\;q+q^{-1}=n$};
\end{scope}
\node[anchor=west,align=center,font=\footnotesize] at (7.4,-7.35) {
  $\langle\tilde c|c\rangle = q+q^{-1} = n = 2\cos\frac{\pi}{p}$, so $e^2=n\,e$:\\[2pt]
  a contractible loop turns by $\pm2\pi$ and, summed\\ over its two orientations, has weight $n$};
\end{tikzpicture}

%% file: figs/fig16_tikz.tex
\begin{tikzpicture}[x=1cm,y=1cm,every node/.style={font=\small},
  slot/.style={draw=blue!60!black,fill=blue!8,minimum width=0.6cm,minimum height=0.6cm,inner sep=1pt,font=\small},
  sslot/.style={draw=red!70!black,fill=red!10,minimum width=0.6cm,minimum height=0.6cm,inner sep=1pt,font=\small},
  dslot/.style={draw=blue!60!black,densely dashed,fill=blue!8,minimum width=0.6cm,minimum height=0.6cm,inner sep=1pt,font=\small}]
\begin{scope}
\node[anchor=west] at (-0.3,1.3) {(a) the single-seam ring: site slots $a_j$ and seam slot $c$ (example)};
\newcommand{\wslot}[2]{\draw[gray!60] (#1-0.3,-0.3) rectangle (#1+0.3,0.3); \node[anchor=north,font=\footnotesize] at (#1,-0.33) {$#2$};}
\newcommand{\upst}[1]{\draw[red!70!black,thick,-{Stealth[length=1.6mm]}] (#1,-0.26) -- (#1,0.26);}
\newcommand{\dnst}[1]{\draw[red!70!black,thick,-{Stealth[length=1.6mm]}] (#1,0.26) -- (#1,-0.26);}
\wslot{0}{a_0}\upst{0}
\wslot{0.8}{a_1}
\wslot{1.6}{a_2}\dnst{1.6}
\wslot{2.4}{a_3}
\node at (3.3,0) {$\cdots$};
\wslot{4.2}{a_{L-2}}\upst{4.2}
\wslot{5.3}{a_{L-1}}
\draw[red!70!black,thick] (6.2-0.45,-0.3) rectangle (6.2+0.45,0.3); \node[anchor=north,font=\footnotesize,text=red!70!black] at (6.2,-0.33) {$c$}; \upst{5.97} \node[font=\scriptsize] at (6.2,0) {or}; \dnst{6.43}
\draw[gray!60,densely dashed] (7.5-0.3,-0.3) rectangle (7.5+0.3,0.3); \node[anchor=north,font=\footnotesize] at (7.5,-0.33) {$a_0$}; \upst{7.5}
\foreach \x in {0.4,1.2,2.0,4.75}{\draw[decorate,decoration={brace,amplitude=3pt}] (\x-0.35,0.42) -- (\x+0.35,0.42); \node[anchor=south,font=\footnotesize] at (\x,0.5) {$h$};}
\draw[decorate,decoration={brace,amplitude=3pt}] (4.95,0.42) -- (7.85,0.42); \node[anchor=south,font=\footnotesize,text=red!70!black] at (6.4,0.5) {$h^{\mathrm{seam}}+K\Phi_{\mathrm{cl}}$};
\draw[dashed] (6.9,-0.55) -- (6.9,0.55);
\node[anchor=north west,text width=8.4cm,align=justify,font=\footnotesize] at (-0.3,-0.75) {Empty box $=$ no strand, arrow $=$ oriented strand; the seam slot $c$ always carries a strand, of either orientation; $a_0$ is repeated (periodic ring); dashed line: the cut, a strand crossing it picks up $e^{\pm i\phi}$.};
\end{scope}
\begin{scope}[xshift=10.5cm,yshift=0.6cm]
\node[anchor=west] at (-0.3,0.7) {(b) the three-slot seam term $h^{\mathrm{seam}}=B_{cy}^{-1}\,h_{xc}\,B_{cy}$ (time upwards)};
\def\xl{0.6}\def\xc{1.4}\def\xr{2.2}\def\yb{-3.6}\def\yt{-0.0}
\draw[blue!60!black,thick] (\xl,\yb) -- (\xl,\yt);
\draw[blue!60!black,thick] (\xr,\yb) -- (\xr,-3.1);
\draw[blue!60!black,thick] (\xr,-3.1) .. controls (\xr,-2.6) and (\xc,-2.55) .. (\xc,-2.1);
\draw[red!70!black,line width=2pt] (\xc,\yb) -- (\xc,-3.1);
\draw[white,line width=5pt] (\xc,-3.1) .. controls (\xc,-2.6) and (\xr,-2.55) .. (\xr,-2.1);
\draw[red!70!black,line width=2pt] (\xc,-3.1) .. controls (\xc,-2.6) and (\xr,-2.55) .. (\xr,-2.1);
\fill[blue!10] (\xl-0.3,-1.95) rectangle (\xc+0.3,-1.45); \draw[blue!60!black] (\xl-0.3,-1.95) rectangle (\xc+0.3,-1.45); \node at (1.0,-1.7) {$h$};
\draw[blue!60!black,thick] (\xc,-2.1) -- (\xc,-1.95); \draw[blue!60!black,thick] (\xc,-1.45) -- (\xc,-1.3);
\draw[red!70!black,line width=2pt] (\xr,-2.1) -- (\xr,-1.3);
\draw[blue!60!black,thick] (\xc,-1.3) .. controls (\xc,-0.85) and (\xr,-0.8) .. (\xr,-0.35);
\draw[white,line width=5pt] (\xr,-1.3) .. controls (\xr,-0.85) and (\xc,-0.8) .. (\xc,-0.35);
\draw[red!70!black,line width=2pt] (\xr,-1.3) .. controls (\xr,-0.85) and (\xc,-0.8) .. (\xc,-0.35);
\draw[blue!60!black,thick] (\xr,-0.35) -- (\xr,\yt); \draw[red!70!black,line width=2pt] (\xc,-0.35) -- (\xc,\yt);
\node[anchor=north,font=\footnotesize] at (\xl,\yb-0.05) {$x=a_{L-1}$}; \node[anchor=north,font=\footnotesize,text=red!70!black] at (\xc,\yb-0.05) {$c$}; \node[anchor=north,font=\footnotesize] at (\xr,\yb-0.05) {$y=a_0$};
\node[anchor=north west,text width=5.2cm,align=left,font=\footnotesize] at (2.9,-0.3) {$B_{cy}^{-1}$: braided back, again in front of $y$; the two crossings cancel.\\[6pt] $h_{xc}$: the ordinary pair term on the two physical contents.\\[6pt] $B_{cy}$: strand braided in front of $y$.\\[6pt] Net effect: $x$ and $c$ may change, $y$ is a spectator.};
\end{scope}
\begin{scope}[yshift=-5.0cm]
\node[anchor=west] at (-0.3,0.7) {(c) the braid $B$ on a pair of slots (Kauffman skein)};
\foreach \x in {0,2.0,4.0,6.0,8.0}{\draw[gray!60] (\x,-2.4) rectangle (\x+0.9,-1.0);}
\draw[red!70!black,thick] (2.0+0.25,-2.4) .. controls (2.0+0.25,-1.7) and (2.0+0.65,-1.7) .. (2.0+0.65,-1.0);
\draw[red!70!black,thick] (4.0+0.25,-2.4) -- (4.0+0.65,-1.0); \draw[white,line width=4pt] (4.0+0.65,-2.4) -- (4.0+0.25,-1.0); \draw[red!70!black,thick] (4.0+0.65,-2.4) -- (4.0+0.25,-1.0);
\node at (1.45,-1.7) {,}; \node at (3.45,-1.7) {,};
\node at (5.45,-1.7) {$=\,A$};
\draw[red!70!black,thick] (6.0+0.25,-2.4) -- (6.0+0.25,-1.0); \draw[red!70!black,thick] (6.0+0.65,-2.4) -- (6.0+0.65,-1.0);
\node at (7.45,-1.7) {$+\,A^{-1}$};
\draw[red!70!black,thick] (8.0+0.25,-2.4) .. controls (8.0+0.25,-1.9) and (8.0+0.65,-1.9) .. (8.0+0.65,-2.4);
\draw[red!70!black,thick] (8.0+0.25,-1.0) .. controls (8.0+0.25,-1.5) and (8.0+0.65,-1.5) .. (8.0+0.65,-1.0);
\node[anchor=north,font=\footnotesize] at (0.45,-2.5) {$B=\mathbf 1$}; \node[anchor=north,font=\footnotesize] at (2.45,-2.5) {$B=P$};
\node[anchor=north,font=\footnotesize] at (6.45,-2.5) {$B=A\,\mathbf 1+A^{-1}e$};
\node[anchor=north,font=\footnotesize] at (4.5,-3.0) {$A=iq^{1/2},\qquad n=-A^2-A^{-2}=q+q^{-1}$};
\node[anchor=north west,text width=9.2cm,align=justify,font=\footnotesize] at (-0.3,-3.5) {The relative phase $A$ between the two-strand and the one-strand block is the skein image of the $R$-symbols of the RSOS seam; it is fixed by requiring the closed two-seam ring to reproduce the bulk vacuum exactly (the seam is topological), and it matters because loops that cross the seam once pick it up.};
\end{scope}
\begin{scope}[xshift=10.5cm,yshift=-5.6cm]
\node[anchor=west] at (-0.3,1.3) {(d) twist $\phi$ and non-contractible loops (space horizontal, periodic)};
\draw[gray!70] (-0.3,-1.1) rectangle (8.9,0.8);
\draw[red!70!black,line width=2pt] (7.7,-1.1) -- (7.7,0.8); \node[anchor=south,font=\footnotesize,text=red!70!black] at (7.7,0.8) {seam};
\draw[dashed] (8.2,-1.1) -- (8.2,0.8); \node[anchor=north,font=\footnotesize] at (8.2,-1.1) {cut};
\node[anchor=north west,text width=9.2cm,align=justify,font=\footnotesize] at (-0.3,-1.5) {A twist $\phi$ multiplies every strand crossing the cut by $e^{\pm i\phi}$; contractible loops cross it an even number of times and keep $n$, a non-contractible loop crosses it once and gets $e^{i\phi}+e^{-i\phi}=2\cos\phi$.};
\draw[red!70!black,thick] (-0.3,0.05) -- (8.9,0.05); \node[anchor=south,font=\footnotesize] at (3.6,0.1) {non-contractible (crosses the cut once): $2\cos\phi$};
\draw[red!70!black,thick] (1.3,-0.6) ellipse (0.9 and 0.32); \node[anchor=west,font=\footnotesize] at (2.4,-0.6) {contractible: $q+q^{-1}=n$};
\end{scope}
\end{tikzpicture}